\documentclass[a4paper,11pt]{article}
\pdfoutput=1 
             
\usepackage{jheppub_mod} 

\usepackage{amsmath}
\usepackage{esvect} 
\usepackage{tikz}
\usetikzlibrary{arrows}
\usetikzlibrary{positioning}
\usepackage{amsfonts}
\usepackage{amsthm}
\usepackage{epic}
\usepackage{mathtools}
\usepackage{url}
\usetikzlibrary{calc,patterns,angles,quotes}
\usepackage{graphicx}
\usepackage{float}
\usepackage{amssymb}
\usepackage{caption}
\usepackage{subcaption}
\usepackage{xcolor}
\usepackage{bm}
\usepackage{multicol} 
\usepackage{seqsplit}
\usepackage{mathrsfs}
\usepackage{comment}
\usetikzlibrary{decorations.pathmorphing}
\usepackage{tikz-cd}
\numberwithin{equation}{section}
\usepackage{braket}
\usepackage{mathtools}
\allowdisplaybreaks[4]

\usetikzlibrary{decorations.pathmorphing}
\usetikzlibrary{arrows.meta}
\usetikzlibrary{decorations.markings}

\definecolor{darkpastelgreen}{rgb}{0.01, 0.75, 0.24 }
\definecolor{hooker\'sgreen}{rgb}{0.0, 0.44, 0.0}
\definecolor{indiagreen}{rgb}{0.07, 0.53, 0.03}
\definecolor{islamicgreen}{rgb}{0.0, 0.56, 0.0}

\title{(A)dS$\mathbf{_4}$ in Bondi gauge}

\author{Aaron Poole,}
\author{Kostas Skenderis}
\author{and Marika Taylor}

\affiliation{Mathematical Sciences and STAG Research Centre, University of Southampton \\
Highfield, Southampton SO17 1BJ, United Kingdom.}
\emailAdd{a.poole@soton.ac.uk}
\emailAdd{k.skenderis@soton.ac.uk}
\emailAdd{m.m.taylor@soton.ac.uk}

\abstract{We obtain the general asymptotic solutions of Einstein gravity with or without cosmological constant in Bondi gauge. The Bondi gauge was originally introduced in the context of gravitational radiation in asymptotically flat gravity. In the original work, initial conditions were prescribed at a null hypersurface and the Einstein equations were shown to take a nested form, which may be used to explicitly 
 integrate them asymptotically. We streamline the derivation of the general asymptotic solution in the asymptotically flat case, and derive the most general asymptotic solutions for the case of non-zero cosmological constant of either sign (asymptotically locally AdS and dS solutions).  With non-zero cosmological constant, we present integration schemes which rely on either prescribing data on the conformal boundary or on a null hypersurface and part of the conformal boundary. We explicitly work out the transformation to Fefferman-Graham gauge and identity how to extract the holographic data  directly in Bondi coordinates. We illustrate the discussion with a number of examples and show that for asymptotically AdS${}_4$ spacetimes the Bondi mass is constant.}

\begin{document}

\maketitle
\flushbottom


\section{Introduction and summary of results}

The Bondi metric was introduced in the seminal works of Bondi, Sachs and others on gravitational radiation \cite{Bondi:1962px, Sachs:1962wk}. While all gauges are equivalent
a convenient choice of a coordinate system may bring in simplifications and make the physical properties of spacetimes most transparent. In the case of gravitational radiation 
the objective was to examine the behaviour of the gravitational field far from the isolated object generating the radiation, and to obtain and use asymptotic solutions of Einstein equations to characterise radiating spacetime. 

In asymptotically flat gravity, gravitational waves travel to future null infinity and the task becomes that of obtaining asymptotic solutions near future null infinity. It was shown in \cite{Bondi:1962px,Sachs:1962wk} that in Bondi gauge the Einstein equations take a nested form and they  can be readily integrated near null infinity. If one specifies initial data on an outgoing null hypersurface then the Einstein equations tell us how to propagate this data forward in time to a nearby outgoing null hypersurface. The asymptotic solution involves  a number of data that are not determined by the asymptotic analysis alone: such data will be fixed in any given exact solution of the field equations. This undetermined data consists of
a scalar function (the Bondi mass aspect);  a vector (the angular momentum aspect) and a tensor (the Bondi news). The mass and angular momentum aspects integrated over a cut at null infinity define the total mass and total angular momentum\footnote{This definition of angular momentum suffers from supertranslation ambiguities. This issue will not play a role here.} of the system at that time  and the news tensor controls how these quantities change in time.  In particular, one can show that if the news tensor vanishes (and the matter stress energy tensor goes to zero fast enough at future null infinity)  the total mass is constant, while if the news tensor is non-vanishing the total Bondi mass monotonically decreases in time capturing the fact that the system loses mass by emitting gravitational waves.

In the presence of a cosmological constant the nature of infinity changes: with negative cosmological constant conformal infinity is timelike while with positive cosmological constant infinity is spacelike. As there is no null infinity in either case one may question whether analyzing Einstein equations with non-zero cosmological constant in Bondi gauge would be useful. There are however several reasons to do this. In the case of a negative cosmological constant, as we review below, asymptotic solutions in Fefferman-Graham gauge \cite{Fefferman:1985zza} have a clear holographic meaning \cite{deHaro:2000vlm} and one would like to understand the holographic meaning of the data in Bondi gauge. This may then be used to get insight into a possible holographic structure of asymptotically flat gravity. In addition, Bondi-like gauges where Einstein equations take a nested form have been in the used already in the holography literature (see \cite{Chesler:2013lia} and references therein) and it would be desirable to understand how to extract the holographic data directly in this gauge. Furthermore, an analogue of  Bondi mass with many interesting properties has already been defined for a class of asymptotically locally AdS spacetimes \cite{Bakas:2014kfa} and one would like to understand whether such a quantity exists more generally in asymptotically locally AdS spacetimes. 

In the case of positive cosmological constant such results are needed even more urgently: current observations indicate that we live in a Universe with a positive cosmological constant and we have also observed gravitational waves. Yet a satisfactory discussion of gravitational waves in de Sitter spacetime is still missing. Recent works addressing these issues include \cite{Ashtekar:2014zfa, Szabados:2015wqa, Ashtekar:2015lla,  Ashtekar:2015ooa, Ashtekar:2015lxa,  He:2015wfa, Chrusciel:2016oux, Saw:2016isu, Saw:2017amv, He:2018ikd, Szabados:2018erf}.

With negative cosmological constant, the appropriate boundary conditions are to fix a conformal class of metrics on the conformal boundary, and a natural coordinate system to use is Gaussian normal coordinates centred at the conformal boundary, the Fefferman-Graham gauge \cite{Fefferman:1985zza}. One may then obtain the general asymptotic solution to Einstein equations by treating the radial coordinate as a small parameter. The Einstein equations become algebraic in this gauge (i.e. they are solved by algebraic manipulation rather than by integrating differential equations) and the pieces of data needed that are left undetermined by the asymptotic analysis are the conformal class and a covariantly conserved symmetric traceless tensor (in even dimensions, in odd dimensions the tensor has a trace). In holography, the boundary metric is the background for the dual CFT and the tensor is (the quantum expectation value) of the energy momentum tensor \cite{Henningson:1998gx,deHaro:2000vlm}. The same tensor can be used to obtain the bulk conserved changes when the spacetime possesses asymptotic Killing vectors \cite{Papadimitriou:2005ii}. 

With positive cosmological constant, one may similarly use Gaussian normal coordinates centred at future infinity and work out the asymptotic expansion \cite{Starobinsky:1982mr} and the data are again a conformal class of metrics and a covariantly conserved symmetric traceless tensor. Actually, the asymptotic solutions for positive and negative cosmological are related by simple analytic continuation \cite{Skenderis:2002wp}.

With non-zero cosmological constant, one may  foliate infinity with null hypersurfaces, now ending either at timelike infinity (negative $\Lambda$) or spacelike infinity (positive $\Lambda$).
The structure of the Einstein equations in Bondi gauge and in the presence of a cosmological constant is very similar to that with no cosmological constant.  To explain the similarities and differences relative to the case of no cosmological constant we first briefly review the latter.

In this paper for simplicity we restrict ourselves to $d=4$ and axial and reflection symmetry.  It would be straightforward but tedious to relax these conditions.
The metric in Bondi gauge (for any cosmological constant) then takes the form
\begin{align} \label{intro:Bondi}
\begin{split}
ds^2=&-(Wr^{2}e^{2\beta}-U^2r^2e^{2\gamma})du^2-2e^{2\beta}dudr-\\
&2Ur^2e^{2\gamma}dud\theta+r^2(e^{2\gamma}d\theta^2+e^{-2\gamma}\sin^2\theta d\phi^2).
\end{split}
\end{align}
Here $u$ is retarded Bondi time, $r$ is a radial coordinate and $\theta, \phi$ parametrise the transverse space (which is topologically an $S^2$) and  $W, U, \beta, \gamma$ are functions to be determined by solving Einstein equations.

We find it useful also to define the coordinate $z=1/r$, which brings infinity to $z=0$. Inserting (\ref{intro:Bondi}) in the Einstein equations leads to four main equations and three supplementary conditions. One can then show that the coefficients appearing in these equations are regular as $z \to 0$.
This means that they admit asymptotic solutions with $W, U, \beta, \gamma$ being regular around $z=0$ and one can obtain the asymptotic solutions by successively differentiating the equations w.r.t. $z$, setting $z=0$ and solving the resulting equations (as was done for AdS gravity in Fefferman-Graham gauge in \cite{deHaro:2000vlm}). In all cases we solve the resulting equations in the most general way, so we obtain the most general asymptotic solutions of Einstein equations with the only assumption being that the functions $W, U, \beta, \gamma$  are four times differentiable. 
   
With no cosmological constant,  one provides as initial condition the value of $\gamma$ at a null hypersurface $u=u_0 = const$. Imposing the ``out-going gauge condition'' 
$\gamma_{,zz}=0$ \footnote{Indices after comma indicate differentiation, i.e. $\gamma_{,z}=\partial \gamma/\partial z$ etc.} (as in \cite{Bondi:1962px}) one finds that all functions admit regular Taylor expansions around $z=0$, and one can iteratively solve for all coefficients, except that the coefficient $W_{,zzz}, U_{,zzz}, \gamma_{,z}$ are left undetermined by the asymptotic analysis (apart from two equations that link their $u$ and $\theta$ derivatives).  These three functions are essentially the Bondi mass aspect, the angular momentum aspect and the Bondi news mentioned earlier and the relation of their derivatives is linked to the monotonicity of the total Bondi mass. This data is then enough to determine $\gamma_{,u}$ which allows us to obtain $\gamma$ at $u=u_0 + \delta u$ and thus continue the iterative construction of the solution. Note that if one is to relax the ``out-going gauge condition'' then the solution will also contain logarithmic terms in $z$ \cite{Chrusciel:1993hx}.
 
In the presence of a cosmological constant (with any sign), three of the four main equations can be solved in exactly the same way as in the $\Lambda=0$ case but the fourth equation couples the coefficients in such a way that the integration scheme we used for $\Lambda=0$ does not work any more. We have found however two alternative integration schemes. First, we note that the ``out-going gauge condition''  $\gamma_{,zz}=0$ is now implied by the field equations, so there is no possibility for logarithms in the case of the vacuum Einstein equations with cosmological constant (in four dimensions). In the presence of matter such terms can arise and they always have a meaning in the AdS/CFT correspondence: they are related to conformal anomalies of the dual CFT. 
The cases of $\Lambda > 0$ is related to $\Lambda <0$ by analytic continuation. We will phrase our discussion using the AdS language, but the same integrations schemes also apply to the dS case (but one should note that  $\partial_u$ now becomes spacelike at future infinity). 

The first integration scheme, which we call the ``boundary scheme'', requires as initial data the values of  $U, \beta, \gamma$ and $\gamma_{,zzz}, U_{,zzz}, W_{,zzz}$ at $z=0$ (i.e. at the conformal boundary).
One can understand the meaning of this data by transforming to the Fefferman-Graham gauge. Recall that in Fefferman-Graham gauge ($l$ is the AdS radius)
\begin{equation} \label{intro: FG gauge}
ds^2=l^2\left[\frac{d\rho^2}{\rho^2}+\frac{1}{\rho^2}(g_{(0)ab}+\rho^2g_{(2)ab}+\rho^3g_{(3)ab}+\ldots)dx^adx^b\right],
\end{equation} 
where now the free data is $g_{(0)}$ and $g_{(3)}$ (with $g_{(3)}$ traceless and divergenceless), with $g_{(0)}$  being a representative of the conformal class and the background metric of the dual CFT and $g_{(3)}$
is related to the energy momentum tensor of the dual CFT.  Now $U, \beta, \gamma$ at $z=0$ determine $g_{(0)}$, while $\gamma_{,zzz}, U_{,zzz}, W_{,zzz}$ at $z=0$ 
determine $g_{(3)}$. So the analysis in Bondi gauge reproduces the salient features  of the asymptotic solutions in Fefferman-Graham gauge. 

As mentioned earlier, one can obtain the bulk conserved charges from $g_{(3)}$ and thus as in the asymptotically flat case $U_{,zzz}, W_{,zzz}$ are related to conserved charges, and so is 
$\gamma_{,zzz}$ which was not related to a conserved charge in the asymptotically flat case.  In contrast to the asymptotically flat case $\gamma_{,z}$ is now fully determined in terms $U, \beta, \gamma$ at $z=0$ , {\it  i.e.} the analogue of the news is now fixed. If we further restrict to Asymptotically AdS solutions, $\gamma_{,z}$ actually vanishes and the Bondi mass is constant. Similar observations were made in \cite{Ashtekar:2014zfa, He:2015wfa, Saw:2016isu, Saw:2017amv, He:2018ikd} (mostly for the dS case).
One can understand this result as follows. Since AdS and dS do not have a null infinity, any gravitational radiation will have to be absorbed at the conformal boundary and this would make the boundary metric time dependent. If we fix the boundary metric to be time independent as in the case of  Asymptotically AdS solutions then there is no possibility for gravitational radiation. A class of radiating spacetimes in AdS, the Robinson-Trautman spacetimes are indeed asymptotically locally AdS and have a time dependent boundary metric \cite{deFreitas:2014lia, Bakas:2014kfa}.

The second integration scheme is a hybrid version of the flat scheme and the previous one: one fixes now $\gamma, W_{,zzz}, U_{,zzz}$ at a null hypersurface $u=u_0 = const$
and $U, \beta, \gamma$ at $z=0$, for all times $u \geq u_0$. With this data one can recursively construct the solution to the future of the initial hypersurface.  


The rest of this paper is organised as follows. In section \ref{sec: Bondi-Sachs_metrics} we present background material needed in order to understand this paper: we introduce null hypersurfaces and the Bondi gauge and we present a brief review of asymptotic flatness and  of asymptotically locally AdS and dS spacetimes. Section \ref{sec: Field_equations} contains the detailed derivation of the asymptotic solutions and in section \ref{sec: Integration Scheme - Minkowski Vs. $AdS$} we compare and contrast the different integration schemes used in section  \ref{sec: Field_equations}. In section \ref{sec: Holographic_interpretation} we derive the transformation from Bondi gauge to Fefferman-Graham gauge and discuss the holographic interpretation of the functions appearing in the asymptotic solution in Bondi gauge. In this section we also illustrate the discussion using $AdS_4$, Schwarzschild $AdS_4$ and $AdS_4$ black branes as examples and discuss the properties of Bondi mass for asymptotically $AdS_4$ solutions. We conclude in section \ref{sec:conclusions}. The paper contains a number of appendices: in appendix \ref{sec: SCapp} we present the solution of the supplementary conditions for asymptotically locally (A)dS solutions, in appendix \ref{sec: FG_appendix} we provide technical details about the coordinate transformation from Bondi gauge to Fefferman-Graham gauge, in appendix \ref{logs_appendix} we discuss the presence of logarithmic terms in the asymptotic solutions when appropriate matter is present and in appendix \ref{AD_mass_appendix} we show the equivalence of the Bondi and Abbott-Deser masses in asymptotically AdS spacetimes.

\section{Bondi gauge metrics} \label{sec: Bondi-Sachs_metrics}

We begin this section with an introduction to the Bondi gauge and explain its advantages in studying asymptotically flat space-times. We then review essential features of anti-de Sitter and de Sitter asymptotics, as a precursor to analysing such spacetimes in Bondi gauge. 

\subsection{Null hypersurfaces}

Bondi gauge metrics were introduced and studied in \cite{Bondi:1962px, Sachs:1962wk} in the context of studying gravitational waves. The Bondi approach involves foliating the spacetime manifold by null hypersurfaces. Following  \cite{Sachs:1962wk}, one chooses the coordinate system as follows. Consider a Lorentzian $4$-manifold, $\mathcal{M}$, equipped with a metric $g_{\mu \nu}(x^{\rho})$ of signature $(-+++)$ and assume the existence of a scalar field $F=F(x^{\mu})$ such that the normal co-vector to $F$, $\partial_{\mu} F,$ is null: 
\begin{equation}
g^{\mu \nu}( \partial_{\mu} F)( \partial_{\nu} F)=0.
\end{equation} 
This criterion means null hypersurfaces, $\mathcal{N}_a$, can be described in terms of the level sets of $F$ i.e. 
\begin{equation}
\mathcal{N}_a=\{x^{\mu} \in \mathcal{M} \, | \, F(x^{\mu})=a\} 
\end{equation} 
and the spacetime $(\mathcal{M}, g_{\mu \nu})$ can be foliated, at least locally, using the null hypersurfaces, namely
\begin{equation}
\mathcal{M}=\{\mathcal{N}_a \, | \, a \in \text{Range}(F)\}
\end{equation}
where $\text{Range}(F)$ denotes all possible values of the function $F$. \par

The motivation for choosing null hypersurfaces can best be illustrated by looking at their interesting geometrical properties. Let us consider an arbitrary surface $\mathcal{N}_a \subset \mathcal{M}$ and the integral curves in the spacetime of the vector field $t^{\mu}=g^{\mu \nu} \partial_{\nu} F$; such curves are clearly null and normal to $\mathcal{N}_a$ and are commonly referred to as {null rays}. Null rays are also geodesic curves contained within $\mathcal{N}_a$: 
\begin{equation}
t^{\mu} \nabla_{\mu} t^{\nu}= \lambda(x^{\rho}) t^{\nu}.
\end{equation}
By choosing a suitable (affine) parametrisation we can set $\lambda=0$ and thus the null rays are also null generators of $\mathcal{N}_a$. This outlines the overall picture of this procedure as being a way to work from space-time $\rightarrow$ null hypersurface $\rightarrow$ null ray $\rightarrow$ null geodesic.

An adapted coordinate system can be chosen to describe such a situation. Typically, one works in \textit{retarded Bondi coordinates} $(u,r,\Theta^1,\Theta^2)$. The coordinate $u$ is a retarded time coordinate which labels the null hypersurfaces $\mathcal{N}_a$ ($u=F$ from above equations);  this coordinate is commonly referred to as the \textit{Bondi time} and takes values in $\mathbb{R}$. The $\Theta^A$ are angular coordinates which are defined to be constant along null rays:
\begin{equation}
t^{\mu} \partial_{\mu} \Theta^1=t^{\mu} \partial_{\mu} \Theta^2 = 0.
\end{equation}
This condition means that rays take the form $c^{\mu}(\lambda)=(u_0, r(\lambda),\Theta^1_0, \Theta^2_0)$ and thus the coordinate $r$ can be interpreted as a radial distance coordinate measuring the distance along a null ray.

\begin{figure}[H]
\begin{center}
	\includegraphics[width=0.54\linewidth]{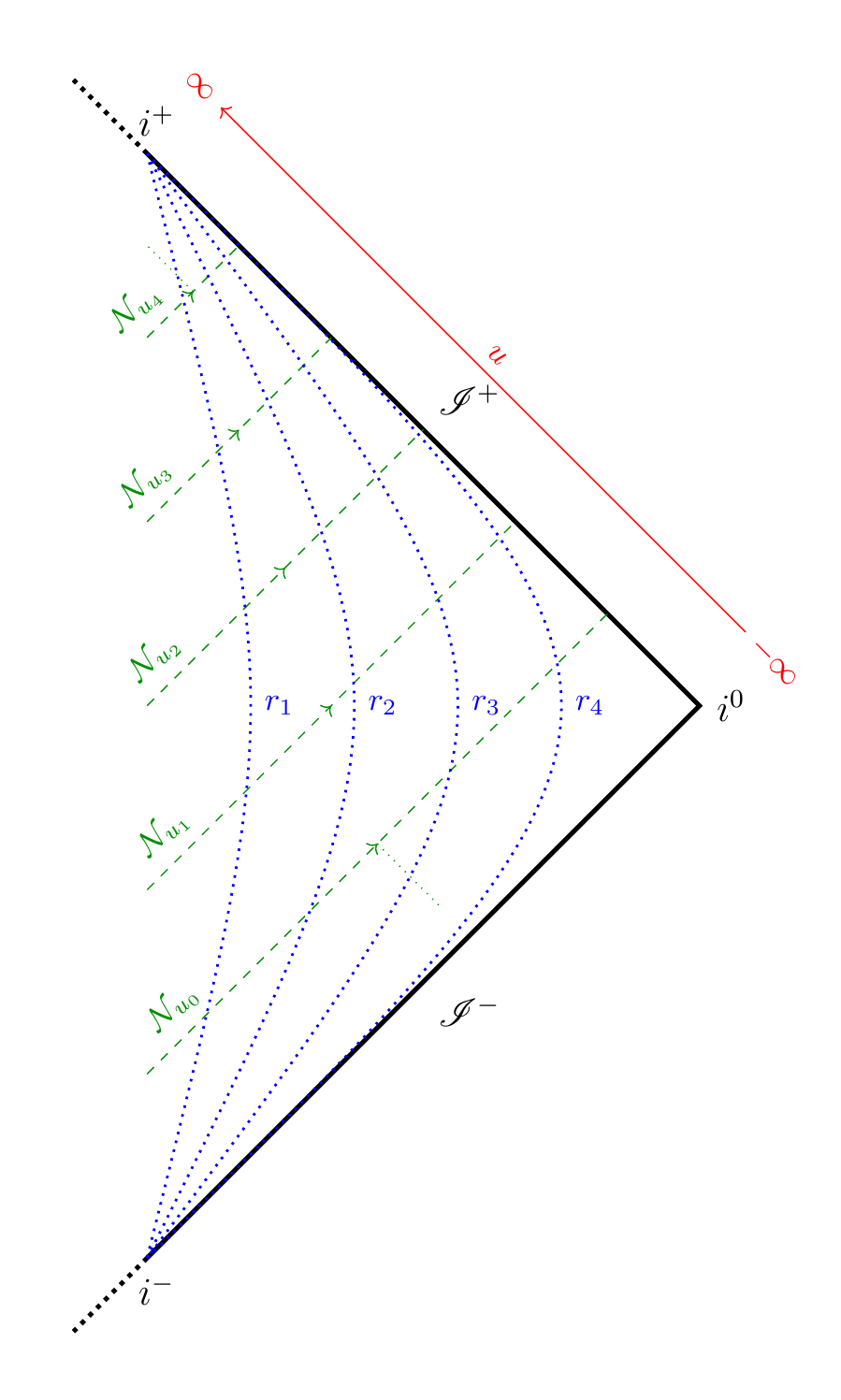}
	\caption{Penrose diagram of null hypersurfaces, $\color{islamicgreen}{\mathcal{N}_{u_i}}$, 	foliating future null infinity, $\mathscr{I^+}$, of an asymptotically flat spacetime. As indicated 	by the \textcolor{red}{solid red} axis, the retarded time coordinate $\color{red}{u}$ ranges from $(-	\infty, \infty)$ along $\mathscr{I^+}$ and thus the \textcolor{islamicgreen}{dashed green} lines 		represent the $u=\text{constant}$ hypersurfaces. (The arrows show the direction of increasing 	radial coordinate $r$). The \textcolor{blue}{dotted blue} curves represent timelike hypersurfaces of 	constant $\textcolor{blue}{r}$. }
\end{center}	
\end{figure}

\subsection{The Bondi gauge}
Following closely the notation of \cite{Strominger:2017zoo}, the most general line element that satisfies the previously discussed coordinate conditions is
\begin{equation}
ds^2=-X du^2 -2e^{2\beta}dudr+ h_{AB}\left(d\Theta^A+\frac{1}{2}U^{A}du\right)\left(d\Theta^B+\frac{1}{2}U^{B}du\right).
\end{equation} 
It is usual to impose in addition the following four gauge conditions:
\begin{equation} \label{eq:Bondi_gauge}
\partial_r \det\left(\frac{h_{AB}}{r^2}\right)=0, \qquad g_{rr}=g_{rA}=0.
\end{equation}
This metric together with the gauge conditions is known as the \textit{Bondi gauge} (or Bondi-Sachs gauge) and any spacetime metric can be locally written in this form. It is the most commonly used approach to analyse foliations by null hypersurfaces, although there are alternative approaches based on the Newman-Penrose formalism \cite{Newman:1961qr} using a null tetrad instead of a metric e.g. \cite{Barnich:2016lyg}.

The capital Roman indices $A,B$ take values $\{1,2\}$ which together with the symmetry of $h_{AB}$, gives seven functions in the line element: $(X, \beta, h_{AB}, U^A)$, all of which depend upon the spacetime coordinates $(u, r, \Theta^1, \Theta^2)$. The gauge condition on the determinant of $h_{AB}$ reduces the number of unknown functions in the metric to six. The latter are determined by the Einstein equations, subject to asymptotic data ($ r \rightarrow \infty$).  

One may choose to retain general covariance in the angular coordinates as in \cite{Flanagan:2015pxa} but it is often useful to consider a local choice. In this paper we will commonly utilise the usual $(\theta, \phi)$ of the spherical coordinate system as well as complex coordinates  $(\zeta, \bar{\zeta})$, related by
\begin{equation}
\zeta=e^{i\phi}\cot\left(\frac{\theta}{2}\right), \qquad \bar{\zeta}=e^{-i\phi}\cot\left(\frac{\theta}{2}\right).
\end{equation}

\subsection{Asymptotic flatness}

The Bondi gauge has frequently been used to study asymptotically flat spacetimes and their symmetries. Asymptotic flatness may be viewed as the property that the spacetime tends to Minkowski spacetime as $r \rightarrow \infty$. This imprecise statement can be given a rigorous definition, which we will briefly touch upon referring to  \cite{Wald:1984rg, Hawking:1973uf} for a detailed discussion, before seeing how to implement asymptotic flatness in a coordinate dependent manner by imposing suitable fall-off conditions upon the metric components. 

Let us first recall the notion of conformal compactification \cite{Penrose:1962ij}. Consider a manifold with boundary $\bar{\mathcal{M}}=\mathcal{M} \cup \partial \mathcal{M}$ where $\partial \mathcal{M}$ is the boundary. A metric $g_{\mu \nu}$ is \textit{conformally compact} if  there exists a defining function $\Omega$ which satisfies 
\begin{equation} \label{def_fun}
\Omega(\partial \mathcal{M})=0, \qquad d\Omega(\partial \mathcal{M}) \neq 0, \qquad \Omega(\mathcal{M}) > 0. 
\end{equation} 
and the metric $\bar{g}$ defined by 
\begin{equation}
\bar{g}_{\mu \nu}=\Omega^2 g_{\mu \nu} 
 \label{eq: double_pole_metric}
\end{equation}
extends smoothly to $\partial \mathcal{M}$. Let us also consider  an embedding $f: \mathcal{M} \rightarrow \tilde{\mathcal{M}}$ such that $f$ embeds $\mathcal{M}$ as a manifold with smooth boundary $\partial \mathcal{M}$ in $\tilde{\mathcal{M}}$ and such that 
\begin{equation}
\bar{g}=f_{*}(\tilde{g}),
\end{equation}
where $f_{*}$ denotes ``pull back'' of the embedding function $f$. The unphysical spacetime $(\tilde{\mathcal{M}}, \tilde{g})$ is often referred to as the \textit{conformal compactification} of the spacetime $(\mathcal{M}, g)$ and $\partial \mathcal{M}$ is the \textit{conformal boundary} of the spacetime. 

Asymptotic flatness is now defined by putting further conditions on the conformal compactification. Different definitions have been proposed through the years, see \cite{Wald:1984rg, Hawking:1973uf} (and references therein). The precise details also depend on whether one would like to consider asymptotic flatness at spatial infinity, null infinity or both. We will not need these details here. For our purposes it suffices to say that we will consider cases with  $R_{\mu \nu}=0$ in an open neighbourhood of $\partial \mathcal{M}$ in $\bar{\mathcal{M}}=\mathcal{M} \cup \partial \mathcal{M}$\footnote{Such conformal compactification is called \textit{asymptotically empty}.}.


Let us now implement asymptotic flatness in a coordinate dependent manner. The Minkowski metric in \textit{retarded} coordinates $(u,r,\zeta,\bar{\zeta})$ is given by 
\begin{equation}
ds_{M}^2=-du^2-2dudr+2r^2\gamma_{\zeta \bar{\zeta}}d\zeta d\bar{\zeta}
\end{equation} 
where 
\begin{equation}
u=t-r , \qquad \gamma_{\zeta \bar{\zeta}}=\frac{2}{(1+\zeta\bar{\zeta})^2}.
\end{equation}
Here $u$ is a retarded time coordinate and $\gamma_{\zeta \bar{\zeta}}$ is the round metric on $S^2$. This metric is in Bondi gauge with function choices  $h_{\zeta \zeta}=h_{\bar{\zeta} \bar{\zeta}}= \beta= U^A=0$, $X=1$, $h_{\zeta \bar{\zeta}}=r^2 \gamma_{\zeta \bar{\zeta}}$. Note that this choice of coordinates is suitable for analysis near $\mathscr{I^+}$. To analyse neighbourhoods of $\mathscr{I^-}$  the metric can be expressed in \textit{advanced} coordinates $(v,r,\zeta,\bar{\zeta})$ where $v=t+r$ and thus 
\begin{equation}
ds_M^2=-dv^2+2dvdr+2r^2\gamma_{\zeta \bar{\zeta}}d\zeta d\bar{\zeta}.
\end{equation}
For this paper, we will use retarded coordinates and thus restrict our attention to neighbourhoods of $\mathscr{I}^+$. \par

For a general asymptotically flat metric the metric functions admit power series expansions in $1/r$ with the leading order term being that of the Minkowski metric, as we will re-derive here. The review \cite{Strominger:2017zoo} discusses suitable fall-off conditions for the subleading terms in the series: the fall-off should include gravitational wave emitting solutions, as was the motivation in \cite{Bondi:1962px}. These criteria were imposed in \cite{Bondi:1962px, Sachs:1962wk} and if we combine this with the following fall-off of the Weyl curvature tensor components at large $r$
\begin{equation}
C_{r\zeta r \zeta} \sim O(r^{-3}), \quad C_{rur\zeta} \sim O(r^{-3}), \quad C_{rur\bar{\zeta}} \sim O(r^{-3})
\end{equation}
as in \cite{Strominger:2017zoo} then we obtain the class of asymptotically flat metrics in Bondi gauge as 
\begin{align}\label{eq:mink_BS}
\begin{split}
ds^2=&\, ds_{M}^2+\frac{2m_B}{r}du^2+rC_{\zeta \zeta}d\zeta^2+rC_{\bar{\zeta} \bar{\zeta}}d\bar{\zeta}^2+D^\zeta C_{\zeta \zeta}dud\bar{\zeta}+D^{\bar{\zeta}}C_{\bar{\zeta}\bar{\zeta}}dud\bar{\zeta} \\
& +\frac{1}{r}\left(\frac{4}{3}(N_\zeta+u\partial_\zeta m_B -\frac{1}{4}\partial_\zeta(C_{\zeta \zeta}C^{\zeta \zeta})\right)dud\zeta + c.c.+ \ldots.
\end{split}
\end{align}
where $D_A$ is the covariant derivative with respect to the metric of the round sphere $\gamma_{AB}$ and the first term in the equation is just the Minkowski metric. The rest of the terms in the first line are the first order subleading terms in powers of $r$. Notice that although these terms have different powers of $r$ preceding them, they are all subleading as $r \rightarrow \infty$ when compared to the Minkowski metric. The second line of the equation contains second order subleading terms, included here as these terms contain physically interesting functions. \par

At $\mathcal{O}(1/r)$ in $g_{uu}$ is a function $m_B=m_B(u,\zeta,\bar{\zeta})$ is known as the \textit{Bondi mass aspect}. One of the key results of \cite{Bondi:1962px} is that the Bondi mass aspect can be integreated over the unit $S^2$ to give the total \textit{Bondi mass}\footnote{This quantity is sometimes referred to in the literature as the \textit{Trautman-Bondi} mass, as it was also discussed by Trautman in \cite{Trautman:2016xic}, see also the lecture notes \cite{Trautman:2002zz} for further comments and references.} $\mathcal{M}_B$ of the system at time $u$ 
\begin{equation} \label{eq: Bondi_mass_time_u}
\mathcal{M}_B=\frac{1}{4\pi}\int_{S^2} m_B=\frac{1}{4\pi}\int d^2z \, \gamma_{\zeta \bar{\zeta}} \, m_B. 
\end{equation}
The Bondi mass is a natural way to define the mass of a system at $\mathscr{I^+}$, and is an alternative to the ADM mass which is defined as an integral at spatial infinity $i^0$.  

Contained in the $1/r$ suppressed terms relative to the Minkowski metric is the shear tensor $C_{AB}(u,\zeta,\bar{\zeta})$; a symmetric and traceless tensor of type $[0,2]$. This tensor describes the gravitational waves in the spacetime (recall we wanted the fall-off conditions to include these solutions) and it motivates the definition of another key concept in the Bondi gauge, the \textit{Bondi news tensor}, $N_{AB}$, 
\begin{equation}
N_{AB}(u,\zeta,\bar{\zeta})=\partial_{u}C_{AB}(u,\zeta,\bar{\zeta}).
\end{equation}
The news tensor is again a symmetric and traceless tensor of type $[0,2]$. The name ``news" for this tensor can be best explained by imposing Einstein's equations (with $\Lambda=0$) upon the metric 
\begin{equation}
R_{\mu \nu}-\frac{1}{2}g_{\mu \nu}R=8\pi T_{\mu \nu}, \qquad \lim_{r \rightarrow \infty} T_{\mu \nu} =0
\end{equation}
where the limit condition on the stress energy tensor is typically enforced such that $\Omega^{-1} T_{ab}$ has a smooth conformal completion to $\mathscr{I^+}= \{\Omega = 0\}$ 
(where $\Omega \sim 1/r$ for the case at hand). 
This condition is a requirement for asymptotic flatness as it forces the asymptotically empty condition mentioned above.  The authors of \cite{Bondi:1962px} solved the field equations by expanding in large $r$ and solving the equations that arise at each order and we will streamline this derivation here. The leading order ($\mathcal{O}(r^{-2})$) of the $(uu)$ component of the Einstein equations then reads (see discussions in \cite{Strominger:2017zoo,Flanagan:2015pxa}) 
\begin{equation}
\partial_u m_B=\frac{1}{4}[D^2_\zeta N^{\zeta \zeta}+D^2_{\bar{\zeta}}N^{\bar{\zeta}\bar{\zeta}}-N_{\zeta \zeta}N^{\zeta \zeta}]-4\pi  \lim_{r\rightarrow \infty} r^2T_{uu}.
\end{equation}
Thus the news tensor, along with the stress tensor, governs the change in the Bondi mass aspect - it provides the ``news'' regarding the change in the mass aspect. If the spacetime under consideration is vacuum (as in \cite{Bondi:1962px}) then the news entirely governs the change in mass.

The final interesting term is the $N^A$ which appears in the subleading terms in the second line of (\ref{eq:mink_BS}). This vector is named the \textit{angular momentum aspect} and - in a similar fashion to the mass aspect - can be used to define the total angular momentum at $\mathscr{I^+}$ via a suitable integral. Both the mass aspect and angular momentum aspect arise as functions of integration in the full set of Einstein field equations, although the field equations do contain evolution equations for these \cite{Flanagan:2015pxa, Strominger:2013jfa, Pasterski:2015tva, Barnich:2010eb} which we will discuss in detail later.

Comparing the general Bondi gauge metric with the asymptotically flat metric, the fall-off conditions on the metric functions are
\begin{align}
\begin{split}
&X=1-\frac{2m_B}{r}+O(r^{-2}), \qquad \beta=O(r^{-2}), \\ 
&g_{AB}=r^2\gamma_{AB}+rC_{AB}+O_{1}, \qquad U_A=\frac{1}{r^2}D^{B}C_{AB}+O(r^{-3}).
\end{split}
\end{align}
The infinite dimensional symmetry group of all coordinate transformations that preserves these conditions as well as the gauge itself is known as the BMS group \cite{Sachs:1962zza}.

\subsection{Anti-de Sitter and de Sitter asymptotics}  \label{subseq: AdS_asymptotics}

Let us now consider spacetimes that satisfy the Einstein field equations with $\Lambda \neq 0$
\begin{equation} \label{eq: EE_lambda}
R_{\mu \nu} - \frac{1}{2}R g_{\mu \nu} + \Lambda g_{\mu \nu} = 8\pi T_{\mu \nu}.
\end{equation}
We will focus mainly on the case of anti-de Sitter asymptotics ($\Lambda < 0$) although the discussion generalises straightforwardly to the de Sitter case ($\Lambda > 0$). Throughout this paper we will concentrate on vacuum spacetimes, i.e. $T_{\mu \nu} = 0$. 

AdS$_4$ is the maximally symmetric solution to the vacuum Einstein equations with negative cosmological constant. The AdS$_4$ metric can be written in Bondi gauge as 
\begin{equation}
ds_{AdS}^2=-\left(1+\frac{r^2}{l^2}\right)du^2-2dudr+r^2d\Omega^2
\end{equation}
where $l^2=-3/\Lambda$; $l$ is the \textit{AdS radius} or \textit{curvature radius} of the spacetime as the Riemann tensor for AdS$_4$ takes the form 
\begin{equation}
R_{\mu \nu \rho \sigma} = \frac{1}{l^2} (g_{\mu \sigma} g_{\nu \rho}-g_{\mu \rho} g_{\nu \sigma}).
\end{equation}

We define an \textit{asymptotically locally AdS} metric to be a \textit{conformally compact Einstein metric} of negative cosmological constant. In what follows we briefly review the key features relevant for this paper,  see \cite{Fefferman:1985zza,penrose_rindler_1986, Graham:1999jg, Skenderis:2002wp} for more details. 
Consider a manifold with boundary $\bar{\mathcal{M}}=\mathcal{M} \cup \partial \mathcal{M}$, equipped with a conformally compact metric $g_{\mu \nu}$,
as in (\ref{def_fun})-(\ref{eq: double_pole_metric}).  We further require 
 \begin{equation}
g_{(0)} = \bar{g}|_{\partial \mathcal{M}}
 \end{equation}
 is non-degenerate. Note that $g_{(0)}$ is not unique since  the choice of defining function is non-unique: if $\Omega$ is a suitable defining function, then so is $\Omega e^{w}$, where $w$ is a function with no zeroes or poles on $\partial \mathcal{M}$. Thus the induced metric at $\partial \mathcal{M}$, $g_{(0)}$, is also non-unique. This procedure defines a conformal class of metric and  $g_{(0)}$ is a representative of the conformal class of metrics.

\begin{figure}[h]
\begin{center}
	\includegraphics[height=8cm]{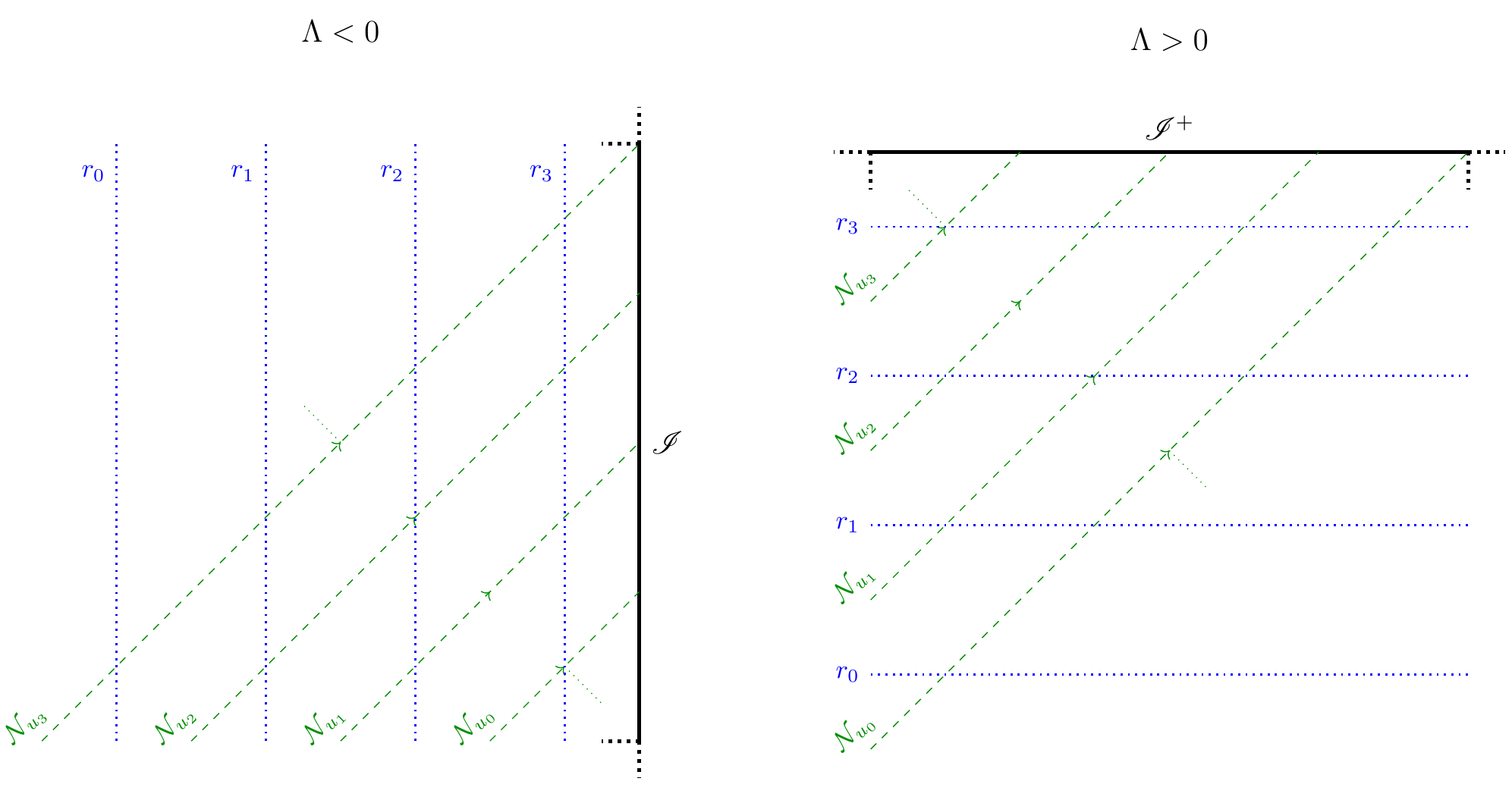}
	\caption{{\it Left panel.} Penrose diagram of the asymptotic region of an asymptotically locally AdS spacetime, where the \textit{timelike} boundary manifold $\partial \mathcal{M}$ is denoted $\mathscr{I}$. The \textcolor{islamicgreen}{dashed green} curves represent null hypersurfaces $\color{islamicgreen}{\mathcal{N}_{u_i}} =\{ u= u_i \, | \, u_i = \text{constant} \}$  	and the \textcolor{blue}{dotted blue} curves timelike surfaces of constant $\color{blue}{r}.$
	{\it Right panel.} Penrose diagram of the asymptotic region of an asymptotically locally dS space-time, where we have chosen to foliate the future spacelike boundary $\partial \mathcal{M}=\mathscr{I}^+$. The \textcolor{islamicgreen}{dashed green} curves represent null hypersurfaces $\color{islamicgreen}{\mathcal{N}_{u_i}} =\{ u= u_i \, | \, u_i = \text{constant} \}$  	and the \textcolor{blue}{dotted blue} curves spacelike surfaces of constant $\color{blue}{r}$. The difference in the properties of  constant $r$ surfaces between AdS (timelike) and dS (spacelike) is due to the presence of a cosmological horizon in asymptotically locally $dS$ spacetimes.} \label{Penr_AdS}
\end{center}	
\end{figure}

Using (\ref{eq: double_pole_metric})  the Riemann tensor of $g_{\mu \nu}$ takes the form
\begin{equation} \label{eq: Riemann_g}
R_{\alpha \beta \gamma \delta}[g]=|d\Omega|^2_{\bar{g}}(g_{\alpha \delta}g_{\beta \gamma} - g_{\alpha \gamma} g_{\beta \delta})+\mathcal{O}(\Omega^{-3})
\end{equation}
where the leading order term is $\mathcal{O}(\Omega^{-4})$. One can also define
\begin{equation}
|d\Omega|^2_{\bar{g}}=\bar{g}^{\mu \nu} (\partial_{\mu} \Omega)( \partial_{\nu} \Omega),
\end{equation}
a quantity which smoothly extends to $\bar{\mathcal{M}}$; its restriction to $\partial \mathcal{M}$ is a conformal invariant \cite{Skenderis:2002wp}.
The metric $g_{\mu \nu}$ should be Einstein, {\it i.e.} it should satisfy (\ref{eq:  EE_lambda}). As in the asymptotically flat case, $\Omega^{-1}T_{\mu \nu}$ should have a smooth conformal completion to $\partial \mathcal{M}=\{\Omega=0\}$. Enforcing (\ref{eq:  EE_lambda}) upon (\ref{eq: Riemann_g}) gives 
\begin{equation}
3g_{\mu \nu} |d\Omega|^2_{\bar{g}}+\Lambda g_{\mu \nu} + \mathcal{O}(\Omega^{-1})=8\pi T_{\mu \nu}
\end{equation} 
and thus as $\Omega \rightarrow 0$ (after rearrangement)
\begin{equation} \label{eq: one_form_norm}
|d\Omega|^2_{\bar{g}}\big\rvert_{\partial \mathcal{M}}=\frac{1}{l^2}.
\end{equation}
Thus near the boundary $\partial \mathcal{M}$, the Riemann curvature tensor of the metric $g_{\mu \nu}$ is to leading order the same as that of the AdS$_4$ metric.    

We emphasise that this definition does not enforce any restriction on the topology of $\partial \mathcal{M}$ or the metric $g_{(0)}$ induced at $\partial \mathcal{M}$. For global AdS$_4$ the conformal boundary has the topology of $\mathbb{R} \times S^2$ and the metric $g_{(0)}$ is conformally flat.  Asymptotically locally AdS spacetimes for which $g_{(0)}$ is conformally flat are called \textit{asymptotically AdS spacetimes}. (Thus asymptotically AdS spacetimes are a subset of asymptotically locally AdS spacetimes). Holographically $g_{(0)}$ corresponds to the background metric for the dual quantum field theory and it is thus essential to consider generic $g_{(0)}$.

The discussion of asymptotically locally AdS metrics extends to the case of a positive cosmological constant in a very straightforward manner. The Einstein equations for asymptotically locally dS spacetimes are related to those of AdS via the simple transformation 
\begin{equation}
l^2_{AdS}\rightarrow -l^2_{dS}
\end{equation}
and thus to define dS asymptotics, one simply repeats  (\ref{eq: EE_lambda})-(\ref{eq: one_form_norm}) with every occurrence of $l^2$ being replaced by $-l^2$. 

In preparation for the discussion in the next section we indicate in figure \ref{Penr_AdS}  how asymptotically locally $\Lambda \neq 0$ spacetimes are locally foliated by null hypersurfaces. 

\section{The Einstein field equations} \label{sec: Field_equations}

In this section we will compute the vacuum Einstein equations in the presence of a cosmological constant for an axisymmetric, $\phi$-reflection symmetric Bondi gauge metric. The techniques employed in doing this are very similar to those of \cite{Bondi:1962px} and many of the properties of the original method carry over.

\subsection{General considerations}
We first apply some simplifications to the general Bondi gauge metric. Working in coordinates $(u , r, \theta, \phi)$, we enforce both axi-symmetry ($\partial/\partial \phi$ a Killing vector field) and reflection symmetry in $\phi$ (so the metric is invariant under $d\phi \rightarrow -d\phi$). In Bondi function notation, this means we set $h_{\theta \phi}=h_{\phi \theta}=U^{\phi}=0$, reducing the number of unknown functions to four. These choices are made entirely for computational simplification in the calculations that follow.

Following \cite{Bondi:1962px}, we now write the remaining functions in the form
\begin{equation}
X=Wr^{2}e^{2\beta}, \qquad h_{\theta \theta}=r^2e^{2\gamma}, \qquad h_{\phi \phi}=r^2\sin^2\theta e^{-2\gamma}, \qquad U^{\theta}=-2U
\end{equation}
giving us the line element 
\begin{align}  \label{eq: Bondi_Metric}
\begin{split}
ds^2=&-(Wr^{2}e^{2\beta}-U^2r^2e^{2\gamma})du^2-2e^{2\beta}dudr-\\
&2Ur^2e^{2\gamma}dud\theta+r^2(e^{2\gamma}d\theta^2+e^{-2\gamma}\sin^2\theta d\phi^2).
\end{split}
\end{align}
This choice of metric has a restriction in the determinant along the sphere ( $\text{det}(h_{AB}/r^2)=\sin^2\theta$); $r$ is a luminosity distance.
The Einstein equations are expressed in terms of the four metric functions $(\gamma(u,r,\theta)$, $\beta(u, r, \theta)$, $U(u,r,\theta)$, $W(u,r,\theta))$. \par

In this paper we will analyse the Einstein vacuum equations, 
\begin{equation} \label{eq:2.1}
R_{\mu \nu}=\Lambda g_{\mu \nu}. 
\end{equation}
The generalization to include matter would be straightforward. 
 It is quite common in the relativity literature to solve Einstein's equations with  ``asymptotically vacuum'' matter such that $\lim_{r \rightarrow \infty} T_{\mu \nu} =0$; an example can be found in \cite{Flanagan:2015pxa}, involving an asymptotic power series expansion in negative powers of the radial coordinate. However, as is well known in holography, the presence of matter generically affects the powers arising in the asymptotic expansions and logarithmic terms can arise for matter of specific masses, see the discussion in appendix \ref{logs_appendix} as well as the references \cite{deHaro:2000vlm,Skenderis:2002wp}.
 

Following \cite{Bondi:1962px}, we separate Einstein's equations into the four `main equations'
\begin{equation} \label{eq:2.3}
R_{rr}=R_{r\theta}=0, \quad R_{\theta \theta}=\Lambda r^2e^{2 \gamma}, \quad R_{\phi \phi}=\Lambda r^2e^{-2\gamma}\sin^2\theta; 
\end{equation}
\noindent three `trivial equations' 
\begin{equation}
R_{u \phi}=R_{r \phi}=R_{\theta \phi}=0
\end{equation}
and three `supplementary equations'
\begin{equation}
R_{uu}=-\Lambda(Wr^2e^{2\beta}-U^2r^2e^{2\gamma}) , \quad R_{u\theta}=-\Lambda Ur^2e^{2\gamma}, \quad R_{u r}=-\Lambda e^{2\beta}.
\end{equation}
The main equations are so named because they must be solved in order to generate solutions to the field equations. The trivial equations are automatically satisfied because of the symmetries of the spacetime metric. The supplementary conditions will be shown to provide constraint equations for the functions of integration arising from the main equations. These will be discussed in section \ref{subsec: SC} but first we will focus our attention on the main equations: 
\begin{subequations}
\begin{align} 
0 & =-R_{rr} = -4\left[\beta_{,r}-\frac{1}{2}r(\gamma_{,r})^2\right]r^{-1} \label{eq: AdS_me1} \\
\begin{split}
0 & =2r^2R_{r\theta} =[r^4e^{2(\gamma-\beta)}U_{,r}]_{,r}- \\
&\phantom{aaaaaaaaaaa} 2r^2[\beta_{,r\theta}-\gamma_{,r\theta}+2\gamma_{,r}\gamma_{,\theta}-2\beta_{,\theta}r^{-1}-2\gamma_{,r}\cot \theta] \label{eq: AdS_me2}
 \end{split}
 \\
\begin{split}
-2 \Lambda r^2e^{2\beta}& = -R_{\theta \theta}e^{2(\beta-\gamma)}-r^2R^{\phi}_{\phi}e^{2\beta} = 2(r^3 W)_{,r}+\frac{1}{2}r^4e^{2(\gamma-\beta)}(U_{,r})^2-r^2 U_{,r \theta} - \\ 
&\phantom{= -R_{\theta \theta}e^{2(\beta-\gamma)}-r^2R^{\phi}_{\phi}e^{2\beta} = a}
4rU_{,\theta} -r^2U_{,r}\cot \theta -4rU\cot \theta + \\ 
&\phantom{= -R_{\theta \theta}e^{2(\beta-\gamma)}-r^2R^{\phi}_{\phi}e^{2\beta} = a} 2e^{2(\beta-\gamma)}[-1-(3\gamma_{,\theta}-\beta_{,\theta})\cot \theta - \\ 
&\phantom{= -R_{\theta \theta}e^{2(\beta-\gamma)}-r^2R^{\phi}_{\phi}e^{2\beta} = a} \gamma_{,\theta \theta}+\beta_{,\theta \theta} +(\beta_{,\theta})^2+2\gamma_{,\theta}(\gamma_{,\theta}-\beta_{,\theta})] \label{eq: AdS_me3}  
\end{split}
\\
\begin{split}
-\Lambda r^2e^{2\beta} & =- r^2R^{\phi}_{\phi}e^{2\beta} =2r(r \gamma)_{,u r}+(1-r\gamma_{,r})(r^3W)_{,r}-r^3(r\gamma_{,r r}+\gamma_{,r}) W-\\
&\phantom{= -r^2R^{\phi}_{\phi}e^{2\beta} =  a} r(1-r\gamma_{,r})U_{,\theta} -r^2(\cot \theta - \gamma_{,\theta})U_{,r}+\\
&\phantom{= -r^2R^{\phi}_{\phi}e^{2\beta} =  a}r(2r\gamma_{,r \theta}+2\gamma_{,\theta}+r\gamma_{,r}\cot \theta - 3\cot \theta)U \\
&\phantom{= -r^2R^{\phi}_{\phi}e^{2\beta} =  a} +e^{2(\beta-\gamma)}[-1-(3\gamma_{,\theta}-2\beta_{,\theta})\cot \theta-\\
&\phantom{= -r^2R^{\phi}_{\phi}e^{2\beta} =  a}\gamma_{,\theta \theta}+2\gamma_{,\theta}(\gamma_{,\theta}-\beta_{,\theta})]. \label{eq: AdS_me4}
\end{split}
\end{align}
\end{subequations}
Notice that the first two equations agree with the first two main equations in \cite{Bondi:1962px}. The second two are altered by the inclusion of the cosmological constant but they manifestly reduce to the original equations in the $\Lambda \rightarrow 0$ limit. We will now follow closely the integration scheme of \cite{Bondi:1962px} to see how this alters the solutions to the equations above. 

\begin{figure}[H]
\begin{center}
\includegraphics[width=0.35\linewidth]{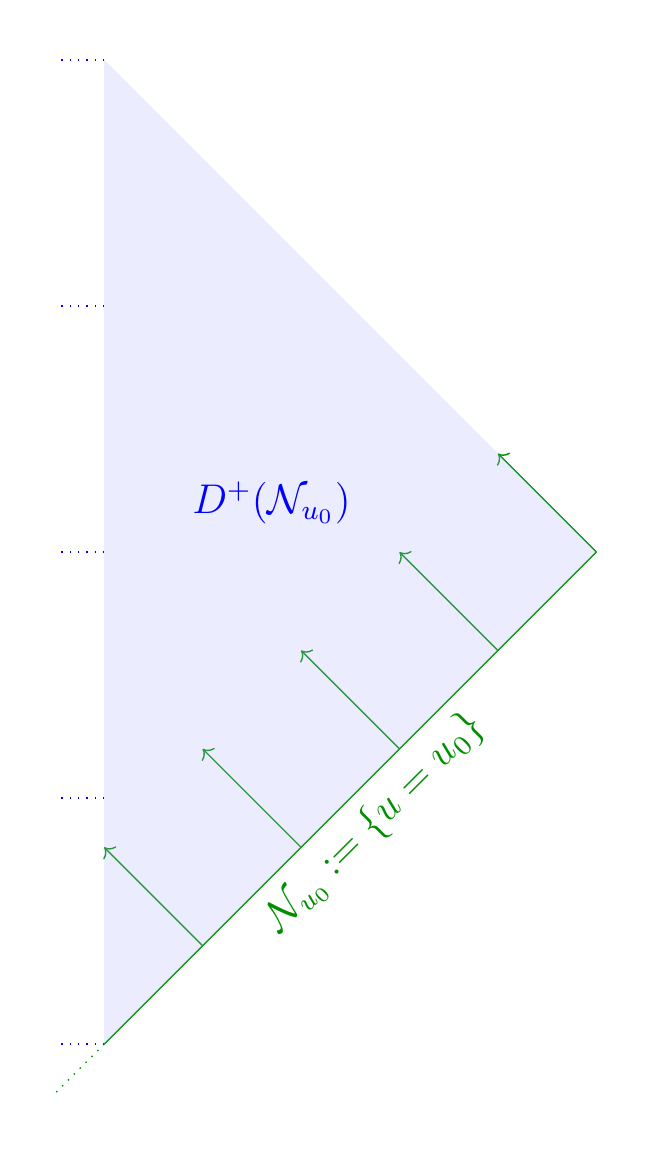} 
\end{center}
\caption{Causal diagram illustrating how one applies the BMS scheme when given suitable initial data on a null hypersurface $\mathcal{N}_{u_0}$} \label{flat_scheme}
\label{fig: fig_3}
\end{figure}

We will first solve the main equations following the same approach as the original analysis  \cite{Bondi:1962px}:
\begin{itemize}
\item[1)] Specify $\gamma(u, r, \theta)$ on an initial null hypersurface $\mathcal{N}_{u_0}$ i.e. $\gamma(u_0, r, \theta)$.
\item[2)] Solve (\ref{eq: AdS_me1}) on the null hypersurface $\mathcal{N}_{u_0}$ to compute $\beta(u_0, r,\theta)$. This is possible as only $\gamma(u_0, r, \theta)$ appears in the equation,
\item[3)] Solve (\ref{eq: AdS_me2})  for $U(u_0, r, \theta)$. This is possible as only $\gamma(u_0, r, \theta)$ and $\beta(u_0, r, \theta)$ appear in the equation. 
\item[4)] Solve (\ref{eq: AdS_me3})  for  $W(u_0, r, \theta)$. Only $\gamma(u_0, r, \theta)$, $\beta(u_0, r, \theta)$ and $U(u_0, r, \theta)$ appear in the equation.
\item[5)] Solve equation (\ref{eq: AdS_me4}) for $\gamma_{,u}(u_0, r, \theta)$ i.e. to obtain $\gamma$ on the next null hypersurface $\mathcal{N}_{u_0 + \delta u}$.
\item[6)] Repeat from step 1 with the new Bondi time $u_0+ \delta u$. Iteration gives the Einstein solution for the future domain of dependence of $\mathcal{N}_{u_0}$, $D^+(\mathcal{N}_{u_0})$, see Fig. \ref{flat_scheme}.
\end{itemize}  

Specialising briefly to the case of AdS, we observe that unlike asymptotically flat space-time we have $D^+(\mathcal{N}_{u_0}) \neq J^+(\mathcal{N}_{u_0})$, where $J^+$ indicates the causal future. To solve the equations in $J^+(\mathcal{N}_{u_0})$ we would need to specify extra data on a new hypersurface (e.g. the conformal boundary $\mathscr{I}$). We will discuss in detail the integration scheme for AdS asymptotics in section \ref{sec: Integration Scheme - Minkowski Vs. $AdS$}. 

In the case of asymptotically locally dS spacetimes the situation is slightly different, firstly because we now have two boundaries: future spacelike infinity, $\mathscr{I}^+$, and past spacelike infinity, $\mathscr{I}^{-}$. We will restrict consideration to a retarded null foliation of the future spacelike boundary, $\mathscr{I}^+$, when we discuss this case in greater detail in section \ref{subsec: dS_schemes}. We will see that this has a number of different subtleties when compared with the flat and AdS cases.  

\subsection{Solving the main equations asymptotically}

We are interested in solving the Einstein equations in the \textit{asymptotic region} (large $r$) of the spacetime in the most general manner possible. It is convenient to implement an inversion map
\begin{equation}
r=\frac{1}{z}
\end{equation} 
\noindent so that solving the equations as $r \rightarrow \infty$ is reduced to the analytically simpler procedure of solving around $z=0$. Carrying out this substitution in the main equations gives 
\begin{subequations}
\begin{align}
0&=2 \beta_{, z} +z (\gamma_{,z})^2 \label{eq: reg_me1} \\ 
\begin{split}
0&= 4 \beta_{, \theta} - 2 e^{2\gamma - 2\beta} U_{,z}-2ze^{2\gamma-2\beta} U_{,z}\beta_{,z}-4z\cot \theta \gamma_{,z}+ 4z\gamma_{, \theta}\gamma_{,z}+ \label{eq: main_eq_2} \\
& \phantom{aaa}2z e^{2\gamma- 2\beta} U_{,z} \gamma_{,z} +2z \beta_{,z\theta}-2z\gamma_{, z \theta}+ze^{2\gamma-2\beta} U_{, zz} 
\end{split}
\\
\begin{split}
-2\Lambda e^{2\beta} & = 6W+4z \cot \theta U - 2z W_{, z} - 4z U_{, \theta} - 2z^2e^{2\beta -2\gamma}+ \label{eq: reg_me3} \\
& \phantom{aaa} 2z^2e^{2\beta-2\gamma}\cot \theta \beta_{, \theta}+ 2z^2e^{2\beta-2\gamma}(\beta_{, \theta})^2-6z^2e^{2\beta-2\gamma}\cot \theta \gamma_{, \theta}-\\
& \phantom{aaa} 4z^2e^{2\beta-2\gamma}\beta_{, \theta} \gamma_{, \theta}+4z^2e^{2\beta-2\gamma}(\gamma_{,\theta})^2+ 2z^2e^{2\beta-2\gamma}\beta_{, \theta \theta}-\\
& \phantom{aaa} 2z^2e^{2\beta-2\gamma}\gamma_{, \theta \theta}+z^2\cot \theta U_{,z}+\frac{z^2}{2}e^{2\gamma-2\beta}(U_{,z})^2+z^2U_{,z \theta} 
\end{split}
\\
\begin{split}
-\Lambda e^{2\beta} & =3W-3z\cot \theta U- zU_{,\theta}+2zU \gamma_{, \theta}-zW_{,z}+2zW \gamma_{,z}+2z\gamma_{,u}- \label{eq: reg_me4} \\
& \phantom{aaa} z^2e^{2\beta-2\gamma}+ 2z^2 e^{2\beta - 2\gamma} \cot \theta \beta_{, \theta} - 3z^2 e^{2\beta -2\gamma}\cot \theta \gamma_{, \theta}-\\
& \phantom{aaa} 2z^2e^{2\beta -2\gamma}\beta_{, \theta} \gamma_{, \theta} + 2z^2e^{2\beta -2\gamma}(\gamma_{, \theta})^2-z^2e^{2\beta -2\gamma}\gamma_{, \theta \theta}+z^2\cot \theta U_{, z} -\\
& \phantom{aaa} z^2\gamma_{, \theta} U_{,z}-z^2\cot \theta U \gamma_{, z}- z^2 U_{, \theta} \gamma_{, z} -z^2 W_{,z}\gamma_{,z}-\\
& \phantom{aaa} 2z^2U \gamma_{,z \theta} - z^2W\gamma_{,zz}-2z^2\gamma_{, u z}.
\end{split}
\end{align}
\end{subequations}
where we have multiplied the second equation through by $z$, and the last two through by $z^2$ in order to obtain expressions that are regular at $z=0$. We have also rescaled the first equation by dividing through by $2z^3$; this is not necessary to make the equation regular at $z=0$ but enables iterative differentiation of the field equation.

We assume that the metric functions $(\gamma, \beta, U, W)$ are suitably differentiable (at least $C^4$) at $z=0$ and derive asymptotic solutions to the field equations via the following procedure:

\begin{itemize}
\item[1)] Evaluate the field equations at $z=0$ and solve the resulting algebraic equations.
\item[2)] Differentiate the field equations with respect to $z$.
\item[3)] Return to step 1) for the differentiated field equation.
\end{itemize}
We will follow this procedure equation by equation, making use of the nested structure to move from one to the next. 

\subsubsection{The first equation}
\noindent The first main equation is \eqref{eq: reg_me1}. Applying the differentiation procedure and solving at each order we obtain 
\begin{align}
\begin{split}
&\beta_{,z}(u_0,0, \theta)=0 \\
&\beta_{,zz}(u_0, 0 ,\theta)=-\frac{1}{2} [\gamma_{,z}(u_0,0,\theta)]^2\\
& \beta_{,zzz}(u_0, 0, \theta)=-2  \gamma_{,z} (u_0, 0, \theta) \gamma_{,zz}(u_0, 0, \theta) \\
& \beta_{,zzzz}(u_0, 0, \theta)=3\left[ (\gamma_{,zz}(u_0, 0, \theta))^2- \gamma_{,z}(u_0, 0, \theta) \gamma_{,zzz}(u_0,0,\theta) \right].
\end{split}
\end{align}
This procedure can be continued to arbitrary order although the terms displayed above will be sufficient for our analysis. In solving the subsequent field equations, it will be left implicit that the equations are evaluated at $(u_0,0,\theta)$. Note that the iterative procedure does not produce an equation for $\beta(u_0, 0, \theta)$; the latter is an integration function, which we denote as $\beta_0(u, \theta)=\beta(u, 0, \theta)$.

\subsubsection{The second equation}
\noindent The second equation is \eqref{eq: main_eq_2}.
Given that we know both $\gamma$ and $\beta$ from the first equation, we can now solve this equation via the recursive differentiation procedure. The first two iterations give
\begin{subequations}
\begin{align}
U_{,z}&= 2 \beta_{0, \theta} e^{2( \beta_0- \gamma)} \label{eq: U_1} \\
U_{,zz}&=-2 e^{2 \beta_0-2 \gamma} (2 \beta_{0,\theta} \gamma_{,z}-2 \gamma_{,\theta}\gamma_{,z}+\gamma_{,z \theta}+2 \cot (\theta ) \gamma_{,z})
\end{align}
\end{subequations}
where all the equations are implicitly evaluated at $z=0$ on the hypersurface $\mathcal{N}_{u_0}$. 
The procedure does not constrain $U(u_0,0, \theta)$ thus giving an integration function $U_0(u, \theta)=U(u, 0, \theta)$.

The third iteration of the differentiation procedure does not give an equation for $U_{,zzz}$ but instead a constraint equation:
\begin{equation}
2\gamma_{,\theta} \gamma_{,zz}- \gamma_{, z z \theta}-2 \cot (\theta ) \gamma_{,z z}=0 \label{con-1}
\end{equation}
This equation was solved in the asymptotically flat case of \cite{Bondi:1962px} by setting $\gamma_{,zz}=0$, the `outgoing wave condition'. 
It has since been argued, most notably in \cite{Chrusciel:1993hx, Andersson:1994ng, Kroon:1998tu}, that this equation implies the existence of \textit{polyhomogeneous} asymptotic solutions for asymptotically flat spacetimes i.e. series involving terms of the form $z^i \log^j(z), \; i,j \in \mathbb{N}$. 

We will leave this equation unsolved for now and return to discuss it after we solve the fourth of the main equations; we will argue that the solution to that equation forbids the possibility of a polyhomogeneous form of the solution for non-zero cosmological constant (in the absence of matter). For now, we merely note that this equation indicates the presence of another integration function, as $U_{,zzz}(u_0, 0, \theta)$ remains undetermined by the iterative procedure. We name this function $U_3(u, \theta)= U_{,zzz}(u, 0, \theta)/3!$ where the choice of normalisation will become clearer as we continue to solve the main equations. In the asymptotically flat literature, this function is related to the \textit{Bondi angular momentum} of the spacetime \cite{Bondi:1962px, Madler:2016xju}. \par
 
The fourth iteration of the differentiation procedure produces the following equation 
\begin{align}
\begin{split}
U_{,zzzz}=&-2 e^{-2 \gamma} (-16 e^{2 \beta _{0}} \beta _{0, \theta} \gamma _{,z}^3+ 30 e^{2 \beta _{0}} \gamma _{, \theta} \gamma _{,z}^3-15 e^{2 \beta _{0}} \gamma _{,z \theta} \gamma _{,z}^2+ 4 e^{2 \beta _{0}} \beta _{0, \theta} \gamma _{,zz} \gamma _{,z}+\\
&10 e^{2 \beta _{0}} \gamma _{,zz} \gamma _{, \theta} \gamma _{,z}- 3 e^{2 \beta _{0}} \gamma _{,zz \theta} \gamma _{,z}+2 e^{2 \beta _{0}} \beta _{0, \theta} \gamma _{,zzz}+ 6 e^{2 \beta _{0}} \gamma _{,zzz} \gamma _{, \theta}-\\
 & 4 e^{2 \beta _{0}} \gamma _{,zz} \gamma _{,z \theta}- 3 e^{2 \beta _{0}} \gamma _{,zzz\theta}-30 \cot (\theta ) e^{2 \beta _{0}} \gamma _{,z}^3- \\
 & 10 \cot (\theta ) e^{2 \beta _{0}} \gamma _{,zz} \gamma _{,z}-6 \cot (\theta ) e^{2 \beta _{0}} \gamma _{,zzz}+3 e^{2 \gamma} \gamma _{,z} U_{,zzz}) 
\end{split}
\end{align}
which is an algebraic equation for $U_{,zzzz}$ in terms of $U_{,zzz}$. The presence of this equation makes sense because of the structure of the integration functions for equation (\ref{eq: main_eq_2}). If we were to repeat the differentiation procedure we would see that $\partial^{(n+1)}_z U$ would be given algebraically in terms of $\partial^{(n)}_z U$ for $ n  \geq 3$  so we observe that knowledge of $U_{,zzz}(u,0,\theta)$ would allow us to compute all higher derivatives at $z=0$. We will later see via the supplementary conditions that one does arrive at an evolution equation for $U_{,zzz}$. 

\subsubsection{The third equation}
The third equation is \eqref{eq: reg_me3};
this is the first equation that explicitly includes the cosmological constant $\Lambda$ and thus it will have different solutions from 
\cite{Bondi:1962px}.  \par

The equations are again solved by applying the iterative differentiation procedure:
\begin{subequations}
\begin{align}
W& = -\frac{1}{3} e^{2 \beta_0 } \Lambda \\
W_{,z}& =\cot (\theta ) U_0+U_{0,\theta} \\
\begin{split}
W_{,zz}&=e^{2(\beta_{0}-\gamma)}(2+\Lambda e^{2\gamma}(\gamma_{,z})^2+4\cot(\theta)\beta_{0,\theta}+8(\beta_{0,\theta})^2+ \\ 
& \phantom{\frac{1}{2}e^{2(\beta_{0}-\gamma_{0})}(=}6\cot(\theta)\gamma_{,\theta}-8\beta_{0,\theta}\gamma_{,\theta}-4(\gamma_{,\theta})^2+4\beta_{0,\theta \theta}+2\gamma_{,\theta \theta}).
\end{split}
\end{align}
\end{subequations}
The third equation does not give an algebraic equation for $W_{,zzz}$ but rather another constraint equation
\begin{equation}
\Lambda e^{2\beta_0} \gamma_{,z} \gamma_{,zz}=0 \quad \implies \quad  \gamma_{,z} \gamma_{,zz} =0. \label{con-2}
\end{equation} 
Note that this equation is unique to $\Lambda \neq 0$. We will not yet solve this constraint: the solution is determined by the fourth main equation. This constraint equation again implies an integration function for the third equation as the differentiation procedure does not produce an equation for $W_{,zzz}$. We will name this integration function $W_3(u, \theta)=W_{,zzz}(u, 0, \theta)/3!$. In the asymptotically flat case, this function is related to the \textit{Bondi mass aspect} of the spacetime, a concept we will examine in more detail in the asymptotically $AdS$ case in section \ref{sec: Holographic_interpretation}. \par

The structure of the higher order equations in the recursive differentiation procedure is similar to that of the second equation. The result of the procedure is that $\partial^{(n+1)}_z W$ is determined algebraically in terms of $\partial^{(n)}_z W$. We again remark that once we know the integration function $W_3$ we can then compute all derivatives of third order and higher. 

\subsubsection{The fourth equation}
\noindent The fourth and final main equation, which we consider as an equation for $\gamma_{,u}$, is \eqref{eq: reg_me4}.
We again apply the recursive scheme to solve for $\gamma_{,u}$. Using the solutions to the previous equations, the first 
non-trivial equation is 
\begin{equation} \label{eq: gamma_1}
\Lambda \gamma_{,z}=- \frac{3}{2}e^{-2\beta_0}(\cot(\theta)U_0-U_{0,\theta}-2U_0\gamma_{,\theta}-2\gamma_{,u})
\end{equation}
This equation is presented slightly differently to the previous main equations; we will discuss this further in section \ref{sec: Integration Scheme - Minkowski Vs. $AdS$}. The key point here is that the presence of the cosmological constant couples the equation for $\gamma_{,u}$ to $\gamma_{,z}$.  

The next non-trivial equation is  
\begin{equation} \label{eq: gamma2=0}
\Lambda e^{2\beta_0}\gamma_{,zz}=0 \implies \gamma_{,zz}=0.
\end{equation}
This constraint automatically solves the two previous constraint equations \eqref{con-1} and \eqref{con-2}. 
This is precisely the \textit{outgoing wave condition} that was enforced {\it a priori} in \cite{Bondi:1962px} and has since been understood in more generality in a Bondi type set up (see e.g. \cite{dInverno:1997xzo}). In the case of non-zero cosmological constant, $\gamma_{,zz} = 0$ is required by the field equations {\it i.e.} it is not an assumption.

At the next order of the recursive differentiation procedure, we find the equation 
\begin{equation}
e^{2\gamma}\gamma_{,uzz}(u_0,0, \theta)=0 \implies \gamma_{,uzz}(u_0,0, \theta)=0
\end{equation}
which implies that the form of $\gamma_{,zz}(u_0,0, \theta)$ is preserved on hypersurfaces $\mathcal{N}_{u}$ for $u > u_0$. Since 
 $\gamma_{,zz}(u_0,0,\theta)=0$ from (\ref{eq: gamma2=0}), the outgoing wave equation is propagated into $D^+(\mathcal{N}_0)$. \par

When $\Lambda=0$ equation  (\ref{eq: gamma_1}) implies $\gamma(u_0,0, \theta)=0$ (as we will shortly discuss in detail in section \ref{sub:general}), $\gamma_{,zz}=0$ for $u>u_0$ as we just discussed, and we are left with one integration function $\gamma_{,z}(u,0,\theta) =\gamma_1(u, \theta)$.  (The $u$-derivative of) this integration function is essentially the Bondi news.

Returning to the $\Lambda \neq 0$ case we note that the procedure of differentiation did not produce an equation for $\gamma_{,uz}(u_0, 0, \theta)$, again implying the presence of an integration functions   $\gamma_{,z}(u,0,\theta) =\gamma_1(u, \theta)$ (as in the $\Lambda=0$ case). Finally, to determine  $\gamma_{,u}$ (so that we can move to the next null hypersurface $\mathcal{N}_{u_0+\delta u}$) 
we also need to know non-trivial integration functions $(U_0, \beta_0)$. We will discuss in more detail (A)dS integration schemes in section \ref{sec: Integration Scheme - Minkowski Vs. $AdS$}.

As a final comment we note that the next non-trivial equation produced by the iterative procedure is  
\begin{align}
\begin{split}  \label{eq: gamma_4}
\Lambda \gamma_{,zzzz}=
-\frac{3}{2} e^{-2 (\beta _{0}+\gamma)} (&48 e^{2 \beta _{0}} \beta _{0, \theta}^2 (\gamma _{,z})^2-96 e^{2 \beta _{0}} \beta _{0, \theta} \gamma _{, \theta} (\gamma _{,z})^2+6e^{2 \beta _{0}} \beta _{0, \theta \theta} (\gamma _{,z})^2-\\
&24 e^{2 \beta _{0}} \gamma _{, \theta \theta} (\gamma _{,z})^2+108 e^{2 \beta _{0}} \beta _{0, \theta} \gamma _{,z \theta} \gamma _{,z}-48 e^{2 \beta _{0}} \gamma _{, \theta} \gamma _{,z \theta} \gamma _{,z}+\\
&18 e^{2 \beta _{0}} \gamma _{,z \theta \theta} \gamma _{,z}+18 e^{2 \beta _{0}} (\gamma _{,z \theta})^2-24 \cot ^2(\theta ) e^{2 \beta _{0}} (\gamma _{,z})^2+\\
&90 \cot (\theta ) e^{2 \beta _{0}} \beta _{0, \theta} (\gamma _{,z})^2+24 \cot (\theta ) e^{2 \beta _{0}} \gamma _{, \theta} (\gamma _{,z})^2+\\
&30 \cot (\theta ) e^{2 \beta _{0}} \gamma _{,z \theta} \gamma _{,z}-24 \csc ^2(\theta ) e^{2 \beta _{0}} (\gamma _{,z})^2-8 e^{2 \gamma} \gamma _{,uzzz}+\\
&e^{2 \gamma} U_{,zzz} (-6 \beta _{0, \theta}- 2\gamma _{, \theta}+\cot (\theta ))-12 e^{2 \gamma } \gamma _{,zzz} U_{0, \theta}-\\
&e^{2 \gamma } U_{,zzz\theta}-8 e^{2 \gamma} \gamma _{,zzz\theta} U_{0}-\\
&12 \cot (\theta ) e^{2 \gamma} \gamma _{,zzz} U_{0}- 2e^{2 \gamma } \gamma _{,z} W_{,zzz})
\end{split}
\end{align}
which shows that the evolution equation for $\gamma_{,uzzz}$ is coupled to $\gamma_{,zzzz}$ via the cosmological constant $\Lambda$. This coupling is a general feature of this field equation at higher orders, namely the equation for $\partial^{(n)}_{z} \gamma_{,u}$ is given in terms of $\partial^{(n+1)}_{z} \gamma$. So if we provide a new integration function 
$\gamma_{,zzz}(u, 0, \theta)/3!=\gamma_3(u, \theta)$ then all higher order terms are determined. A more detailed discussion will be given in section \ref{sec: Integration Scheme - Minkowski Vs. $AdS$}.

\subsection{General form of the asymptotic solutions} \label{sub:general}

Using the procedure of recursive differentiation we have obtained a general form for the asymptotic solution to the field equations. The key to this structure is that $\gamma_{,zz}(u,0,\theta)=0$ which results in the vanishing of potential polyhomogeneous terms in the asymptotic solution (as discussed in \cite{Chrusciel:1993hx}). Note that the vanishing of this term is forced by equation (\ref{eq: gamma2=0}) rather than being assumed as it was in \cite{Bondi:1962px}. 

In previous literature it has been found that the metric function expansions can contain logarithmic terms of the form $\log^j(r)r^{-i}$, both for the asymptotically flat case in \cite{Andersson:1994ng, 0264-9381-16-5-314} and for arbitrary $\Lambda$ but with matter in \cite{Chrusciel:2016oux}.  These cases are qualitatively different. In the asymptotically flat case there is no analogue of (\ref{eq: gamma2=0}). In the presence of a negative cosmological constant, logarithmic terms in asymptotic expansions arise whenever the coupled matter is of specific masses, see for example \cite{ deHaro:2000vlm, Skenderis:2002wp}; such matter is associated with matter conformal anomalies in the dual CFT. 
 \par

For pure cosmological constant, the most general form of the asymptotic solutions take the form of power series about $z=0$, consistent with the boundary conditions of asymptotically locally AdS$_4$ and dS$_4$ in the absence of matter. Specifically, $\gamma$ admits an expansion of the form 
\begin{equation} \label{eq: Taylor_Series}
\gamma(u,r,\theta)=\sum_{n=0}^{\infty}\gamma_n(u, \theta) z^n=\sum_{n=0}^{\infty} \frac{\gamma_n(u,\theta)}{r^n}, \qquad \gamma_{n}=\frac{\partial^{(n)}_z  \gamma}{n!}\bigg\rvert_{z=0}
\end{equation}
and the other functions admit analogous expansions. These conditions ensure that the metric coefficients do not grow exponentially with $r$ and that the metric has a pole of order two at the conformal boundary $\mathscr{I}$; this will be discussed in greater detail in section \ref{sec: Holographic_interpretation}. \par

We also note how the presence of the cosmological constant in the Einstein equations modifies the solutions compared with the asymptotically flat case considered in \cite{Bondi:1962px}, even though the asymptotic series form of the equations initially seems to be the same. At this point it will be helpful to consider the AdS and dS cases separately, as there are subtle differences in the two cases. 

The key assumption made in \cite{Bondi:1962px} which results in this discrepancy is that the vector field $\partial_u$ is everywhere timelike $\iff g_{uu}<0$. Physically, this is a reasonable condition to impose on asymptotic solutions for $\Lambda \leq 0$, as the neighbourhood of the conformal boundary in these cases is exterior to any potential region where $\partial_u$ ceases to be timelike (e.g inside a horizon). We note that the leading order terms $(\gamma_0, \beta_0, U_0)$ are not present in the asymptotically flat case and are forced to vanish due to this condition. 

These choices are overly restrictive in the AdS case as the cosmological constant allows for freedom in these functions. To see this, consider the limit
\begin{equation}
\lim_{r \rightarrow \infty} \frac{g_{uu}}{r^2} = -(W_0 e^{2\beta_0} - U_0^2 e^{2\gamma_0}) < 0 
\end{equation}
where the inequality on the right hand side follows from the condition that $\partial_u$ is timelike. In the flat case, $W_0=0$ and so the above equation reduces to $U_0=0$. It was then argued in \cite{Bondi:1962px} that $U_0=0$ implies $\gamma_{0,u}=0$ (use (\ref{eq: gamma_1}) with $\Lambda=0$) and this may be reduced further to $\gamma_0=0$ by using a coordinate transformation (for details see \cite{Bondi:1962px}). 

In the AdS case $W_0=-\Lambda e^{2\beta_0} /3$ so the inequality is different, namely 

\begin{equation}
\frac{\Lambda e^{4\beta_0}}{3}+ U_0^2 e^{2\gamma_0} < 0 \Rightarrow |U_0|< \sqrt{-\frac{\Lambda}{3}e^{4\beta_0-2\gamma_0}}=\frac{e^{2\beta_0-\gamma_0}}{l}
\end{equation}

\noindent from which we see that $U_0$ can now clearly be non-zero, implying that generically  $\gamma_0 \neq 0$ also. 

The integration function $\beta_0$ is also set to zero in the flat case, using the freedom in the BMS group. Since the BMS group is the asymptotic symmetry group of flat space-time it would be premature to make the same choice before determining the $AdS$ asymptotic structure. For the time being we will choose $\beta \neq 0 $ to retain full generality. \par

Turning now to the dS case of $\Lambda > 0$, the previously imposed condition of $\partial_u$ being timelike is unphysical in the asymptotic region. Using the Bondi gauge in a neighbourhood of $\mathscr{I}^+$, the cosmological horizon in the asymptotically locally dS spacetime must have been crossed, and thus the vector field $\partial_u$ is spacelike in the region of interest, see discussion in \cite{Anninos:2012qw}. Thus one should not impose this condition in the dS case, leaving $(\gamma_0, \beta_0, U_0 )$ generically unconstrained. 

A  second important difference to note, for any non-zero cosmological constant, is that the cosmological constant couples the fourth equation at each order in $z$. We find equations which give $\partial^{(n)}_z \gamma_{,u}$ in terms of $\partial^{(n+1)}_z \gamma$, {\it e.g.} (\ref{eq: gamma_1}) and (\ref{eq: gamma_4}). This coupling of orders together with the structure of the other main equations implies that if we are given suitable seed coefficients then we can obtain all the other expansion coefficients. The initial coefficients are $(\gamma_0, \beta_0, U_0)$ together with $(\gamma_3, U_3, W_3)$; from these the entire solution can be determined algebraically. We will  see below that these coefficients have an important holographic interpretation but first we analyse the remaining Einstein equations, the so-called {\it supplementary conditions}.

\subsection{The supplementary conditions} \label{subsec: SC}
Although the main equations give equations for the four metric functions, they do not form the complete set of field equations. The remaining three supplementary equations are:
\begin{eqnarray}
R_{uu} &=&\Lambda g_{uu}=-\Lambda(Wr^2e^{2\beta}-U^2r^2e^{2\gamma}); \\ 
R_{u\theta} &=& \Lambda g_{u \theta}=-\Lambda Ur^2e^{2\gamma}, \quad R_{u r}=\Lambda g_{u r}=-\Lambda e^{2\beta}. \nonumber 
\end{eqnarray} 
In the asymptotically flat case, these equations were denoted as {supplementary conditions} as they are automatically satisfied provided they hold on a particular hypersurface of constant radius and the main equations are satisfied \cite{Sachs:1962wk}. In this section we will discuss how  this property carries over to the $\Lambda \neq 0$ case. 

Following the original work, the supplementary conditions are derived from the contracted Bianchi identity
\begin{equation}
\nabla^{\nu}G_{\nu \mu}=g^{\nu \sigma}\nabla_{\sigma }\left(R_{\nu \mu}-\frac{1}{2}g_{\nu \mu}R\right)=0.
\end{equation}
We can expand the Bianchi identity as
\begin{equation}
g^{\nu \sigma}\left(R_{\mu \nu, \sigma}- \Gamma^{\beta}_{\sigma \nu}R_{\beta \mu}\right) - g^{\nu \sigma}\Gamma^{\beta}_{\sigma \mu} R_{\beta \nu}-\frac{1}{2} R_{,\mu}=0 
\end{equation}
and using $R_{,\mu}=\nabla_{\mu}(g^{\nu \sigma}R_{\nu \sigma})=g^{\nu \sigma} \nabla_{\mu} R_{\nu \sigma}=g^{\nu \sigma} R_{\nu \sigma, \mu}-2g^{\nu \sigma} \Gamma^{\beta}_{\sigma \mu} R_{\beta \nu}$ allows us to write the contracted Bianchi identity as 
\begin{equation}  \label{eq: Contracted_Bianchi_Identity}
g^{\nu \sigma} \left(R_{\mu \nu, \sigma}-\frac{1}{2} R_{\nu \sigma, \mu} -\Gamma^{\beta}_{\nu \sigma} R_{\beta \mu}\right)=0.
\end{equation} 
To analyse the components of the contracted Bianchi identity we use the inverse metric 
\begin{equation} \label{eq: inverse_metric}
g^{\mu \nu}= \begin{pmatrix} 
0 & -e^{-2\beta} & 0 & 0 \\
-e^{-2\beta}  & We^{-2\beta}r^2 & -Ue^{-2\beta} & 0 \\
0 & -Ue^{-2\beta} & e^{-2\gamma}r^{-2} & 0 \\
0 & 0 & 0 & e^{2\gamma}r^{-2} \sin^{-2} \theta 
\end{pmatrix}
\end{equation}
where we use the coordinates $(u, r, \theta, \phi)$. The following identity is also useful:
\begin{equation} \label{eq: Christoffel_Identity}
g^{\mu \nu} \Gamma^{u}_{\mu \nu}= 2e^{-2\beta} r^{-1}
\end{equation}
This identity is computed using the inverse metric above and the metric (\ref{eq: Bondi_Metric}); the same identity was given in \cite{Bondi:1962px}, up to a sign change due to different signature conventions. \par

We will now examine the components of the contracted Bianchi identity (\ref{eq: Contracted_Bianchi_Identity}) and show that they lead to the supplementary equations. When doing this, we enforce the main equations, expressed as 
\begin{equation}
R_{rr}=R_{r \theta}=0, \quad R_{\theta \theta}=\Lambda g_{\theta \theta}, \quad R_{\phi \phi}=\Lambda g_{\phi \phi}
\end{equation}
as well as the trivial equations $R_{u \phi}=R_{r \phi}=R_{\theta \phi}=0$.

Let us consider first the $\mu=r$ component of (\ref{eq: Contracted_Bianchi_Identity}) 
\begin{equation}
g^{\nu \sigma} \left(R_{r \nu, \sigma}-\frac{1}{2} R_{\nu \sigma, r} -\Gamma^{\beta}_{\nu \sigma} R_{\beta r}\right)=0.
\end{equation}
Using the main and trivial equations this reduces to 
\begin{equation}
-\frac{1}{2}\left(g^{\theta \theta}\Lambda g_{\theta \theta, r}+g^{\phi \phi} \Lambda g_{\phi \phi, r}\right)-g^{\nu \sigma} \Gamma^{u}_{\nu \sigma} R_{u r}=0. 
\end{equation}
The latter term can be processed using the identity (\ref{eq: Christoffel_Identity}) and after algebraic manipulation we obtain
\begin{equation}
R_{u r}=-\Lambda e^{2\beta}=\Lambda g_{u r}.
\end{equation}
which is precisely the $\{u r \}$ component of the field equations. Thus we conclude that if the main equations hold then the $\{u r \}$ equation is automatically satisfied. 

Next consider the $\mu=\theta$ component of (\ref{eq: Contracted_Bianchi_Identity}) 
\begin{equation}
g^{\nu \sigma} \left(R_{\theta \nu, \sigma}-\frac{1}{2} R_{\nu \sigma, \theta} -\Gamma^{\beta}_{\nu \sigma} R_{\beta \theta}\right)=0
\end{equation}
Uising the main and trivial equations we obtain
\begin{align} \label{eq: SC1_midpoint}
\begin{split}
&g^{u r}R_{u \theta, r}-g^{\nu \sigma} \Gamma^u_{\nu \sigma} R_{u \theta}+ \\
&\phantom{aaa}\Lambda \left( g^{r \theta}g_{\theta \theta,r}+\frac{1}{2}g^{\theta \theta}g_{\theta \theta, \theta} - g^{u r}g_{u r, \theta}-\frac{1}{2}g^{\phi \phi}g_{\phi \phi, \theta}-g^{\nu \sigma}\Gamma^{\theta}_{\nu \sigma} g_{\theta \theta}\right)=0.
\end{split}
\end{align}
Applying equation (\ref{eq: Christoffel_Identity}) to the second term on the first line and using equations (\ref{eq: Bondi_Metric}) and (\ref{eq: inverse_metric}) to write the second line in terms of metric functions we obtain
\begin{equation}
-r^{-2}e^{-2\beta}\frac{\partial}{\partial r}(r^2 R_{u \theta})=\Lambda e^{2\gamma- 2\beta} r (r U_{,r}+2 U(2+\gamma_{,r})) 
\end{equation} 
which can be integrated to give 
\begin{equation}
r^2 R_{u \theta}=- \Lambda Ur^4 e^{2\gamma} + f(u, \theta)
\end{equation}
where $f(u,\theta)$ is an integration function. Dividing through by $r^2$ gives
\begin{equation}
R_{u \theta}=-\Lambda Ur^2 e^{2\gamma}+\frac{f(u, \theta)}{r^2}=\Lambda g_{u \theta}+\frac{f(u,\theta)}{r^2}.
\end{equation}
which implies that the $\{u \theta \}$ component of the Einstein equations is only satisfied if $f(u, \theta)=0$; this is our first supplementary condition. \par

Finally consider the $\mu=u$ component of the contracted Bianchi identity
\begin{equation}
g^{\nu \sigma} \left(R_{u \nu, \sigma}-\frac{1}{2} R_{\nu \sigma, u} -\Gamma^{\beta}_{\nu \sigma} R_{\beta u}\right)=0.
\end{equation}
Applying the field equations (including $f(u,\theta) = 0$), we obtain
\begin{align}
\begin{split}
&g^{u r}R_{u u , r}-g^{\nu \sigma}\Gamma^{u}_{\nu \sigma} R_{u u}+ \\
&\phantom{a} \Lambda \bigg[ g^{rr} g_{u r, r}+g^{r \theta} (g_{u r, \theta}+  g_{u \theta, r}) + g^{\theta \theta}\left( g_{u \theta, \theta}-\frac{1}{2} g_{\theta \theta, u}\right)-\\
&\phantom{aaa} \frac{1}{2}g^{\phi \phi} g_{\phi \phi, u}-g^{\nu \sigma}\Gamma^r_{\nu \sigma} g_{u r}- g^{\nu \sigma}\Gamma^{\theta}_{\nu \sigma}g_{u \theta} \bigg] = 0.
\end{split}
\end{align} 
The structure of this equation is similar to that of (\ref{eq: SC1_midpoint}): the first line contains the Ricci tensor terms of interest and all other terms can be written explicitly using (\ref{eq: Bondi_Metric}) and (\ref{eq: inverse_metric}). Doing this gives 
\begin{equation}
-r^{-2}e^{-2\beta}\frac{\partial}{\partial r}(r^2 R_{uu})=\Lambda r[2W(2+r \beta_{,r})+rW_{,r}]-2\Lambda e^{2\gamma-2\beta}r U [2U+rU_{,r}+rU\gamma_{,r}]
\end{equation}
which can be integrated to give
\begin{equation}
r^2R_{uu}=\Lambda r^4 (-W e^{2\beta} + U^2 e^{2\gamma})+g(u, \theta)
\end{equation}
with $g(u,\theta)$ an integration function. Thus
\begin{equation}
R_{uu}=\Lambda r^2 (-W e^{2\beta} + U^2 e^{2\gamma}) + \frac{g(u, \theta)}{r^2}= \Lambda g_{uu}+\frac{g(u,\theta)}{r^2}
\end{equation}
implying that the second supplementary condition is $g(u, \theta)=0$. 

Explicit expressions for the supplementary conditions may be derived using the solutions to the main equations up to $\mathcal{O}(1/r^4)$ for $(\gamma, \beta, U, W)$ and then inputting these into the above equations to derive expressions for $(f, g)$. 
The resulting equations take the form of evolution equations for $U_3$ and $W_3$ and they will be discussed further in section \ref{subsec: hybrid_scheme}. The explicit expressions 
for these equations can be found in appendix \ref{sec: SCapp}. Here we present the much simpler expressions for asymptotically (A)dS and flat spacetimes.

Asymptotically (A)dS spacetimes in Bondi coordinates have $\gamma_0=\beta_0=U_0=0$ (this will be shown explicitly in section \ref{sec: Holographic_interpretation}) which gives $\gamma_1=0$ by equation (\ref{eq: gamma_1}). Setting these values in the supplementary equations gives us
\begin{subequations}
\begin{align}
U_{3,u}&=\frac{1}{3}(4\Lambda \cot(\theta) \gamma_3 + W_{3, \theta}+ 2\Lambda \gamma_{3,\theta})  \\
W_{3,u}&=-\frac{1}{2}\Lambda (\cot(\theta) U_3 + U_{3, \theta}) 
\end{align}
\end{subequations}
For the asymptotically flat supplementary conditions, we again have $\gamma_0=\beta_0=U_0=0$ as well as $\Lambda=0$ but now $\gamma_1 \neq 0$. Then 
\begin{subequations}
\begin{align}
U_{3,u}&=\frac{1}{3} \left(7 \gamma_{1,\theta} \gamma_{1,u}+\gamma_1 \left(3 \gamma_{1,u\theta}+16 \cot (\theta ) \gamma_{1,u}\right)+W_{3, \theta}\right)  \\
W_{3,u}&=2 \left(\gamma_{1,u}\right)^2+2 \gamma_{1,u}-\gamma_{1,u\theta\theta}-3 \cot (\theta ) \gamma_{1,u\theta}.
\end{align} 
\end{subequations}
in agreement with the expressions given in \cite{Bondi:1962px}.


\section{Integration scheme} \label{sec: Integration Scheme - Minkowski Vs. $AdS$}

In this section we will discuss the integration scheme used in the previous section in order to solve the Einstein equations. We will begin with a reminder of the Bondi integration scheme in asymptotically flat spacetime before focusing specifically on the $\Lambda < 0$ case of asymptotically locally AdS spacetime. We will propose two modified integration schemes for the AdS case which will be compared and contrasted to the flat scheme. Much of what we will discuss for the AdS case has a corresponding description in the $\Lambda > 0$ case of asymptotically locally dS spacetime, a topic we will discuss briefly here and elaborate upon in future work.

\subsection{The flat scheme}

Let us briefly review the integration scheme in the asymptotically flat case as presented in \cite{Bondi:1962px}. The basic quantity necessary to solve the field equations for all $u$ was the knowledge of $\gamma$ on some initial null hypersurface $\mathcal{N}_{u_0}$; this allows us to solve the main equations up to the undetermined integration functions. In the Ricci flat case we can reapply the field equations (\ref{eq: AdS_me1}-\ref{eq: AdS_me4}) although we now set $\Lambda=0$ in those equations. For the remainder of this subsection we have $\Lambda =0$. 

Knowledge of $\gamma |_{\mathcal{N}_{u_0}}$ allows us to solve for the other functions. Disregarding integration functions, (\ref{eq: AdS_me1}) determines $\beta |_{\mathcal{N}_{u_0}}$; (\ref{eq: AdS_me2}) determines $U |_{\mathcal{N}_{u_0}}$, (\ref{eq: AdS_me3}) gives $W|_{\mathcal{N}_{u_0}}$ and (\ref{eq: AdS_me4}) allows us to compute our $\gamma$ at the next time step i.e $\gamma  |_{\mathcal{N}_{u_0+\delta}}$. Iterating this process allows us to determine all metric functions at time $u > u_0$, i.e. the functions in the future domain of dependence of $\mathcal{N}_{u_0}$, $D^+(\mathcal{N}_{u_0})$.

\begin{figure}[h]
\begin{center}
\includegraphics[width=0.5\linewidth]{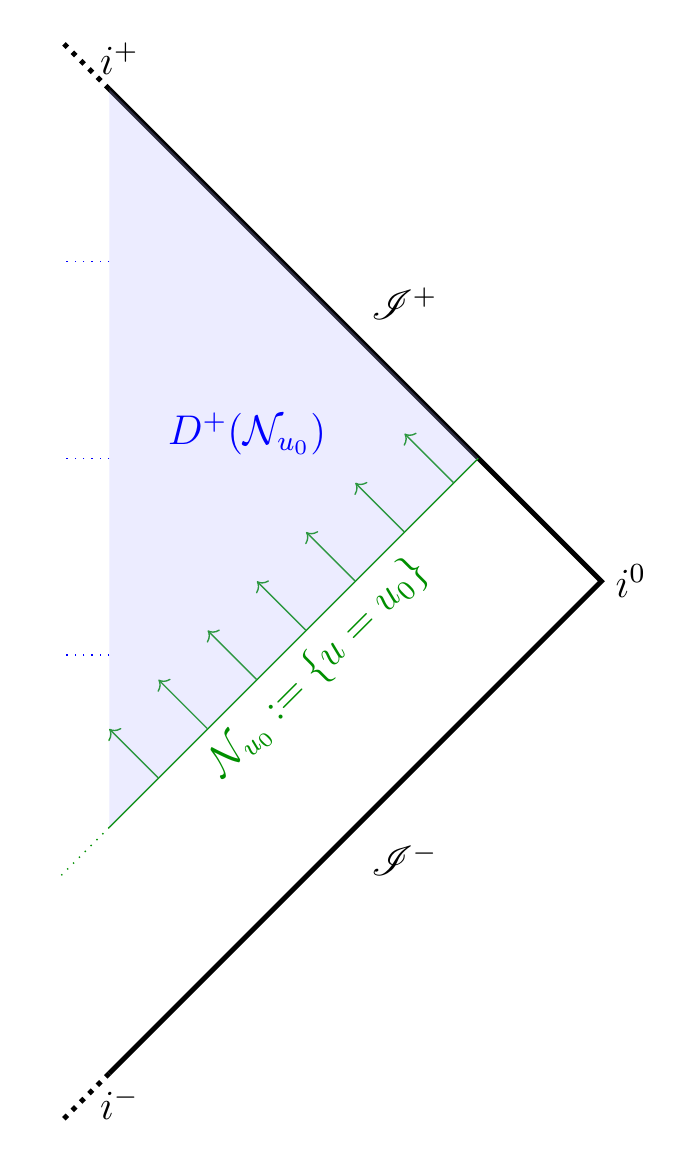}
\end{center}
\caption{Penrose diagram illustrating the integration scheme for asymptotically flat space time}
\end{figure}

Turning to the integration functions, we recall that the main equations in the flat case admit five such functions; $(\beta_0, U_0, U_3, \gamma_1, W_3)$. The original argument of \cite{Bondi:1962px} was that $U_0$ and $\beta_0$ could be set to zero. $U_0$ is set to zero to preserve the condition that the vector field $\partial_u$ is everywhere timelike and $\beta_0$ can be fixed to zero using the freedom of the BMS group. These restrictions also give $\gamma_{0,u}=0$ and thus we can also set $\gamma_0=0$ by a suitable BMS transformation. \par

Such considerations reduce the number of unknown functions to three: $(\gamma_1, U_3, W_3)$, all of which are functions of $u$ and $\theta$. These integration functions have well understood physical meaning: $\gamma_1$ plays the role of the Bondi news function, $U_3$ the Bondi angular momentum aspect and $W_3$ the Bondi mass aspect \cite{Bondi:1962px}. If we know the values of these three functions and we know $\gamma |_{\mathcal{N}_{u_0}}$, then from the main equations we can obtain the full solution to the Einstein equations in the region $D^+(\mathcal{N}_{u_0})$. The integration scheme runs as follows (for the no-log case with $\gamma_2=0$)
\begin{equation}
\gamma_1(u, \theta) \xrightarrow{(\ref{eq: AdS_me1})} \beta_1, \beta_2, \beta_3 \xrightarrow{(\ref{eq: AdS_me2})} U_1, U_2 \xrightarrow{(\ref{eq: AdS_me3})} W_0, W_1, W_2, W_4  \tag{4.1a}
\end{equation} 
so $\gamma_1$ gives us these functions. The rest of the scheme is 
\begin{figure}[H]
\begin{center}
\includegraphics[width=1.005\linewidth]{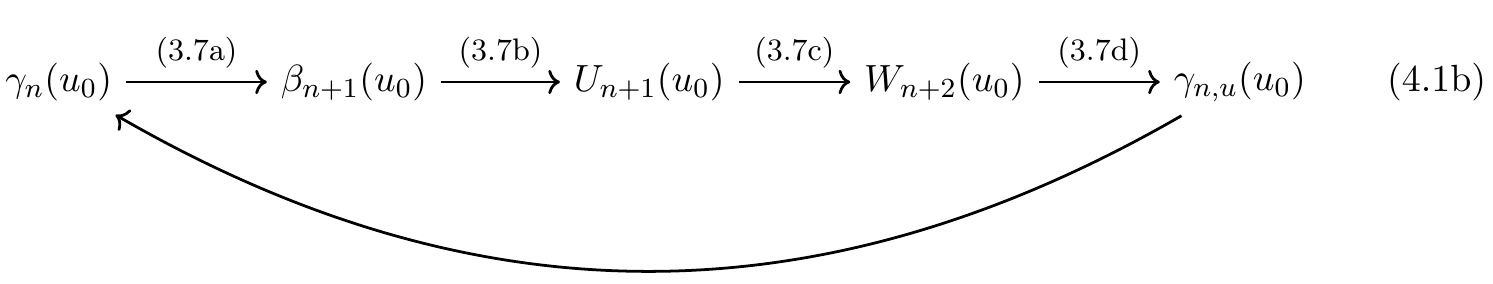}
\end{center}
\end{figure}
\noindent where the subscript $n>2$. The final arrow going back to the original function indicates that we are solving for $\gamma$ at the next instant of time i.e. $u_0+\delta u_0$, so iteration gives us the future evolution. We note that if the functions $U_3, W_3$ are not specified for all $u$ a priori, this scheme treats them as integration functions which are constrained by the supplementary conditions (\ref{eq: SC1}), (\ref{eq: SC2}) respectively. We will return to discuss these equations in the context of the AdS integration schemes to follow but for now these steps outline the procedure of the integration scheme in the asymptotically flat case, using some of the simplifications that BMS originally applied (namely $\gamma_2=0$).  

\subsection{The AdS integration schemes} \label{subsec: AdS_Scheme}

In order to understand how one needs to modify the specified data in the case of asymptotically locally AdS spacetimes it is convenient to first observe the results when one na\"ively applies the flat scheme as described in the previous section to asymptotically locally AdS spacetime. 

To repeat the steps of the flat scheme we again specify $\gamma$ on an initial null hypersurface $\mathcal{N}_{u_0}$ as well as $\gamma_1, U_3, W_3$ over the whole spacetime. The issue with applying this procedure to an asymptotically AdS spacetime is that we now have three additional integration functions $(\gamma_0(u,\theta), \beta_0(u, \theta), U_0(u, \theta))$ and in particular $\beta_0, U_0$ will not be determined using the Einstein equations (\ref{eq: AdS_me1}-\ref{eq: AdS_me4}) and the specified data ($\gamma_0$ would be determined using $\gamma_0$ on $\mathcal{N}_{u_0}$ and equation (\ref{eq: gamma_1})). These functions will also appear in the expressions for the higher order coefficients (e.g (\ref{eq: U_1})) and can be seen in the evolution equation for $\gamma_0$ (\ref{eq: gamma_1}). Clearly we will need an alternative integration scheme which specifies these functions and thus generates a fully determined solution to the field equations.

This issue can also be framed in terms of a causal picture as in figure \ref{fig: fig_3}. In this figure we see that specifying $\gamma$ on an initial null hypersurface $\mathcal{N}_{u_0}$ and $\gamma_1, U_3, W_3$ for $u \geq u_0$ and following the flat scheme will give us the solution in $D^+(\mathcal{N}_{u_0})$. In the AdS case (unlike the flat case) this region is not equivalent to the causal future of the null hypersurface, $J^+(\mathcal{N}_{u_0})$ (as shown in figure \ref{fig: AdS_dom_of_dep} below). In order to solve the Einstein equations for $J^+(\mathcal{N}_{u_0})$ in asymptotically locally AdS space-time, one either has to specify extra data on an additional hypersurface or different data to that of $\gamma$ on the null slice $\mathcal{N}_{u_0}$. We will now present two different integration scheme for asymptotically locally AdS spacetimes which will allow one to solve the field equations in $J^+(\mathcal{N}_{u_0})$.

\begin{figure}[h] 
\begin{center} 
\includegraphics[width=0.45\linewidth]{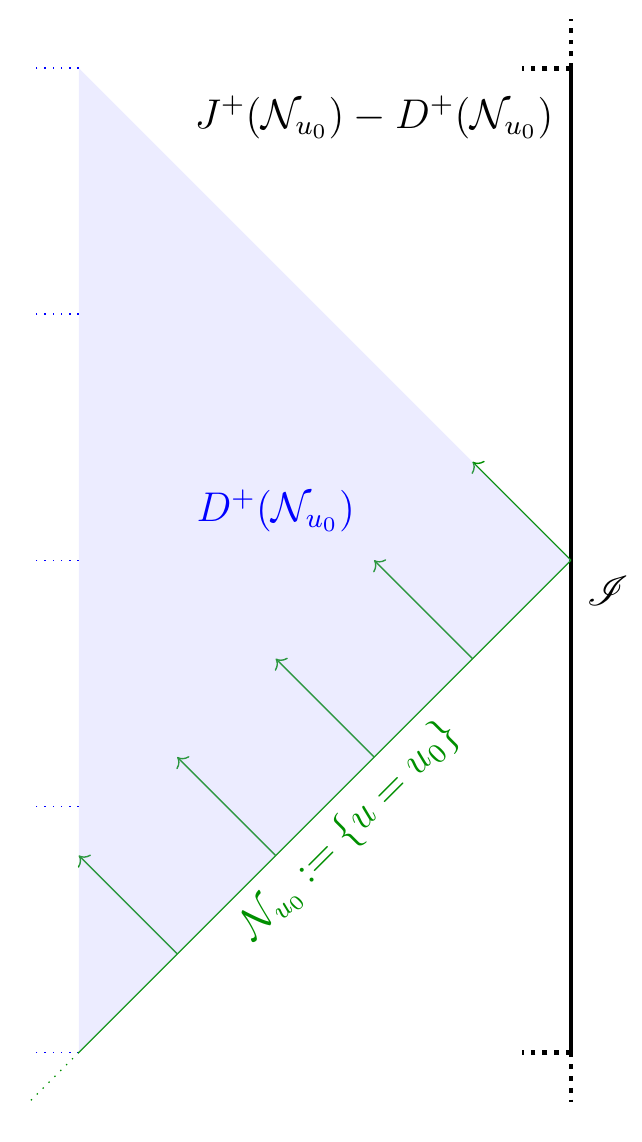}
\end{center}
\caption{Penrose diagram illustrating the difference between $D^+(\mathcal{N}_{u_0})$ and $J^+(\mathcal{N}_{u_0})$ in asymptotically locally AdS spacetime.}
\label{fig: AdS_dom_of_dep}
\end{figure}

\subsubsection{The ``boundary'' scheme} \label{subsec: AdS_boundary_scheme}

The first scheme we present is one which we will refer to as the ``boundary'' scheme where instead of specifying all coefficients $\gamma_i$ on an initial null hypersurface, one should specify certain coefficients (of our metric functions) for all available Bondi time, and use these coefficients in order to make the equations algebraic. The coefficients that should be specified are
\begin{equation}
\gamma_0, \quad  \beta_0, \quad U_0, \quad \gamma_3, \quad U_3, \quad W_3. \tag{4.2}
\end{equation}



We will see later that these particular coefficients admit a natural holographic interpretation. Even before relating them to coefficients in the Fefferman-Graham expansion, one can note that the coefficients $(\gamma_0,  \beta_0, U_0)$ clearly specify the values of the metric functions $(\gamma, \beta, U)$ at the conformal boundary $\mathscr{I}$;
\begin{equation} 
\lim_{r \rightarrow \infty} \gamma(u,r,\theta) = \gamma_0(u, \theta), \quad \lim_{r \rightarrow \infty} \beta(u,r,\theta) = \beta_0(u, \theta), \quad \lim_{r \rightarrow \infty} U(u,r,\theta) = U_0(u, \theta) \tag{4.3}
\end{equation}
and thus define the boundary metric for the dual conformal field theory.
We will understand the precise physical meaning of the components $(\gamma_3, U_3, W_3)$ in section \ref{sec: Holographic_interpretation}, when we discuss the relation to the asymptotic expansion in Fefferman-Graham gauge. In particular, we will see that the $(\gamma_0,  \beta_0, U_0)$ and $(\gamma_3, U_3, W_3)$ are conjugate variables in a radial Hamiltonian formalism, thus explaining why they provide a good set of initial data.


The scheme works in two parts. Given the boundary data $(\gamma_0, \beta_0, U_0)$ we see that the first part of the integration scheme is 
\begin{align} 
\begin{split}
\gamma_0, \beta_0, U_0 & \xrightarrow{(\ref{eq: AdS_me1}) } \beta_1  \xrightarrow{(\ref{eq: AdS_me2}) } U_1 \xrightarrow{(\ref{eq: AdS_me3}) } W_0, W_1 \xrightarrow{(\ref{eq: AdS_me4}) } \gamma_1 \ldots \\
\ldots & \xrightarrow{(\ref{eq: AdS_me1}) } \beta_2 \xrightarrow{(\ref{eq: AdS_me2}) } U_2 \xrightarrow{(\ref{eq: AdS_me3}) } W_2 \xrightarrow{(\ref{eq: AdS_me4}) } \gamma_2 \ldots \\
\ldots & \xrightarrow{(\ref{eq: AdS_me1}) } \beta_3.
\end{split}
\tag{4.4}
\end{align}
In words: we specify the data $(\gamma_0, \beta_0, U_0)$ at $\mathscr{I}$; (shown in figure \ref{fig: AdS_scheme_part_1} below) at the 2-surface where a particular null hypersurface $\color{islamicgreen}{\mathcal{N}_{u_0}}$ meets the conformal boundary. We can solve equations (\ref{eq: AdS_me1})-(\ref{eq: AdS_me3}) algebraically for the coefficients $\color{islamicgreen}{\beta_1, U_1, W_0, W_1}$.  This is indicated in the figure by the solid \textcolor{islamicgreen}{green} arrow in the diagram which points from $\mathscr{I}$ to the timelike surface $r=r_1$. 

\begin{figure}[h]
\begin{center}
\includegraphics{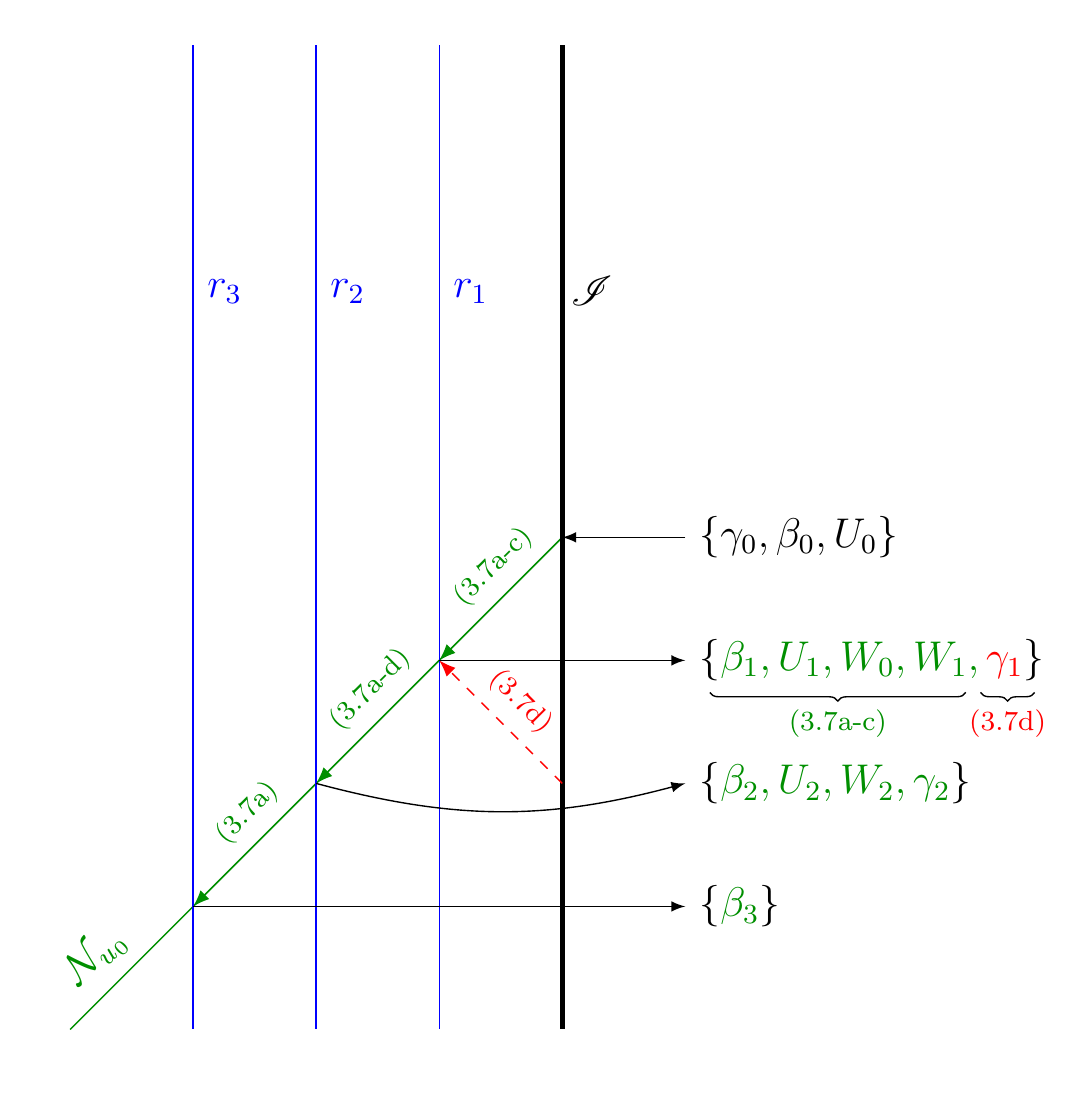}
\end{center}
\caption{Penrose diagram for AdS indicating discretely how the first part of the scheme is solved. This figure only includes one hypersurface, $\color{islamicgreen}{\mathcal{N}_{u_0}}$, for clarity; when solving the equations explicitly we would consider all null surfaces $\mathcal{N}_i$ in the foliation.}
\label{fig: AdS_scheme_part_1}
\end{figure}
 
In order to continue the scheme, we need to know $\gamma_{0,u}$ as this function will allow us to algebraically solve equation (\ref{eq: AdS_me4}) at the lowest non-trivial order for $\color{red}{\gamma_1}$ (equation (\ref{eq: gamma_1})). 
Since we know all values of $\gamma_0$ on $\mathscr{I}$ and we know $\gamma_{0,u}$.
The knowledge of this derivative is indicated in the diagram by the dotted \textcolor{red}{red} arrow which points into the bulk spacetime, again ending on the timelike surface $r=r_1$. In order to implement this step in a numerical scheme, one would want to know $\gamma_0(u_0)$ and $\gamma_0(u_0-\delta u_0)$ and construct a backward difference. This explains why the dotted \textcolor{red}{red} arrow starts at a different cut of $\mathscr{I}$, simply to indicate that we have used the extra information of $\gamma_0(u_0-\delta u_0)$ (and thus $\gamma_{0,u}$ discretely) in order to solve (\ref{eq: AdS_me4}). 
 
The arrows point towards smaller values of $\color{blue}{r}$ as we solve the Einstein equations. The purpose of this is to show that as we solve the Einstein equations, we obtain the values of higher order coefficients in the metric functions $\gamma, \beta, U, W$. Obtaining these higher order coefficients extends the series expansions (\ref{eq: Taylor_Series}) to higher powers of $1/r$, hence our solution includes contributions from smaller (but still asymptotic) values of $\color{blue}{r}$.
 
After these first steps have been performed, we solve (\ref{eq: AdS_me1}-\ref{eq: AdS_me4}) algebraically  for $\color{islamicgreen}{\beta_2, U_2, W_2, \gamma_2, \beta_3}$ (no extra evolution equation is needed as the field equation imply $\gamma_2=0$, as noted earlier). Knowledge of these functions is not enough to continue the integration scheme as the next unknown function in the field equations is $U_3$, an integration function which cannot be determined by the iteration process. 

We now give the second piece of the scheme: now the functions $\gamma_3, U_3, W_3$ are specified for all Bondi time $u$. This allows us to compute the higher order metric function coefficients via the following application of the Einstein equations     
\begin{align}
\begin{split}
\gamma_3, U_3, W_3 & \xrightarrow{(\ref{eq: AdS_me1}) } \beta_4 \xrightarrow{(\ref{eq: AdS_me2}) } U_4 \xrightarrow{(\ref{eq: AdS_me3}) } W_4 \xrightarrow{(\ref{eq: AdS_me4}) } \gamma_4 \ldots \\
\ldots & \xrightarrow{(\ref{eq: AdS_me1}) } \beta_5  \xrightarrow{(\ref{eq: AdS_me2}) } U_5 \xrightarrow{(\ref{eq: AdS_me3}) } W_5 \xrightarrow{(\ref{eq: AdS_me4}) } \gamma_5 \ldots \\
&\phantom{aaaaaaaaaaaaaa} \vdots \\
\ldots & \xrightarrow{(\ref{eq: AdS_me1}) } \beta_{n} \xrightarrow{(\ref{eq: AdS_me2}) } U_n \xrightarrow{(\ref{eq: AdS_me3}) } W_n \xrightarrow{(\ref{eq: AdS_me4}) } \gamma_n.
\end{split}
\tag{4.5}
\end{align} 
\begin{figure}[H]
\begin{center}
\includegraphics{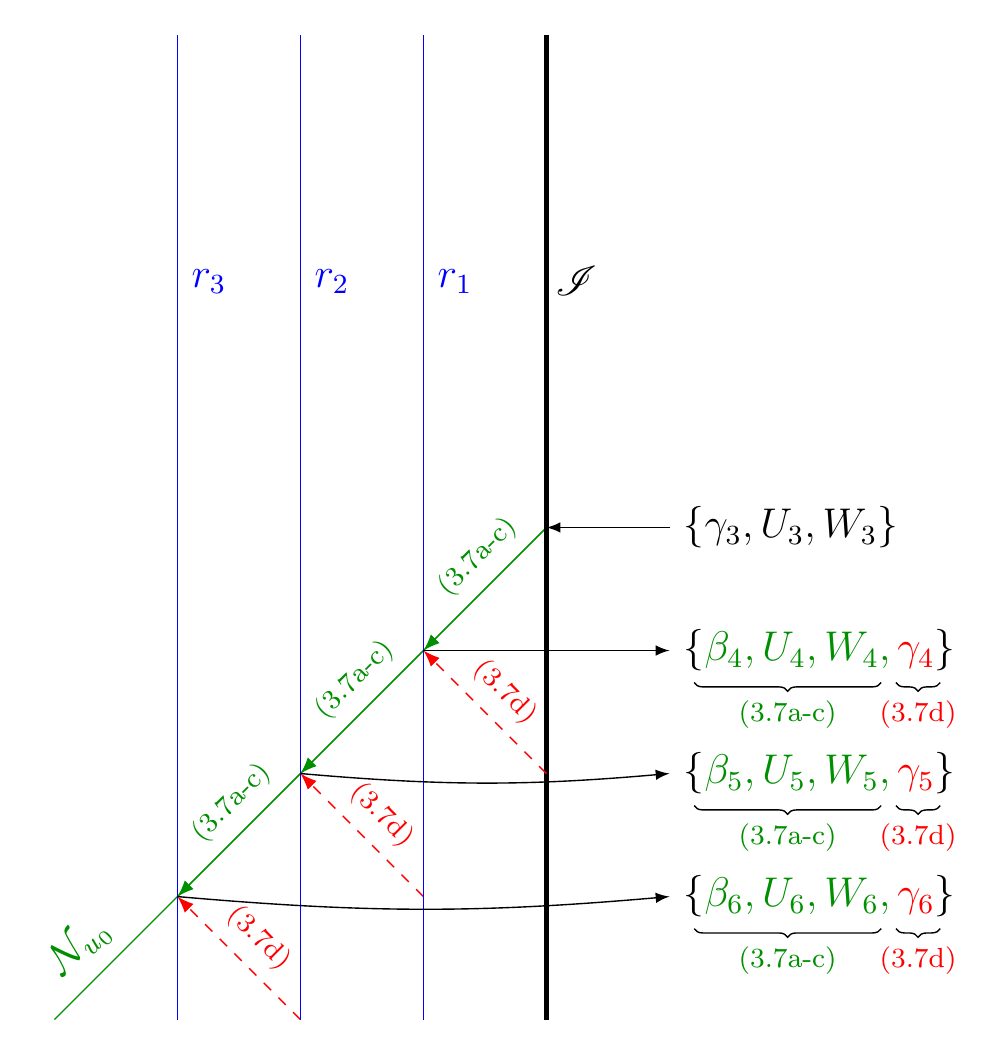}
\end{center}
\caption{Penrose diagram for AdS indicating how the second part of the scheme is implemented. The logic for this scheme is much the same as the one presented on the original diagram of figure \ref{fig: AdS_scheme_part_1}.}
\end{figure}
\noindent  Putting the two parts of the integration scheme together, we observe that knowledge of the six functions $\gamma_0, \beta_0, U_0, U_3, \gamma_3, W_3$ is sufficient to algebraically solve the Einstein equations for all other coefficients. 

Finally, we recall the functions $U_3$ and $W_3$ have close analogies to the angular momentum and mass aspect functions and may be thought of as representatives for these functions. 	We will discuss the holographic interpretation of these functions and gain an extra understanding using the AdS/CFT correspondence in section \ref{sec: Holographic_interpretation}. 

As a final comment upon this procedure, we note that this alternative scheme includes no evolution from one null hypersurface to the next. This algebraic procedure may be somewhat preferable when applied to a numerical scheme as one does not have to worry about errors accumulating in a discretisation scheme when evolving from one null hypersurface to the next. We will now present another new scheme which is based both on null evolution and boundary data.

\subsubsection{The ``hybrid'' scheme} \label{subsec: hybrid_scheme}

It has been shown that asymptotically locally AdS spacetimes admit an integration scheme where one specifies data at the conformal boundary as opposed to an initial null hypersurface (as in asymptotically flat spacetime). We will now present a ``hybrid'' scheme for asymptotically locally AdS spacetimes, where one specifies a mixture of data on the conformal boundary $\mathscr{I}$ and on an initial null hypersurface $\mathcal{N}_{u_0}$.  

\begin{figure}[h] \begin{center}
\includegraphics{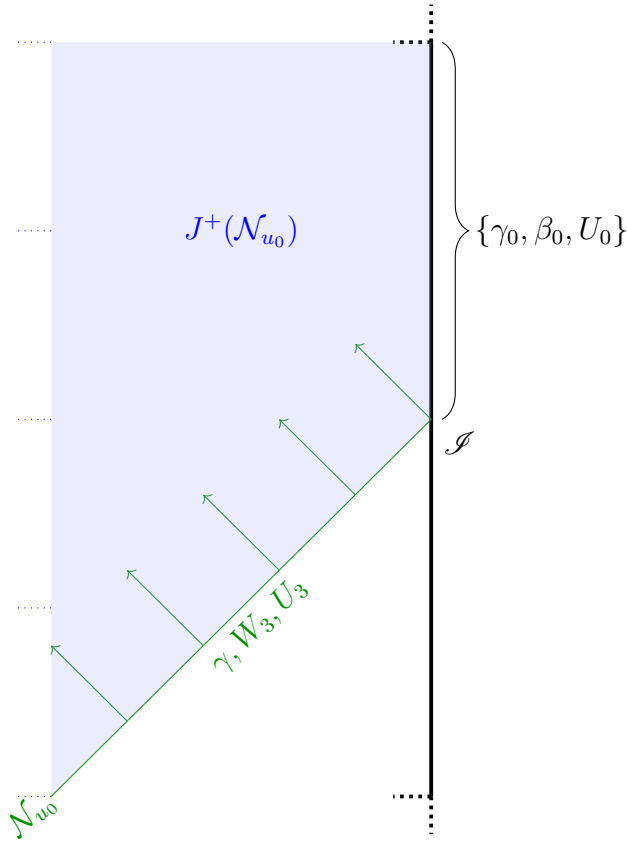}
\end{center}
\caption{Penrose diagram for the ``hybrid scheme''. The data which we specify is indicated on the hypersurfaces $\mathcal{N}_{u_0}$ and $\mathscr{I}$ for $u \geq u_0$. Specifying these coefficients allows one to solve the field equations in the causal future of $\mathcal{N}_{u_0}$, $J^+(\mathcal{N}_{u_0})$.} \label{fig: hybrid}
\label{fig: Fig_Hybrid}
\end{figure}

This scheme consists of the following data which one must specify before solving the field equations: $\{\gamma, W_3, U_3\}$ on $\mathcal{N}_{u_0}$ and $\{\gamma_0, U_0, \beta_0\}$ on $\mathscr{I}$ $\forall$ $u \geq u_0$. This is illustrated in the asymptotic Penrose diagram below. As we will see in the section \ref{sec: Holographic_interpretation},  $\{\gamma_0, U_0, \beta_0\}$ are related to positions and $\{\gamma_3, W_3, U_3\}$ to (radial) canonical momenta in the covariant phase space of the theory, thus we effectively specify 
momenta on $\mathcal{N}_{u_0}$  and positions at the conformal boundary, as shown in figure \ref{fig: Fig_Hybrid}. 

It remains to explain how this data is sufficient to solve for all coefficients of the series expansions of the metric functions in $J^+(\mathcal{N}_{u_0})$. We will first show that one is able to obtain all coefficients of $\gamma$ and then use this to show that one can obtain all coefficients of the other metric functions, including $U_3$ and $W_3$. 

Specifying $\{\gamma_0, U_0, \beta_0\}$ at $\mathscr{I}$ gives us these functions as well as all spatial and time derivatives of these functions $\forall$ $u \geq u_0$. This information gives $\gamma_1$ and all spatial and time derivatives of $\gamma_1$ $\forall$ $u \geq u_0$ via the Einstein equation (\ref{eq: gamma_1}). The higher order coefficients of $\gamma$ are obtained by evolving from the initial null hypersurface instead. As we saw in equation (\ref{eq: gamma2=0}), $\gamma_2=0$, so the first coefficient to consider is $\gamma_3$. In order to do this we apply equation (\ref{eq: gamma_4}) to obtain $\gamma_{3,u}$ on the null hypersurface $\mathcal{N}_{u_0}$  (this can be done because the scheme specifies $\gamma, U_3, W_3$ on $\mathcal{N}_{u_0}$). As was the case in asymptotically flat spacetime, when applied to a numerical scheme this will correspond to knowing $\gamma_3$ on the next null hypersurface $\mathcal{N}_{u_0+\delta u}$. This procedure will repeat for the higher order coefficients in $\gamma$ in that the Einstein equations will produce expressions for $\gamma_{n,u}$ on $\mathcal{N}_{u_0}$ and thus will determine $\gamma_n$ on $\mathcal{N}_{u_0+\delta u}$ $\forall$ $n \geq 3$.

Using the main equations (\ref{eq: AdS_me1}-\ref{eq: AdS_me4}) we know that knowledge of $\gamma(u_0+\delta u)$ is of course sufficient to give us $\beta(u_0+\delta u)$, as well as $U(u_0+\delta u)$ and $W(u_0+\delta u)$ up to the coefficients $W_3$ and $U_3$ which are of course not determined by the main equations (higher coefficients are also determined by these). To solve for these coefficients we will need to consider the supplementary conditions (\ref{eq: SC1}), (\ref{eq: SC2}) which take the schematic form
\begin{align}
U_{3,u} & =\mathcal{F}(\tilde{\gamma}_{0}, \tilde{\beta}_0, \tilde{U}_0, \tilde{\gamma}_{0,u}, \tilde{\beta}_{0,u}, \tilde{\gamma}_1, \tilde{\gamma}_{1,u}, \tilde{\gamma}_3, \tilde{\gamma}_{3,u}, \tilde{U}_3, \tilde{W}_3, \tilde{\gamma}_4) \label{eq: SC1_schematic} \tag{4.6a} \\
W_{3,u} & =\mathcal{H}(\tilde{\gamma}_{0}, \tilde{\beta}_0, \tilde{U}_0, \tilde{\gamma}_{0,u}, \tilde{\beta}_{0,u}, \tilde{\gamma}_1, \tilde{\gamma}_{1,u}, \tilde{\gamma}_3, \tilde{\gamma}_{3,u}, \tilde{U}_3, \tilde{W}_3, \tilde{\gamma}_4, \tilde{U}_{3,u}) \label{eq: SC2_schematic} \tag{4.6b}
\end{align}
where the tildes indicate that spatial derivatives of these functions may also be present. 

These are $u$-evolution equations for the functions $W_3$ and $U_3$. 
Note that all of the functions on the right hand side of each equation are known on $\mathcal{N}_{u_0}$. Starting with equation (\ref{eq: SC1_schematic}): $\tilde{\gamma}_{0}, \tilde{\beta}_0, \tilde{U}_0, \tilde{\gamma}_{0,u}, \tilde{\beta}_{0,u}, \tilde{\gamma}_3, \tilde{U}_3, \tilde{W}_3, \tilde{\gamma}_4$ are all given on $\mathcal{N}_{u_0}$ as part of the specified data and the remaining functions $\tilde{\gamma}_1, \tilde{\gamma}_{1,u}, \tilde{\gamma}_{3,u}$ can all be determined on $\mathcal{N}_{u_0}$ by using the Einstein equations (\ref{eq: gamma_1}) and (\ref{eq: gamma_4}) as discussed above. This means that we are able to obtain $U_{3,u}$ on $\mathcal{N}_{u_0}$ and thus $U_3$ on the next hypersurface $\mathcal{N}_{u_0+\delta u}$. An identical argument holds for (\ref{eq: SC2_schematic}), although now there is the extra requirement of knowing $U_{3,u}$ on $\mathcal{N}_{u_0}$, which is of course obtained from (\ref{eq: SC1_schematic}). 

Putting all of this together, we conclude that the specified data, along with iteration of both the main equations and supplementary conditions is an alternative way of constructing solutions to the field equations for asymptotically locally AdS metrics in the Bondi gauge 
for $J^+(\mathcal{N}_{u_0})$. 

\subsection{dS schemes} \label{subsec: dS_schemes}

Much of the previous discussion for asymptotically locally AdS spacetimes has a parallel discussion in the case of asymptotically dS spacetimes. The two new integration schemes that we have introduced are only dependent upon $\Lambda \neq 0$ in the field equations (\ref{eq: AdS_me1}-\ref{eq: AdS_me4}), and are insensitive to the sign of $\Lambda$. Due to this, we will now provide a brief description of the Bondi scheme applied to asymptotically locally dS spacetimes, as well as an analogue of the two AdS schemes that we have introduced. 

Firstly, we must mention that will restrict our attention to a retarded null foliation of $\mathscr{I}^+$ when discussing the Bondi approach to dS. If we consider applying the asymptotically flat integration scheme of specifying $\gamma$ on $\mathcal{N}_{u_0}$ as well as $(\gamma_1, U_3, W_3)$ for all $u$ and $\theta$, then in a similar fashion to the AdS case we will not be able to construct a fully determined solution to the field equations in a neighbourhood of $\mathscr{I}^+$. In the dS case (as in AdS) we will still have the undetermined functions $(\beta_0, U_0)$ which will propagate into solutions at later retarded times via the null hypersurface evolution. 

In order to remedy this problem we can adjust the two AdS integration schemes that we introduced in the previous section in order to describe asymptotically locally $dS$ spacetimes and solve the field equations in precisely the same order as before. The ``boundary'' scheme now consists of specifying the data $\{\gamma_0, \beta_0, U_0, \gamma_3, U_3, W_3\}$ on $\mathscr{I}^{+}$ and then solving the field equations in the same order as described in section \ref{subsec: AdS_boundary_scheme}.
\begin{figure}[H]
\begin{center}
\includegraphics{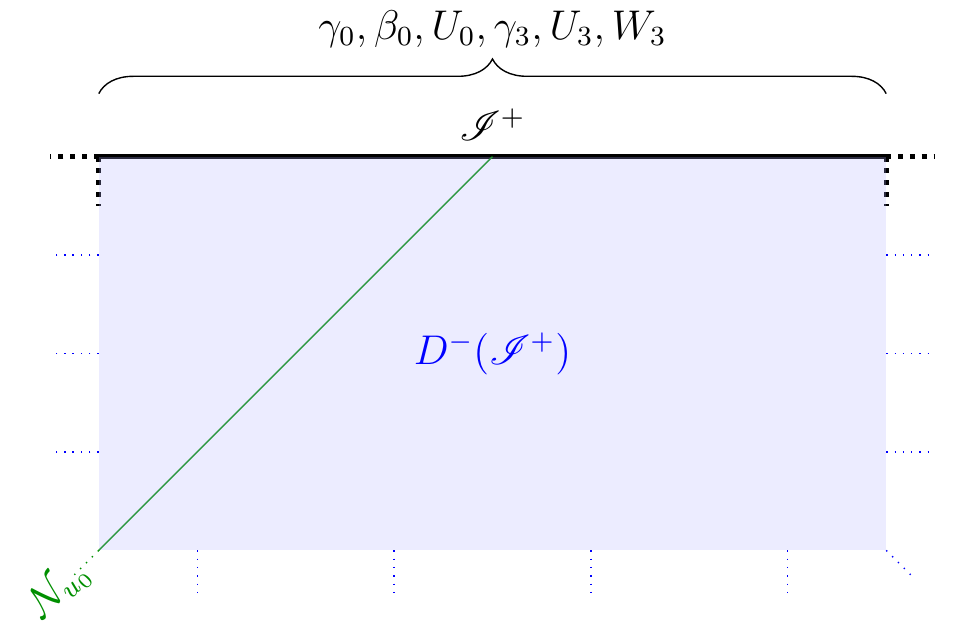}
\end{center}
\caption{Penrose diagram for the boundary scheme applied to an asymptotically locally dS spacetime. Notice that giving the data over the whole boundary $\mathscr{I}^+$ gives us the solution in $J^+(\mathcal{N}_{u_0})$}
\end{figure}
The ``hybrid'' scheme is again a scheme which involves specifying data on $\mathscr{I}^+$ and $\mathcal{N}_{u_0}$. As in the AdS hybrid scheme we specify ($\gamma_0, U_0, \beta_0$) on $\mathscr{I}^+$ for $u \geq u_0$ and ($\gamma, W_3, U_3$) on $\mathcal{N}_{u_0}$, solving the field equations in the same manner as described in section \ref{subsec: hybrid_scheme}.
\begin{figure}[H]
\begin{center}
\includegraphics{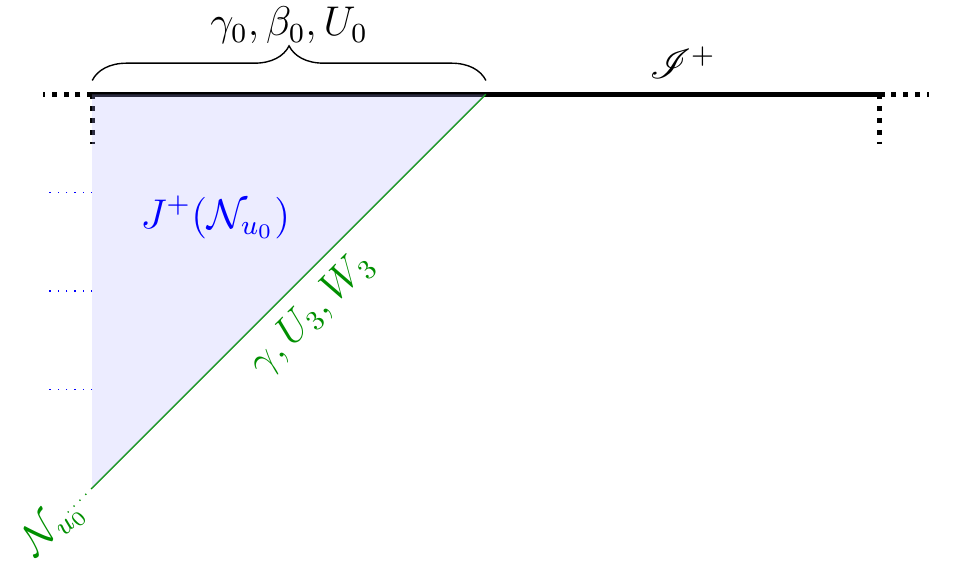}
\end{center}
\caption{Penrose diagram for the hybrid scheme applied to an asymptotically locally dS spacetime. This scheme generates solutions to the field equations in $J^+(\mathcal{N}_{u_0})$}
\end{figure}

It seems that the hybrid scheme applied to dS generates a smaller portion of the spacetime when compared with the scheme applied to AdS as shown in figure \ref{fig: hybrid}. This discrepancy is simply due to the causal differences between the respective cases and not an issue with either class of spacetimes. We note that in both cases the hybrid scheme generates the solutions to the field equations in $J^+(\mathcal{N}_{u_0})$ and thus the solutions in the neighbourhood of the conformal boundary to the future of $\mathcal{N}_{u_0}$. This method of specifying data agrees with similar Bondi type integration schemes for asymptotically dS spacetimes as discussed in \cite{Anninos:2012qw, 1981AcPP...12..719C}.


\section{Holographic interpretation} \label{sec: Holographic_interpretation}

In this section we will study the Bondi gauge metric from the perspective of holography, connecting with 
 \cite{Maldacena:1997re, Gubser:1998bc, Witten:1998qj, Henningson:1998gx, Henningson:1998ey, Balasubramanian:1999re, deHaro:2000vlm, Skenderis:2000in, Skenderis:2002wp, Papadimitriou:2005ii}. We begin with a review of the Fefferman-Graham coordinate system before deriving the coordinate transformation from Bondi gauge to Fefferman-Graham gauge. 
 This would allow us to give a holographic interpretation to the metric functions used in the integration scheme of section \ref{sec: Integration Scheme - Minkowski Vs. $AdS$}.

\subsection{Fefferman-Graham gauge} \label{subseq: FG_gauge}

Asymptotically locally AdS spacetimes can be described in Fefferman-Graham gauge in the neighbourhood of the conformal boundary
$\partial \mathcal{M}=\mathscr{I}$; see the review \cite{Skenderis:2002wp}. In this gauge the metric can be expressed as
\begin{equation} \label{eq: FG gauge}
ds^2=l^2\left[\frac{d\rho^2}{\rho^2}+\frac{1}{\rho^2}(g_{(0)ab}+\rho^2g_{(2)ab}+\rho^3g_{(3)ab}+\ldots)dx^adx^b\right],
\end{equation} 
where $l=\sqrt{-3/\Lambda}$ is the AdS radius. Following the discussion of section \ref{subseq: AdS_asymptotics}, $\rho$ is a coordinate which describes the location of the conformal boundary, specifically $\mathscr{I}=\{ \rho = 0 \}$. 
The lower case Roman indices $(a,b)$ run from 1 to 3 for asymptotically locally AdS$_4$ spacetimes. 

Comparing with (\ref{eq: double_pole_metric}) and choosing $\rho$ as the defining function, we see that the term $g_{(0)}$ in the FG expansion is a representative of the conformal class of metrics induced on $\mathscr{I}$. If the metric $g_{(0)ab}$ is conformally flat i.e. the Cotton tensor vanishes, then the spacetime is \textit{Asymptotically} AdS; otherwise it is \textit{Asymptotically locally} AdS.

Holographically, $g_{(0)}$ is viewed as the background metric for the 3-dimensional conformal field theory dual to the 4-dimensional spacetime. 
The coefficients of even powers of $\rho$ in the asymptotic expansion are determined locally in terms of derivatives of $g_{(0)}$; see \cite{deHaro:2000vlm} for explicit expressions.  The coefficient $g_{(3)}$ is constrained to be divergenceless and traceless with respect to $g_{(0)}$, but is otherwise undetermined. This coefficient
corresponds to the energy momentum tensor in the dual 3-dimensional field theory, which is defined as 
\cite{Henningson:1998gx, Balasubramanian:1999re, deHaro:2000vlm, Skenderis:2002wp}
\begin{equation}
\braket{T_{ab}}=\frac{2}{\sqrt{-\text{det}g_{(0)}}}\frac{\delta S_{r}}{\delta g_{(0)}^{ab}}
\end{equation}
where $S_r$ is the renormalised on-shell gravitational action. For asympotically locally AdS$_4$ spacetimes
\begin{equation} \label{eq: energy_momentum_tensor}
\braket{T_{ab}}=-\frac{3l^2}{2\kappa^2}g_{(3)ab}
\end{equation}
where $2 \kappa^2 = 16 \pi G$ and $G$ is Newton's constant.
This energy momentum tensor satisfies tracelessness and conservation properties with respect to $g_{(0)}$
\begin{equation} \label{eq: g_3_conditions}
g_{(0)}^{ab}\braket {T_{ab}}=0, \qquad \phantom{aaaaaaaaaaa} \nabla_{(0)}^a \braket{ T_{ab}} =0.
\end{equation}

Finally, we note that the pair $(g_{(0)}, T_{ij})$ or equivalently $(g_{(0)}, g_{(3)})$ provide local coordinates on the covariant phase space \cite{Crnkovic:1986ex, Lee:1990nz} 
of the theory; in a radial Hamiltonian formalism, where the radial coordinate plays the role of time, $g_{(0)}$ is the position and $g_{(3)}$ the corresponding canonical momentum \cite{Papadimitriou:2005ii}.

\subsection{Coordinate transformations}
In order to extract holographic data from spacetimes expressed in Bondi gauge, we need to determine the coordinate transformation from our asymptotically locally AdS metric in Bondi gauge
\begin{align}
\begin{split} \label{eq: Bondi_Sachs_Metric}
ds^2=&-(Wr^{2}e^{2\beta}-U^2r^2e^{2\gamma})du^2-2e^{2\beta}dudr-\\
&2Ur^2e^{2\gamma}dud\theta+r^2(e^{2\gamma}d\theta^2+e^{-2\gamma}\sin^2\theta d\phi^2) 
\end{split}
\end{align}
to the Fefferman-Graham form of (\ref{eq: FG gauge}). We will derive the transformation up to the coefficient $g_{(3)}$, as this is the highest order term of holographic interest.

\subsubsection{Global AdS$\mathbf{_4}$}

A useful first step in performing this computation is to recall the transformation of global AdS$_{4}$ spacetime into Fefferman-Graham form. We begin with the metric of AdS$_4$ in Bondi gauge 
\begin{equation} \label{eq: AdS_Bondi}
ds^2=-\left(1+\frac{r^2}{l^2}\right)du^2-2dudr+r^2d\Omega^2,
\end{equation}
where we have reinstated the factors of $l$ for clarity. In Bondi coordinates, the metric for AdS$_4$ corresponds to choosing functions 
\begin{equation}
\beta=\gamma=U=0; \qquad W=\frac{1}{l^2}+\frac{1}{r^2},
\end{equation}
so in the notation of section \ref{sec: Field_equations} this corresponds to $W_{0}=1/l^2$, $W_{2}=1$, with all other coefficients zero. 

We begin by transforming from the retarded time coordinate $u$ into the usual time coordinate $t$. This is achieved by 
\begin{equation} \label{eq: time_transformation}
t=u+r_*
\end{equation}
where the tortoise coordinate $r_*$ is defined by 
\begin{equation} \label{eq: tortoise_def}
dr_{*}=\frac{dr}{f(r)}=\frac{dr}{1+(r/l)^2} \implies r_{*}=l\arctan\left(\frac{r}{l}\right)+c, 
\end{equation}
with $c$ is an integration constant whose value will be fixed later. Applying equations (\ref{eq: time_transformation}) and (\ref{eq: tortoise_def}) transforms (\ref{eq: AdS_Bondi}) into the standard AdS metric of 
\begin{equation} \label{eq: AdS_metric}
ds^2=-\left(1+\frac{r^2}{l^2}\right)dt^2+\left(1+\frac{r^2}{l^2}\right)^{-1}dr^2+r^2d\Omega^2.
\end{equation}

The next step is to transform from our radial distance coordinate $r$ into the tortoise coordinate $r_*$. The motivation for doing this is that we can fix the the conformal boundary to be located at $r_*=0$, providing an immediate comparison with the FG coordinate $\rho$ as the conformal boundary in those coordinates is also located at $\rho=0$. Choosing the integration constant in \eqref{eq: tortoise_def} to be $c=-l\pi/2$ allows us locate the conformal boundary $\mathscr{I}$ at $r_*=0$.  We implement this part of the transformation by only including the leading order term in the large $r$ approximation of $r_*$ 
\begin{equation} \label{eq: leading_order_tortoise}
r_*=-\frac{l^2}{r}+\mathcal{O}(r^{-3})
\end{equation}
which brings the line element (\ref{eq: AdS_metric}) into the form 
\begin{equation} \label{eq: tortoise_AdS_metric}
ds^2=\frac{l^2}{r_{*}^2}\left[-\left(1+\frac{r_{*}^2}{l^2}\right)dt^2+\left(1+\frac{r_{*}^2}{l^2}\right)^{-1}dr_*^2+l^2d\Omega^2\right].
\end{equation}
This metric has similarities with (\ref{eq: FG gauge}); the gauge conditions of $g_{\rho t}=g_{\rho \theta}=g_{\rho \phi}=0$ are all satisfied automatically if $r_{*}=f(\rho)$ for any function $f(\rho)$.  We hence need to solve for $f$ such that $g_{\rho \rho}=l^2/\rho^2$. Carrying out this procedure we derive the defining equation for $f(\rho)$
\begin{equation} \label{eq: FG_ODE}
\frac{l^2 f'^2}{f^2 [l^2+f^2]}=\frac{1}{\rho^2}
\end{equation}
which admits two solutions 
\begin{equation}
f_1=\frac{2k l \rho}{1-(k\rho)^2}, \qquad \qquad f_2=\frac{2kl\rho}{\rho^2-k^2}
\end{equation}
where in both cases $k$ is an integration constant. These two solutions are related via the map $k \rightarrow -1/k$ so it is unimportant which is chosen to be $f$. 

Picking $f=f_1$ we observe that in a neighbourhood of $\mathscr{I}$ we have 
\begin{equation} \label{eq: tortoise_nbhd}
r_*=\frac{2 k \rho l}{1-k^2 \rho^2}\approx 2k\rho l 
\end{equation}
The metric (\ref{eq: tortoise_AdS_metric}) transforms to 
\begin{equation}
ds^2=\frac{l^2}{\rho^2}d\rho^2-\frac{(1+k^2\rho^2)^2}{4k^2\rho^2}dt^2+\frac{l^2(k^2\rho^2-1)^2}{4k^2\rho^2} d\Omega^2. 
\end{equation}
We can now read off $g_{(0)}$, which is conformally equivalent to the Einstein metric on $\mathbb{R} \times S^2$ 
\begin{equation} \label{Ein_Un}
ds_{(0)}^2=-dt^2+d\Omega^2.
\end{equation}
Notice that the leading order truncations of the Taylor series for our transformations (\ref{eq: leading_order_tortoise}), (\ref{eq: tortoise_nbhd}) allow us to compute only $g_{(0)}$ correctly. To compute higher order $g_{(i)}$ we need to include higher order terms in the transformation, giving
\begin{equation} \label{eq: pure_AdS_g_2}
ds_{(2)}^2=\frac{1}{2}(-dt^2-d\theta^2-\sin^2\theta d\phi^2)
\end{equation} 
as well as $g_{(3)ab}=0$. The latter is the expected result for the energy momentum tensor of the CFT state dual to global AdS$_4$.

In generalising this procedure to asymptotically locally AdS$_4$ spacetimes we repeat the steps of this procedure, namely using series expansions to transform the coordinates and truncating at the necessary point to compute each $g_{(i)}$ coefficient. 

\subsubsection{Computing $\mathbf{g_{(0)ab}}$} \label{sec:g0}

For computational and notational simplicity we will from here onwards fix the AdS radius $l=1$ ($\Lambda=-3$).  Factors of the radius
may be reinstated using the following dimensional considerations. The Fefferman-Graham coordinates $(t, \rho, \theta, \phi)$ are dimensionless coordinates, and thus the only dimensions are those of the Bondi metric functions $(\gamma, \beta, U, W)$. Working with dimensional conventions of $[\text{length}]=+1$ we first compute the dimensions of the functions in the Bondi gauge metric (\ref{eq: Bondi_Sachs_Metric}). Using the standard definitions of the Bondi coordinates we have 
\begin{equation}
[u]=1, \quad [r]=1, \quad [\theta]=0, \quad [\phi]=0
\end{equation}
    and the line element has dimension $[ds^2]=2$. Using the length dimensions of the coordinates, the dimensions of the Bondi functions are 
\begin{equation}
[\gamma]=0, \quad [\beta]=0, \quad [U]=-1, \quad [W]=-2.
\end{equation}
Each of these functions is expanded in negative powers of $r$ in the asymptotic region of the spacetime (\ref{eq: Taylor_Series}). Using this, we can determine the dimension of each of the coefficients in the asymptotic expansions as
\begin{equation} \label{eq: Bondi_dimensions}
[\gamma_i]=i, \quad [\beta_i]=i, \quad [U_i]=i-1, \quad [W_i]=i-2.
\end{equation} 
To reinstate all the factors of $l$ in the transformation formulae one simply needs to match the dimensions of each side of the equations by multiplying the Bondi functions by suitable powers of $l$ as determined by (\ref{eq: Bondi_dimensions}). 

\bigskip

To compute $g_{(0)}$ we need to impose the vacuum Einstein equations to leading order; this corresponds to switching on
the leading coefficients in the metric functions $\beta_{0}, \gamma_{0}, U_{0}$ and imposing $W_{0}=e^{2\beta_{0}}$). 
The leading order line element (\ref{eq: Bondi_Sachs_Metric}) takes the form 
\begin{align}
\begin{split} \label{eq: Bondi_metric_zero_functions}
ds^2=&-\left(e^{4\beta_{0}}r^2-U_{0}^2r^2e^{2\gamma_0}\right)du^2-2e^{2\beta_{0}}dudr- \\
&2U_{0}r^2e^{2\gamma_{0}}dud\theta+r^2(e^{2\gamma_{0}}d\theta^2+e^{-2\gamma_{0}}\sin^2\theta d\phi^2). 
\end{split}
\end{align}
\noindent We now carry out the coordinate transformations (\ref{eq: time_transformation}) $\rightarrow$ (\ref{eq: leading_order_tortoise}) $\rightarrow$ (\ref{eq: tortoise_nbhd}) using the form $r_*=\rho$ as we are for now only concerned about computing $g_{(0)}$. This sequence of transformations gives the metric components at order $1/\rho^2$ as 
\begin{subequations}
\begin{align}
g_{\rho \rho}&=\frac{1}{\rho^2}(2e^{2\beta_{0}}-e^{4\beta_{0}}+ e^{2\gamma_{0}} U_{0}^2) \label{eq: g_rhorho} \\
g_{\rho t}&=\frac{1}{\rho^2}(e^{4\beta_{0}}-e^{2\beta_{0}}-e^{2\gamma_{0}}U_{0}^2) \\
g_{\rho \theta}&=\frac{e^{2\gamma_{0}}U_{0}}{\rho^2} \label{eq: g_rhotheta} \\
g_{tt}&=\frac{1}{\rho^2}(e^{2\gamma_{0}}U_{0}^2-e^{4\beta_{0}}) \label{eq: g_tt} \\
g_{t \theta}&=-\frac{ e^{2\gamma_{0}} U_{0}}{\rho^2} \\
g_{\theta \theta}&=\frac{e^{2\gamma_{0}}}{\rho^2} \\
g_{\phi \phi}&=\frac{e^{-2\gamma_{0}}\sin^2(\theta)}{\rho^2} \label{eq: g_phiphi}.
\end{align}
\end{subequations}
The resulting coefficients  (\ref{eq: g_rhorho}-\ref{eq: g_rhotheta}) are clearly incompatible with the Fefferman-Graham gauge. We thus carry out further transformations in $\theta$ and $t$, namely 
\begin{equation} \label{eq: theta_t_transformations}
t \rightarrow t+ \alpha_1(t,\theta) \rho, \qquad \theta \rightarrow \theta + \alpha_2(t,\theta) \rho. 
\end{equation}
where $\alpha_{1,2}$ are functions which are fixed by setting $g_{\rho \rho}=1/\rho^2$, $g_{\rho t}=g_{\rho \theta}=0$. When considering the $\mathcal{O}(1/\rho^2)$ pieces of the metric it suffices to transform the forms as 
\begin{equation}
dt \rightarrow dt+ \alpha_1(t,\theta) d\rho + \cdots , \qquad d\theta \rightarrow d\theta + \alpha_2(t,\theta) d\rho  + \cdots 
\end{equation}
as terms involving derivatives of $\alpha_{1,2}$ are subleading in the radial expansion. 

Under this transformation $g_{\rho \rho}$ is given by 
\begin{align}
\begin{split}
g_{\rho \rho}=&\frac{1}{\rho^2}[(-e^{4\hat{\beta}_{0}}+e^{2\hat{\gamma}_{0}}\hat{U}_{0}^2)\alpha_1^2-\alpha_1(2(e^{2\hat{\beta}_{0}}-e^{4\hat{\beta}_{0}}+e^{2\hat{\gamma}_{0}}\hat{U}_{0}^2)+2e^{2\hat{\gamma}_{0}}\hat{U}_{0}\alpha_2) + \\ 
&\phantom{\frac{1}{4\rho^2}}(2e^{2\hat{\beta}_{0}}-e^{4\hat{\beta}_{0}}+e^{2\hat{\gamma}_{0}}\hat{U}_{0}^2+e^{2\hat{\gamma}_{0}}\hat{U}_{0}\alpha_2+e^{2\hat{\gamma}_{0}}\alpha_2^2)]
\end{split}
\end{align}
where the hat symbol  over metric  functions signifies the boundary value e.g.. 
\begin{equation} \label{eq: tortoise_limit}
\hat{\gamma}_0(t,\theta)=\lim_{r_* \rightarrow 0} \gamma_0(u,\theta).
\end{equation}
Let us now solve the equation $g_{\rho \rho}=l^2/\rho^2$, which is regarded as a quadratic equation for $\alpha_1$ (or equivalently $\alpha_2$). Solving this equation gives us two roots:
\begin{subequations}  \label{eq: alphas}
\begin{align}
\alpha_1^+=\frac{1-e^{2\hat{\beta}_{0}}+e^{\hat{\gamma}_{0}}\hat{U}_{0}+e^{\hat{\gamma}_{0}}\alpha_2}{e^{\gamma_{0}}\hat{U}_{0}-e^{2\hat{\beta}_{0}}}  \\
\alpha_1^-=\frac{-1+e^{2\hat{\beta}_{0}}+e^{\hat{\gamma}_{0}}\hat{U}_{0}+e^{\hat{\gamma}_{0}}\alpha_2}{e^{\hat{\gamma}_{0}}\hat{U}_{0}+e^{2\hat{\beta}_{0}}}.
\end{align}
\end{subequations}
There seems to be no particular motivation to choose one or the other so we will proceed by choosing $\alpha_{1}^+$; we will show below that either root could have been chosen. Notice that (\ref{eq: alphas}) gives $\alpha_1$ in terms of $\alpha_2$, which is viewed as a free function. Examining the transformations of the $g_{\rho t}, g_{\rho \theta}$ coefficients fixes $\alpha_2$ and thus $\alpha_1$ also.

Using the transformation with $\alpha_1=\alpha_1^+$, $g_{\rho t}$ reduces to 
\begin{equation}
g_{\rho t}=\frac{e^{\hat{\gamma}_{0}}(\hat{U}_{0}+e^{2\hat{\beta}_{0}}\alpha_2)}{\rho^2}
\end{equation}
so we can set $g_{\rho t}=0$ by choosing $\alpha_2=-\hat{U}_{0} e^{-2 \hat{\beta}_{0}}$. We thus conclude that the coordinate transformations are given by 
\begin{equation} \label{eq: ttheta_trans}
t \rightarrow t+ (1 - e^{-2 \hat{\beta}_0}) \rho, \qquad \theta \rightarrow \theta  -\hat{U}_{0} e^{-2 \hat{\beta}_{0}}\rho.
\end{equation}
Note that this value of $\alpha_2$ automatically sets $\alpha_1^+=\alpha_1^-$. We could have alternatively started by choosing $\alpha_1=\alpha_1^-$; this would have resulted in the same value for $\alpha_2$, showing that the freedom in choosing $\alpha_1$ was actually trivial. As a final check for this part of the transformation, we can show that $g_{\rho \theta}=0$, verifying that the Fefferman-Graham gauge has been reached.  

This transformation illustrates the leading order part of the general procedure to transform from Bondi to FG gauge. Using our solutions of the vacuum Einstein equations, we first transform from the Bondi coordinates $(u, r, \theta, \phi)$ into coordinates $(t, r_*, \theta, \phi)$ and then use transformations of the form 
\begin{equation}
r_* \rightarrow \sum_{j=1}^{i+1} r_{* j}(t, \theta) \rho^j, \quad t \rightarrow t+ \sum_{j=1}^{i+1}  t_{j}(t, \theta) \rho^j, \quad \theta \rightarrow \theta + \sum_{j=1}^{i+1} \theta_{j}(t, \theta) \rho^j
\end{equation}
where the limit of the sum $i+1$ indicates the order necessary to compute the coefficient $g_{(i)}$ (thus we will only be concerned about summing to an upper limit of four). At each order we need to solve for the coefficients $ r_{* j}, t_{j}, \theta_{j}$ to preserve the FG gauge conditions $g_{\rho \rho}=1/\rho^2,\; g_{t \rho}=g_{\theta \rho}=0$ ($g_{\phi \rho}=0$ will be satisfied automatically due to axisymmetry and trivial $\phi \rightarrow \phi$ transformation). More detail and computation of the higher order coefficients is given in appendix \ref{sec: FG_appendix}.

\subsection{Background metric}

The transformation (\ref{eq: theta_t_transformations}) gives the following results for $g_{(0)ab}$:
\begin{equation} \label{eq: g_(0)}
ds_{(0)}^2=(e^{2\hat{\gamma}_{0}}\hat{U}_{0}^2-e^{4\hat{\beta}_{0}})dt^2-2e^{2\hat{\gamma}_{0}}\hat{U}_{0}dtd\theta+e^{2\hat{\gamma}_{0}}d\theta^2+e^{-2\hat{\gamma}_{0}}\sin^2(\theta)d\phi^2.
\end{equation}
Note that the boundary is not necessarily topologically equivalent to $\mathbb{R} \times S^2$ in general; the spacetimes are asymptotically locally AdS rather than asymptotically AdS. 

When $\hat{U}_0$ vanishes, the boundary metric is topologically $\mathbb{R} \times S^2$ but the metric on the $S^2$ is deformed by non-trivial $\hat{\gamma}_0$. The boundary metric retains the determinant condition on the angular part of the metric 
\begin{equation} \label{eq:bdry_det}
d\Omega^2=e^{2\hat{\gamma}_{0}}d\theta^2+e^{-2\hat{\gamma}_{0}}\sin^2(\theta)d\phi^2 \implies |\Omega|=\sin^2 \theta,
\end{equation}
which was part of the definition of the Bondi gauge. This is an unusual restriction on the boundary metric: it is somewhat unnatural to impose a fixed determinant for the metric on the sphere. It would thus be interesting to revisit the Bondi gauge analysis, dropping the determinant condition on the spherical part of the metric. 

\subsection{The energy-momentum tensor}

The final term of physical interest in the Fefferman-Graham expansion is $g_{(3)}$ as this describes the energy-momentum tensor of the dual conformal field theory (\ref{eq: energy_momentum_tensor}). To compute $g_{(3)ab}$ we have to include terms up to $\mathcal{O}(r^{-3})$ in the metric functions
\begin{subequations}
\begin{align}
\gamma(u,r,\theta)&=\gamma_{0}+ \frac{\gamma_1}{r}+\frac{\gamma_3}{r^3}\\
\beta(u,r,\theta)&=\beta_{0}-\frac{\gamma_1^2}{4r^2} \\
\begin{split}
U(u,r,\theta)&=U_{0}+\frac{2}{r}\beta_{0, \theta} e^{2( \beta_0- \gamma_0)} - \\
& \phantom{--} \frac{1}{r^2}e^{2( \beta_0- \gamma_0)} (2 \beta_{0,\theta} \gamma_1-2 \gamma_{0,\theta}\gamma_1+\gamma_{1,\theta}+2 \cot (\theta ) \gamma_1)+\frac{U_3}{r^3}
\end{split}\\
\begin{split}
W(u,r,\theta)&=e^{2\beta_{0}}+\frac{1}{r}[\cot(\theta)U_{0}+U_{0,\theta}]+ \frac{1}{2r^2}e^{2(\beta_{0}-\gamma_{0})}[2-3e^{2\gamma_0}\gamma_1^2+4\cot(\theta)\beta_{0,\theta}+\\
&\phantom{aa}8(\beta_{0,\theta})^2+6\cot(\theta)\gamma_{0,\theta}-8\beta_{0,\theta}\gamma_{0,\theta}-4(\gamma_{0,\theta})^2+4\beta_{0,\theta \theta}+2\gamma_{0,\theta \theta}]+\frac{W_3}{r^3}.
\end{split}
\end{align}
\end{subequations}
As a brief aside, we observe that the integration functions $U_3$ and $W_3$ enter the metric at this order. Recall that $W_3$ has the interpretation in asymptotically flat spacetime as the Bondi mass aspect, $W_3=-2m_B$ \cite{Bondi:1962px}. If we follow \cite{Chrusciel:2016oux} in defining the mass aspect function as the $\mathcal{O}(1/r)$ term in the Bondi metric component $g_{uu}$  then we obtain
\begin{align} \label{eq: mass_aspect_1}
\begin{split}
2m_B=&-e^{-2 (\beta_{0}+\gamma_{0})} (2 \gamma_{0,u}-U_{0} (\cot (\theta )-2 \gamma_{0,\theta})+U_{0,\theta}) (4 e^{4 \beta_{0}} (\beta_{0,\theta})^2- \\
&e^{2 \gamma_{0}} U_{0} (-2 \gamma_{0,u \theta}+4 \gamma_{0,u} (\gamma_{0,\theta}-\cot (\theta ))+U_{0} (4 (\gamma_{0,\theta})^2-\\
&2 \gamma_{0,\theta \theta}-6 \cot (\theta ) \gamma_{0,\theta}+\cot ^2(\theta )-1)-U_{0,\theta \theta}-\cot (\theta ) U_{0,\theta}))+\\
&e^{-2 \gamma_{0}} (2 e^{4 \gamma_{0}} U_{0} U_{3}-2 e^{2 \beta_{0}} \beta_{0,\theta} (-2 \gamma_{0,u \theta}+4 \gamma_{0,u} (\gamma_{0,\theta}-\cot (\theta ))+\\
&U_{0} (4 (\gamma_{0,\theta})^2-2 \gamma_{0,\theta \theta}-6 \cot (\theta ) \gamma_{0,\theta}+\cot ^2(\theta )-1)-U_{0,\theta \theta}-\cot (\theta ) U_{0,\theta}))+\\
&\frac{1}{3} e^{2 \gamma_{0}} U_{0}^2 \left[6 \gamma_{3}+\frac{1}{2} e^{-6 \beta_{0}} (-2 \gamma_{0,u}+U_{0} (\cot (\theta )-2 \gamma_{0,\theta})-U_{0,\theta})^3\right]+\\
&2 e^{-2 \beta_{0}} \beta_{0,\theta} U_{0} (2 \gamma_{0,u}-U_{0} (\cot (\theta )-2 \gamma_{0,\theta})+U_{0,\theta})^2+\\
&\frac{1}{8} e^{-2 \beta_{0}} (U_{0,\theta}+\cot (\theta ) U_{0}) (2 \gamma_{0,u}-U_{0} (\cot (\theta )-2 \gamma_{0,\theta})+U_{0,\theta})^2-e^{2 \beta_{0}} W_{3}.
\end{split}
\end{align}
Here we have used the Einstein equation (\ref{eq: gamma_1}) to express contributions in terms of ($\gamma_0, U_0, \beta_0$) wherever possible. In the asymptotically AdS case $\gamma_0=\beta_0=U_0=0$ we obtain the same definition of the mass aspect, $2 m_B= -W_3$, as in the asymptotically flat case \cite{Bondi:1962px}.

In the asymptotically flat case, the Bondi mass at time $u=u_0$ is obtained by integrating the mass aspect over the $u_0$ cut of $\mathscr{I^+}$ (\ref{eq: Bondi_mass_time_u}). It is natural to suggest that an extension should exist for the AdS case whereby one could obtain the analogue of the Bondi mass in asymptotically locally AdS spacetime by integrating over a cut of $\mathscr{I}$ instead. We will discuss this definition in asymptotically AdS spacetimes in section \ref{sec: Bondi_mass_AAdS} while the more general case of asymptotically locally AdS 
remains ongoing work.

Returning to the discussion of the coordinate transformation in order to obtain $g_{(3)}$, we note that when performing the series transformation into the Fefferman-Graham form we also need to extend our transformation in the coordinates to $\mathcal{O}(\rho^4)$
\begin{align}
\begin{split}
&r_* \rightarrow \rho  + b_1(t,\theta) \rho^2+c_1(t,\theta) \rho^3+d_1(t,\theta) \rho^4 \\
&t \rightarrow t + \alpha_1(t,\theta)\rho + b_2(t,\theta) \rho^2+c_2(t,\theta) \rho^3+d_2(t,\theta) \rho^4  \\
&\theta \rightarrow \theta + \alpha_2(t,\theta)\rho + b_3(t,\theta) \rho^2+c_3(t,\theta) \rho^3+d_3(t,\theta) \rho^4.
\end{split}
\end{align}
where $\alpha_i, b_i, c_i$ are the functions already obtained from previous orders (see appendix \ref{sec: FG_appendix} for $b_i$ and $c_i$).  To obtain $g_{(3)ab}$ we will need to choose $d_{1,2,3}$ suitably in order to force the $d\rho$ terms to vanish at $\mathcal{O}(1/\rho)$. 

Once we have performed this transformation we have to check equations (\ref{eq: g_3_conditions}) are satisfied. First we use the $g_{(0)}$ of equation (\ref{eq: g_(0)}) to check tracelessness
\begin{equation} \label{eq: tracelessness_coordinate_form}
g_{(0)}^{ab}g_{(3)ab}=g_{(0)}^{tt}g_{(3)tt}+2g_{(0)}^{t\theta}g_{(3)t\theta}+g_{(0)}^{\theta \theta}g_{(3)\theta \theta}+g_{(0)}^{\phi \phi}g_{(3)\phi \phi}=0,
\end{equation} 
which is automatically satisfied by $g_{(3)ab}$ without having to apply either the supplementary conditions or the higher order main equations. 

In order to present expressions for the $g_{(3)}$ coefficients, we give formulae for $(U_3, \gamma_3, W_3)$ which have been obtained via rearrangement of the expressions for $(g_{(3)tt}, g_{(3)t\theta}, g_{(3) \theta \theta})$. Although there are four non-zero components of the energy-momentum tensor, the three functions below suffice to read off all components due to the tracelessness equation (\ref{eq: tracelessness_coordinate_form}). 
\begin{eqnarray}
\hat{U}_3 &=& e^{-2 \hat{\gamma}_0} (g_{(3) \theta \theta} \hat{U}_0 + g_{(3) t \theta}) + {\cal U}_3 (\hat{\gamma}_0,\hat{\beta}_0,U_0); \nonumber\\
\hat{W}_3 &=& \frac{3}{2} e^{- 2 \hat{\beta}_0} \left( g_{(3) \theta \theta} \hat{U}_0^2  + 2  g_{(3) t \theta} \hat{U}_0 +   g_{(3) tt}  \right ) + {\cal W}_3 (\hat{\gamma}_0,\hat{\beta}_0,\hat{U}_0);   \label{g3-rel}  \\
\hat{\gamma}_3 &=& \frac{1}{4} \left ( e^{-4 \hat{\beta}_0} (g_{(3) \theta \theta} \hat{U}_0^2 +2 g_{(3) t \theta}\hat{U}_0+ g_{(3) tt} ) - 2 e^{-2 \hat{\gamma}_0} g_{(3) \theta \theta} \right ) + {\cal G}_3 (\hat{\gamma}_0,\hat{\beta}_0,\hat{U}_0), 
\nonumber
\end{eqnarray}
where all of the metric coefficients are functions of $(t, \theta)$, defined at $\mathscr{I}$, and 
explicit expressions for $({\cal U}_3,{\cal W}_3,{\cal G}_3)$ can be found in appendix \ref{g3-expressions}. 

Verification of the conservation condition (\ref{eq: g_3_conditions}) is less straightforward than checking tracelessness. The simplest component to check is the $\phi$ component, for which the required result is obtained using the equations (\ref{eq: U_3_EM_tensor}-\ref{eq: gamma_3_EM_tensor}) above and the tracelessness property (\ref{eq: tracelessness_coordinate_form})  
\begin{equation}
\nabla_{(0)}^a g_{(3)a\phi}=g^{ac}_{(0)}\nabla_{(0)c} g_{(3)a\phi}=-g_{(0)}^{ca}\Gamma^{\phi}_{ca}g_{(3)\phi \phi}-g_{(0)}^{ca}\Gamma^{d}_{c \phi}g_{(3) ad }=0
\end{equation}
where the Christoffel symbols $\Gamma^{a}_{bc}$ are those associated with the metric $g_{(0)ab}$. 

The remaining conservation equations are harder to verify. The Einstein equations (\ref{eq: gamma_1}), (\ref{eq: gamma_4}) for $\hat{\gamma}_1$ and $\hat{\gamma}_{3,t}$ and the supplementary conditions (\ref{eq: SC1}-\ref{eq: SC2}) are required, the latter giving expressions for the functions $\hat{U}_{3,t}$ and $\hat{W}_{3,t}$. These equations, combined with the relations (\ref{eq: U_3_EM_tensor}-\ref{eq: gamma_3_EM_tensor}), are sufficient to show that the $t$ and $\theta$ components of the conservation conditions (\ref{eq: g_3_conditions}) are satisfied. 


\subsection{Asymptotically AdS$\mathbf{_4}$ examples}

The first interesting example to look at is the class of asymptotically AdS$_4$ Bondi gauge pacetimes. Recall that we defined asymptotically AdS$_4$ spacetimes as asymptotically locally AdS$_4$ spacetimes for which $g_{(0)}$ is conformally flat. 
We can choose the representative of this conformal class to be
\begin{equation}
\hat{\gamma}_{0}=\hat{\beta}_{0}=\hat{U}_{0}=0,
\end{equation}
so that the metric $g_{(0)}$ is the standard metric on the Einstein universe. 

Applying these values to (\ref{eq: g_2tt}-\ref{eq: g_2phiphi}) 
and (\ref{eq: U_3_EM_tensor}-\ref{eq: gamma_3_EM_tensor}) to compute $g_{(3)}$ we obtain
\begin{equation} \label{eq: g_2_asym_AdS}
ds_{(2)}^{2}=-\frac{1}{2}[dt^2+d\Omega^2]
\end{equation}
\begin{equation} \label{eq: g_3_asym_AdS}
ds_{(3)}^2=\frac{2}{3}\hat{W}_{3} dt^2 + 2\hat{U}_{3} dt d\theta +\left(\frac{1}{3} \hat{W}_3 - 2\hat{\gamma}_{3}\right) d\theta^2+\left( \frac{1}{3}\sin^2 \theta \hat{W}_{3} +2\sin^2 \theta \hat{\gamma}_{3} \right) d\phi^2
\end{equation}
Notice that $g_{(2)}$ can also be obtained from (\ref{Ein_Un}) using the curvature formula (\ref{eq: AdS_g_2_check})). 

The second of these two formulae gives us the energy-momentum tensor for an asymptotically AdS$_4$ spacetime in terms of Bondi gauge functions. From (\ref{eq: g_3_asym_AdS}) we note that 
\begin{equation} \label{eq: g_3_tt_asym_AdS}
g_{(3)tt}=\frac{2\hat{W}_{3}}{3} = -\frac{4\hat{m}_B}{3}
\end{equation}
which arises from the formula (\ref{eq: mass_aspect_1}) for the Bondi mass aspect, $m_B$, now restricted to the boundary, $\hat{m}_B= m_B |_\mathscr{I}$. Thus, the $g_{(3)tt}$ component of the energy-momentum tensor is determined entirely by the mass aspect function.
This implies in particular that the Bondi mass for asymptotically AdS$\mathbf{_4}$ spacetimes is equal to the mass computed using the holographic energy momentum tensor. 
 Indeed, 
\begin{align}
{\cal M} &= \int_{S^2} dS_\mu \langle T^\mu{}_\nu\rangle  \xi^\nu = -\frac{3}{16 \pi} \int_{S^2} g_{(3)tt} 
=\frac{1}{4 \pi} \int_{S^2} \hat{m}_B ={\cal M}_B
 \end{align}
where in the first equality $\xi^\mu$ is an asymptotic timelike killing vector, which we take to be 
$\xi^\mu=-\left(\frac{\partial}{\partial t}\right)^{\mu}$ and we set $l=G=1$.
This also implies that the Bondi mass for asymptotically AdS$\mathbf{_4}$ spacetimes is equal  with all other definitions of mass for asymptotically AdS$\mathbf{_4}$ spacetimes as all of them are known to agree with the holographic mass (as  they had to since \cite{Papadimitriou:2005ii} provided a first principles 
derivation that the conserved charges for general AlAdS spacetimes are the holographic charges). In appendix \ref{AD_mass_appendix} we demonstrate the equality between the Bondi mass and the Abbott-Deser mass. 

We will now discuss interesting examples of asymptotically AdS$_4$ spacetimes.

\subsubsection{Global AdS$\mathbf{_4}$}

An obvious example of an asymptotically AdS$_4$ spacetime is the case of global AdS$_4$ itself. Using the usual normalisation of $l=1$, the line-element in retarded Bondi coordinates reads 
\begin{equation}
ds^2=-(1+r^2)du^2-2du dr+r^2d\Omega^2.
\end{equation}
Clearly $W_{3}=U_{3}=\gamma_{3}=0$. Applying this to (\ref{eq: g_3_asym_AdS}) we see that $g_{(3)}$ vanishes
and thus the energy-momentum tensor of the CFT state (the vacuum state) dual to global AdS$_{4}$ is zero.

\subsubsection{AdS$\mathbf{_{4}}$ Schwarzschild}  

We now consider the AdS$_4$-Schwarzschild black hole solution whose metric in retarded Bondi coordinates reads 
\begin{equation} \label{eq: AdS_Schwarzchild}
ds^2=-\left(1+r^2-\frac{2m}{r}\right)du^2-2du dr+r^2 d\Omega^2.
\end{equation}
This solution is an example of an asymptotically AdS$_4$ metric and thus it automatically has the same values for $g_{(0)}$ and $g_{(2)}$ as presented above. \par

This solution has metric functions $\beta=\gamma=U=0$ and matching (\ref{eq: AdS_Schwarzchild}) with the general Bondi gauge metric (\ref{eq: Bondi_Metric}) gives $W=1+1/r^2-2m/r^3$ i.e. $W_{3}=-2m$. Using the relation (\ref{eq: g_3_asym_AdS}) we obtain
\begin{equation}
g_{(3) ab}= -\frac{2m}{3} \left( \begin{array}{ccc} 
2 & 0 & 0 \\
0 & 1 & 0 \\
0 & 0 & \sin^2 \theta \\
\end{array} \right)
\end{equation}
which reduces to the case of global AdS$_4$ when $m=0$.

\subsubsection{Flat $\mathbf{g_{(0)}}$}

Let us now consider the case where the metric $g_{(0)}$ is flat. One can show explicitly that the metric on the Einstein universe is conformally flat using the coordinate transformation 
\begin{equation} \label{eq: S2_to_flat_trans}
\tau \pm y = \tan\left[ \frac{1}{2}(t\pm \theta) \right].
\end{equation}
to obtain 
\begin{equation} \label{eq: conformal_flat_g_0}
ds_{(0)}^2=4 \cos^2 \left[ \frac{1}{2}(t+ \theta) \right] \cos^2 \left[ \frac{1}{2}(t- \theta) \right] (-d\tau^2 + dy^2 + y^2d\phi^2 )
\end{equation}
which is clearly conformal to the flat metric on $\mathbb{R}^{2,1}$ in polar coordinates.

Under a conformal transformation $g_{(0)} \rightarrow e^{2\sigma} g_{(0)}$ the coefficients of the Fefferman-Graham expansion transform as
(see discussion in \cite{Skenderis:2000in})
\begin{align}
\begin{split} \label{flat}
g'_{(0)ab}&=e^{2\sigma}g_{(0)ab} \\
g'_{(2)ab}&=g_{(2)ab}+\nabla_{a}\nabla_{b} \sigma - \nabla_{a} \sigma \nabla_{b} \sigma + \frac{1}{2}(\nabla \sigma)^2 g_{(0)ab} \\
g'_{(3)ab}&=e^{-\sigma} g_{(3) ab}
\end{split}
\end{align}
and therefore
\begin{align}
\begin{split} \label{eq: g_i_flat_g_0}
g'_{(0)ab}&=\eta_{ab}, \qquad  ds'^2_{(0)}=-d\tau^2+dy^2+y^2d\phi^2 \\
g'_{(2)ab}&=0 \\
g'_{(3)ab}&=2 \cos \left[ \frac{1}{2}(t+ \theta) \right] \cos \left[ \frac{1}{2}(t- \theta) \right] 
\left(
\begin{array}{ccc} 
\frac{2}{3} \hat{W}_{3} & \hat{U}_{3} & 0 \\
\hat{U}_{3} & \frac{1}{3}\hat{W}_{3} - 2\hat{\gamma}_{3} & 0 \\
0 & 0 & \sin^2 \theta \left(\frac{1}{3}\hat{W}_{3} +2 \hat{\gamma}_{3}\right) \\
\end{array}
\right) \\
ds'^2_{(3)}&=\frac{2}{3}\left[(1+(\tau+y)^2)(1+(\tau-y)^2)\right]^{-5/2}\\
& \phantom{aa} \times \{ [-48 y \tau(1+y^2+\tau^2)\hat{U}_3+8(y^4+(1+\tau^2)^2+y^2(1+4\tau^2))\hat{W}_3-\\
&  \phantom{aaaaa\,}  96y^2\tau^2 \hat{\gamma}_3] d\tau^2 +  [24(y^4+(1+\tau^2)^2+y^2(2+6\tau^2))\hat{U}_3-\\
&  \phantom{aaaaa\,} 48y \tau(1+y^2+\tau^2)\hat{W}_3+96y \tau (1+y^2+\tau^2)\hat{\gamma}_3]dy d\tau+\\
& \phantom{aaaaa\,}  [-48y \tau(1+y^2+\tau^2)\hat{U}_3+ 4(y^4+(1+\tau^2)^2+2y^2(1+5\tau^2))\hat{W}_3-\\
& \phantom{aaaaa\,} 24(1+y^2+\tau^2)^2\hat{\gamma}_3]dy^2+\\
& \phantom{aaaaa\,} [4y^2((1+(\tau+y)^2)(1+(\tau-y)^2))(\hat{W}_3+6\hat{\gamma}_3)]d\phi^2\}.
\end{split}
\end{align}
The equation for $g'_{(0)}$ is presented in the flat coordinates $(\tau, y, \phi)$ and both $g_{(2)}$ and $g_{(3)}$ have been presented in both the old $(t, \theta, \phi)$ coordinates as well as the new coordinates $(\tau, y, \phi)$ ($g_{(2)}$ trivially so). 

We observe that $g'_{(3)}$ is merely (\ref{eq: g_3_asym_AdS}) multiplied by a conformal factor. (\ref{eq: g_i_flat_g_0}) presents the specific factor when we have a flat metric at the boundary $g_{(0) ab}=\eta_{ab}$. We also remark that one could immediately deduce that $g'_{(2)ab}$ vanishes by applying (\ref{eq: AdS_g_2_check}) to the flat metric. 

\subsubsection{AdS$\mathbf{_4}$ black brane}

An example of a vacuum solution with a flat $g_{(0)}$ is the AdS black brane solution.
The black brane is an asymptotically AdS solution to the vacuum Einstein equations with planar horizon topology,
\begin{align}
\begin{split}
ds^2&=\frac{d \rho^2}{4\rho^2 f_b(\rho)}+\frac{-f_b(\rho)dt^2+dx_1^2+dx_2^2}{\rho}, \\
f_b(\rho)&=1-\frac{\rho^{3/2}}{b^3}
\end{split} 
\end{align}
where $b$ is related to the temperature $T$ of the brane via
$b=3/(4\pi T)$.

It is straightforward to transform the black brane solution into the Fefferman-Graham form using a redefintion of the radial coordinate $\rho$ (see for example \cite{Kanitscheider:2009as}), resulting in Fefferman-Graham expansion coefficients:
\begin{align}
\begin{split} \label{eq: g_i_black_brane}
g_{(0)ab}&=\eta_{ab}; \quad ds^2_{(0)}=-d\tau^2+dy^2+y^2d\phi^2;\\
g_{(2)ab}&=0; \\
g_{(3)ab}&=-\frac{1}{3} \left( \frac{4\pi T}{3} \right)^3 \left( \begin{array}{ccc} 
2 & 0 & 0 \\
0 & 1 & 0 \\
0 & 0 & y^2 \\
\end{array} \right),
\end{split}
\end{align}
where we use the flat coordinates $(\tau, y, \phi)$ of (\ref{eq: g_i_flat_g_0}). 

We can calculate the relevant Bondi quantities for the AdS black brane from (\ref{eq: g_i_flat_g_0}) and (\ref{eq: g_i_black_brane}):\begin{align}
\begin{split} \label{eq: Bondi_quantities_black_brane}
\hat{\gamma}_3&=\frac{1}{8} \left( \frac{4\pi T}{3} \right)^3 \tau ^2 y^2 \sqrt{\left((y-\tau )^2+1\right) \left((\tau +y)^2+1\right)} \\
\hat{U}_3&=- \frac{1}{4} \left( \frac{4\pi T}{3} \right)^3 \tau  y \left(\tau ^2+y^2+1\right) \sqrt{\left((y-\tau )^2+1\right) \left((\tau +y)^2+1\right)} \\
\hat{W}_3 &= - \frac{1}{8} \left( \frac{4\pi T}{3} \right)^3 \sqrt{\left((y-\tau )^2+1\right) \left((\tau +y)^2+1\right)} \\
&  \phantom{aaaaa} \times \left(\left(\tau ^2+1\right)^2+y^4+\left(4 \tau ^2+2\right) y^2\right).
\end{split}
\end{align}
Note that $\hat{W}_3$ will be related to the mass aspect if we use (\ref{flat}) to transform the solution so that to boundary metric is $\mathbb{R} \times S^2$. The corresponding mass will then be the conserved charge associated with time translations. However, as the coordinate transformation (\ref{eq: S2_to_flat_trans}) transforms $t$ to $\tau$ and $y$, what was a mass aspect on $\mathbb{R} \times S^2$ is not a mass aspect on $\mathbb{R} ^{1,2}$. Indeed,  it was shown in \cite{Skenderis:2000in} that
\begin{equation}
\partial_t=\frac{1}{2} (P_{\tau}+K_{\tau})
\end{equation}
where $P_{\tau}=\partial_{\tau}$ is the generator of $\tau$-translations and $K_{i}=x^2 \partial_i - 2x_i x^j \partial_j$ the generator of special conformal transformations (see also the discussion in \cite{Aharony:1999ti}) . Thus, $\hat{W}_3$
is related to a linear combination of the  mass and the ``special conformal'' aspects on $\mathbb{R} ^{1,2}$. \footnote{One can explicitly confirm this using (\ref{flat}), (\ref{eq: S2_to_flat_trans})  and
$K_{\tau} 
=(y^2+\tau^2)\partial_{\tau}+2\tau y \partial_y$.}

\subsubsection{Bondi mass} \label{sec: Bondi_mass_AAdS}

In our gauge the Bondi mass (\ref{eq: Bondi_mass_time_u}) reduces to
\begin{equation} \label{eq: mass_formula}
\mathcal{M}_{B}=\frac{1}{4\pi} \int_{S^2} m_B =\frac{1}{2} \int_{0}^{\pi} m_B \sin(\theta) \, d\theta 
\end{equation}
\noindent where $m_B$ is the mass aspect function defined in (\ref{eq: mass_aspect_1}). \par

We would like to examine whether or not the Bondi mass in asymptotically locally AdS spacetimes maintains the monotonicity property of the mass in asymptotically flat spacetime \cite{Bondi:1962px, Sachs:1962wk}, namely
\begin{equation} \label{eq: mass_monotonicity}
\frac{ \partial \mathcal{M}_B }{\partial u} \leq 0.
\end{equation}
Note that for asymptotically flat spacetimes saturation of the bound corresponds to the absence of gravitational radiation.

To examine the AdS analogue of this result, we begin by examining the case of asymptotically AdS space-times for which $\gamma_0=\beta_0=U_0=0$. In this case  the mass aspect coincides with the original definition, $2m_B = -W_3$ and 
\begin{equation}
\frac{ \partial \mathcal{M}_B }{\partial u}=\frac{1}{2} \int_{0}^{\pi} \frac{ \partial m_B}{\partial u} \sin(\theta) \, d\theta = -\frac{1}{4} \int_{0}^{\pi} \frac{ \partial W_3}{\partial u} \sin(\theta) \, d\theta.
\end{equation} 
To analyse this, we use the supplementary condition (\ref{eq: SC2}) (evolution equation for $W_3$), which reduces to
\begin{align}
\begin{split}
W_{3, u} = & \frac{1}{2} [6 \gamma_1^4-\gamma_{1,\theta}^2+4 \gamma_{1,u}^2+4 \gamma_{1,u} -2 \gamma_{1, u \theta \theta}-8 \cot ^2(\theta ) \gamma_{1}^2+ \\
& \phantom{aa} \gamma_{1} (-12 \gamma_{3}+\gamma_{1, \theta \theta}-15 \cot (\theta ) \gamma_{1, \theta})-6 \cot (\theta ) \gamma_{1, u \theta}+3 U_{3, \theta }+3 \cot (\theta ) U_{3}]
\end{split}
\end{align}
in this case. 

To simplify this relation further we use the Einstein equation (\ref{eq: gamma_1}), which implies
\begin{equation} \label{eq: gamma_1=0}
\gamma_1=0
\end{equation}
and thus 
\begin{equation}
W_{3,u}=\frac{3}{2}\left(U_{3, \theta}+ \cot(\theta) U_3\right)
\end{equation}   
and thus substituting into equation (\ref{eq: mass_monotonicity})  gives 
\begin{equation} \label{eq: Bondi_Mass_derivative}
\frac{ \partial \mathcal{M}_B }{\partial u}= -\frac{3}{8} \int_0^{\pi} \left(U_{3, \theta}\sin(\theta)+ \cos(\theta) U_3\right)  \, d\theta = -\frac{3}{8} \left[ U_3 \sin(\theta) \right]^{\pi}_{0}.
\end{equation}
To evaluate the limits of this integral we use the same regularity conditions as in \cite{Bondi:1962px}. At the poles of the 2-sphere, \begin{equation}
\frac{U_3}{\sin(\theta)}= f(\cos(\theta)) 
\end{equation}
where the function $f$ is regular at the poles. Applying this condition in (\ref{eq: Bondi_Mass_derivative}) gives 
\begin{equation}
\frac{ \partial \mathcal{M}_B }{\partial u}=-\frac{3}{8}[\sin^2(\theta) f(\cos(\theta))]^{\pi}_{0}=0
\end{equation}
using the regularity of $f$. 

Thus for asymptotically $AdS$ spacetimes the Bondi mass is constant and does not vary with respect to the Bondi time, $u$. This confirms earlier results in \cite{Ashtekar:2014zfa, He:2015wfa, Saw:2016isu, Saw:2017amv, He:2018ikd} (mostly for the dS case). The result is striking and is what would be expected on physical grounds, as we will now explain.  

Firstly, let us recall the interpretation of equation (\ref{eq: gamma_1=0}) in the language of the original work by BMS. Vanishing of $\gamma_1$ implies there is no news and thus (in the asymptotically flat case) the mass is automatically conserved. This interpretation carries over to the asymptotically AdS case. Note however that it seems less likely that this result will extend trivially to the broader class of asympotically locally AdS spacetimes, as it is possible to have vanishing $\gamma_1$ but non-trivial $(\gamma_0, \beta_0, U_0)$. The latter would play a role in the equation (\ref{eq: SC2}) for the evolution of the mass aspect and could alter the monotonicity properties of the mass. 

Another way to understand why the Bondi mass remains constant for asymptotically AdS space-times is that the boundary metric is unchanging, indicating a lack of gravitational radiation to perturb it. Any outgoing radiation would effect the boundary metric and as the metric is unchanging with time there is no gravitational radiation. The original motivation of BMS was to define a mass which captured radiation escaping at (null) infinity and thus our conclusion is consistent with their approach.  

\subsection{Integration scheme} 

In this section we summarise the relation between the Fefferman-Graham integration scheme, which effectively allows the spacetime to be reconstructed in the neighbourhood of the conformal boundary in terms of CFT data, and the integration scheme in Bondi gauge discussed in 
section \ref{subsec: AdS_Scheme}. In the latter, one specifies the data
\begin{equation}
\{\hat{\gamma}_0(t, \theta), \hat{\beta}_0(t, \theta), \hat{U}_0(t, \theta), \hat{\gamma}_3(t, \theta), \hat{U}_3(t, \theta), \hat{W}_3(t, \theta) \, | \, t \in \mathbb{R} , \, \theta \in (0, 2\pi) \} 
\end{equation}
which has the effect of reducing the Einstein equations to \textit{algebraic} equations from which one construct fully the asymptotic solutions to the Einstein equations without having to evolve betweeen null hypersurfaces.

The holographic interpretation of $(\hat{\gamma}_0, \hat{\beta}_0, \hat{U}_0)$ is given by equation (\ref{eq: g_(0)}): these functions define the metric at the conformal boundary, $g_{(0)ab}$. The commonly imposed determinant constraint on the spherical part of the metric in Bondi gauge translates into a determinant constraint on the spherical part of the boundary metric, a constraint which is unnatural from a CFT perspective. 

The data $(\hat{\gamma}_3, \hat{U}_3, \hat{W}_3)$ defines the energy momentum tensor of the dual theory, $T_{ab}$. More precisely, equation \eqref{g3-rel} gives the relation between $g_{(0)ab}$ and $T_{ab} \, ( \sim g_{(3)ab})$ and the coefficients $(\hat{\gamma}_3, \hat{U}_3, \hat{W}_3 )$. With this holographic interpretation we can rephrase the Bondi integration scheme in the following form:

\noindent \textit{Knowledge of the metric $g_{(0)ab}$ at $\mathscr{I}$ and the energy momentum tensor $T_{ab} \sim g_{(3)ab}$ for the CFT dual of the Bondi gauge spacetime is sufficient to algebraically solve the vacuum Einstein equations in the asymptotic region.}

\section{Conclusions} \label{sec:conclusions}

The main result of this paper is the general asymptotic solution of asymptotically local AdS and dS spacetimes in Bondi gauge. We saw that we can use two different integration schemes:  in the boundary scheme we fix data on the conformal boundary only, while in the hybrid scheme  we give data on a null hypersurface  and a portion of the conformal boundary.  We also presented the coordinate transformation to Fefferman-Graham coordinates and identified how to extract the holographic data/conserved quantities directly in Bondi gauge.

The analysis was done for vacuum Einstein gravity in four dimensions and for solutions that are axially and reflection symmetric. It would be straightforward to relax these conditions, {\it i.e.} to 
consider solutions with no axial and reflection symmetry, add matter and generalise to higher dimensions. In odd dimensions the asymptotic expansion will involve logarithmic terms, and so it will in any dimension with specific types of matter (as discussed for $d=4$ in appendix  \ref{logs_appendix}). These logarithms are related to logarithmic divergences in the on-shell value of the gravitational action \cite{Henningson:1998gx, deHaro:2000vlm}.

One undesirable feature of the Bondi gauge is the determinant condition on the angular part of the metric (\ref{eq:Bondi_gauge}). In the context of (A)dS this implies that the angular part of the boundary metric satisfies a similar condition (\ref{eq:bdry_det}). Via gauge/gravity duality however the boundary metric also has the interpretation of a source for the energy momentum tensor of the dual QFT and in QFT the sources should be unconstrained. It would thus be desirable to relax/replace this condition so that the boundary metric is unconstrained.

We have seen that the Bondi mass is constant for asymptotically (A)dS metrics, reflecting the fact that these boundary conditions do not allow for radiating spacetimes. To accommodate 
radiating spacetimes one needs to consider asymptotically locally (A)dS spacetimes with a time dependent boundary metric. While we now know the general asymptotic solution for such spacetimes, we do not  know yet what is the correct identification of the appropriate notion of mass that accounts for the radiation (but see \cite{Szabados:2015wqa,He:2015wfa, Chrusciel:2016oux, Saw:2016isu, Saw:2017amv, He:2018ikd, Szabados:2018erf}).
Such mass should be monotonic and it is not yet clear whether the Bondi mass defined using (\ref{g3-rel}) has such monotonicity properties. A radiating spacetime which is asymptotically locally AdS and possesses a ``Bondi mass'' with the required properties \cite{Bakas:2014kfa}  is the AdS Robinson-Trautman solution. It would thus be useful to bring this solution to Bondi gauge and use it as a playground.

In this paper we only touched upon the case of positive cosmological constant, only discussing properties that can be directly inferred from those of negative $\Lambda$. There are however important global differences between the two cases and it would be interesting to completely analyse the case of positive cosmological constant in detail, especially given its phenomenological importance. We hope to return to this and related issues in the near future.

The direct analogue of the asymptotically flat case when $\Lambda \neq 0$ is the case of asymptotically (A)dS spacetimes. When $\Lambda \neq 0$, however, we have seen that we can obtain asymptotic solutions more generally for asymptotically locally (A)dS spacetimes. It would be interesting to revisit the case of no cosmological constant and determine the most general boundary conditions allowed by Einstein's equations (and the variational problem) at null infinity and find the corresponding asymptotic solutions. This may be relevant in understanding how holography works in asymptotically flat gravity.

\section*{Acknowledgements} We would like to thank James Vickers for discussions and for comments on the manuscript.
KS and MT are supported in part by the STFC Consolidated Grant
``New Frontiers in Particle Physics and Cosmology", ST/P000711/1. AP is supported by STFC. 
This project has received funding from the European Union's Horizon 2020 research and innovation programme under the Marie Sk\l{}odowska-Curie grant agreement No 690575. This research was supported in part by the National Science Foundation under Grant No. NSF PHY-1748958. MT would like to thank the Banff International Research Centre and KITP for hospitality during the completion of this work. 

\appendix

\section{Supplementary conditions} \label{sec: SCapp}

In section \ref{subsec: SC}, we explained how the $\{u \theta\}$ and $\{u u \}$ Einstein equations can be reduced to $f=0$ and $g=0$ where $f,g$ are functions of $(u, \theta)$. Here we present these equations in the $\Lambda < 0$ case as constraints upon the derivatives of the functions $U_{3, u}$ and $W_{3, u}$. We have used the normalisation of $l=\sqrt{-3/\Lambda}=1$. The formulae for $\Lambda > 0$  can be obtained by using dimensional analysis (see section  \ref{sec:g0}) to reinstate $l$ and then  $\Lambda$. \\

\textbf{$f=0$:}
\begin{align}  \label{eq: SC1}
\begin{split}
U_{3, u}=\,& \frac{4}{3} \gamma_{3} ((2\gamma_{0, \theta}-4 \cot (\theta )) (U_{0})^2+2 (e^{2\beta_{0}}\gamma_{1}-U_{0, \theta}+\gamma_{0,u}) U_{0}- \\
&3 e^{4\beta_{0}-2\gamma_{0}} (\cot (\theta )+\beta_{0,\theta}-\gamma_{0,\theta}))+\frac{1}{9} (28 e^{2\beta_{0}} U_{0} (\gamma_{1})^4-\\
&30 e^{4\beta_{0}-2\gamma_{0}} (\cot (\theta )-\beta_{0,\theta}-\gamma_{0,\theta}) (\gamma_{1})^3-14 (U_{0})^2 (\cot (\theta )-2\gamma_{0, \theta}) (\gamma_{1})^3+\\
&14 U_{0} (U_{0, \theta}+2\gamma_{0,u}) (\gamma_{1})^3+3 e^{2\beta_{0}-2\gamma_{0}} (U_{0, \theta} (7 \cot (\theta )+8\beta_{0,\theta}-8\gamma_{0, \theta})+\\
&e^{2\beta_{0}}\gamma_{1,\theta}-U_{0, \theta \theta}+4 (-4 \cot (\theta )+\beta_{0,\theta}+4\gamma_{0, \theta})\gamma_{0,u}+3\beta_{0, u \theta}-8\gamma_{0, u \theta}) (\gamma_{1})^2-\\
&6 U_{0} W_{3}\gamma_{1}+3 e^{2\beta_{0}-4\gamma_{0}} (2 e^{2\beta_{0}} (-4 (\gamma_{0,\theta})^3+6 \cot (\theta ) (\gamma_{0,\theta})^2+\\
&(3 \csc ^2(\theta )+4\beta_{0, \theta \theta}+6\gamma_{0, \theta \theta}+2)\gamma_{0, \theta}+8 (\beta_{0,\theta})^2 (2 \cot (\theta )-\gamma_{0,\theta})+\\
&2 \cot (\theta )\beta_{0, \theta \theta}-3 \cot (\theta )\gamma_{0, \theta \theta}-2\beta_{0,\theta} (\csc ^2(\theta )-4 (\gamma_{0,\theta})^2+8 \cot (\theta )\gamma_{0, \theta}+\\
&4\beta_{0, \theta \theta}+2\gamma_{0, \theta \theta}+2(-2\beta_{0, \theta \theta \theta}-\gamma_{0, \theta \theta \theta})+e^{2\gamma_{0}} (4\gamma_{1,\theta} (U_{0, \theta}-2\gamma_{0,u})+\\
&2 (8 \cot (\theta )+3\beta_{0,\theta}-8\gamma_{0, \theta})\gamma_{1,u}+3\gamma_{1, u \theta}))\gamma_{1}-24 e^{2\beta_{0}} U_{0}\gamma_4-12 U_{0} U_{3, \theta}+\\
&3 e^{2\beta_{0}-2\gamma_{0}} W_{3, \theta}+12 e^{2\beta_{0}-2\gamma_{0}} W_{3}\beta_{0,\theta}-12 U_{0} U_{3} (\cot (\theta )+\gamma_{0,\theta})-\\
&18 e^{4\beta_{0}-2\gamma_{0}} \gamma_{3, \theta}-24 (U_{0})^2 \gamma_{3, \theta}+\\
&12 e^{4\beta_{0}-4\gamma_{0}} (\gamma_{1, \theta} (2\beta_{0,\theta} (\cot (\theta )+3\beta_{0,\theta}-2\gamma_{0, \theta})+\beta_{0, \theta \theta})+\beta_{0,\theta}\gamma_{1, \theta \theta})+\\
&3 e^{2\beta_{0}-2\gamma_{0}} U_{0} ((\csc ^2(\theta )+16 (\gamma_{0,\theta})^2+8\beta_{0,\theta} (2 \cot (\theta )+\beta_{0,\theta})-\\
&4 (7 \cot (\theta )+3\beta_{0,\theta})\gamma_{0, \theta}+4\beta_{0, \theta \theta}-12 (\gamma_{0, \theta \theta}+1)) (\gamma_{1})^2+\\
&2 ((15 \cot (\theta )+12\beta_{0,\theta}-16\gamma_{0, \theta})\gamma_{1,\theta}+3\gamma_{1, \theta \theta})\gamma_{1}+10 (\gamma_{1, \theta})^2)+\\
&6 U_{3} (2 e^{2\beta_{0}}\gamma_{1}-2 U_{0, \theta}+3\beta_{0,u}-\gamma_{0,u})+21 e^{2\beta_{0}-2\gamma_{0}}\gamma_{1,\theta}\gamma_{1,u}-24 U_{0} \gamma_{3,u})
\end{split}
\end{align}

\textbf{$g=0$:}
\begin{align*}  \label{eq: SC2}
W_{3,u}=\,&3 e^{4 \beta_{0}} \gamma_{1}^4+\frac{1}{2} e^{-2 \gamma_{0}} (e^{2 \gamma_{0}} U_{0, \theta}^2-e^{4 \beta_{0}} (8 \text{ct}^2(\theta )-16 \gamma_{0, \theta} \text{ct}(\theta )+4 \beta_{0, \theta}^2+8 \gamma_{0, \theta}^2+\\
&7 \beta_{0, \theta} (\text{ct}(\theta )-2 \gamma_{0, \theta})+7 \beta_{0, \theta \theta})) (\gamma_{1})^2+\frac{1}{2} e^{-2 \gamma_{0}} (-12 e^{4 \beta_{0}+2 \gamma_{0}} \gamma_{3}-\\
&2 U_{0, \theta} (4 e^{2 \beta_{0}} \text{ct}^2(\theta )-9 e^{2 \beta_{0}} \gamma_{0, \theta} \text{ct}(\theta )-3 e^{2 \beta_{0}} \csc ^2(\theta )+6 e^{2 \beta_{0}} \beta_{0, \theta}^2+6 e^{2 \beta_{0}} \gamma_{0, \theta}^2+\\
&e^{2 \beta_{0}} \beta_{0, \theta} (11 \text{ct}(\theta )-16 \gamma_{0, \theta})+3 e^{2 \beta_{0}} \beta_{0, \theta \theta}-5 e^{2 \beta_{0}} \gamma_{0, \theta \theta}-2 e^{2 \gamma_{0}} \gamma_{1,u})+\\
&e^{2 \beta_{0}} (-e^{2 \beta_{0}} (15 \text{ct}(\theta )-14 \beta_{0, \theta}-14 \gamma_{0, \theta}) \gamma_{1, \theta}+4 (-\text{ct}(\theta )+\beta_{0, \theta}+\gamma_{0, \theta}) U_{0, \theta \theta}+\\
&e^{2 \beta_{0}} \gamma_{1, \theta \theta}+2 U_{0, \theta \theta \theta}+32 \text{ct}(\theta ) \beta_{0, \theta} \gamma_{0,u}-32 \beta_{0, \theta} \gamma_{0, \theta} \gamma_{0,u}-12 \text{ct}(\theta ) \beta_{0, u \theta}-\\
&16 \beta_{0, \theta} \beta_{0, u \theta}+16 \gamma_{0, \theta} \beta_{0, u \theta}+4 \text{ct}(\theta ) \gamma_{0, u \theta}+24 \beta_{0, \theta} \gamma_{0, u \theta}-8 \gamma_{0, \theta} \gamma_{0, u \theta}-4 \beta_{0, u \theta \theta}+\\
&4 \gamma_{0, u \theta \theta})) \gamma_{1}+8 e^{4 \beta_{0}-4 \gamma_{0}} (\gamma_{0, \theta})^4-16 e^{4 \beta_{0}-4 \gamma_{0}} \text{ct}(\theta ) (\beta_{0, \theta})^3+(-14 e^{4 \beta_{0}-4 \gamma_{0}} \text{ct}(\theta )-\\
&8 e^{4 \beta_{0}-4 \gamma_{0}} \beta_{0, \theta}) (\gamma_{0, \theta})^3-\frac{1}{2} e^{4 \beta_{0}-2 \gamma_{0}} (\gamma_{1, \theta})^2+(U_{0})^3 (\frac{7}{3} e^{2 \gamma_{0}-2 \beta_{0}} (\text{ct}(\theta )-2 \gamma_{0, \theta}) (\gamma_{1})^3+\\
&4 e^{2 \gamma_{0}-2 \beta_{0}} (\gamma_{3} (2 \text{ct}(\theta )-\gamma_{0, \theta})+\gamma_{3, \theta}))+4 e^{2 \beta_{0}-4 \gamma_{0}} \beta_{0, \theta}^2 (-e^{2 \beta_{0}} \text{ct}^2(\theta )-2 e^{2 \beta_{0}}+\\
&2 e^{2 \beta_{0}} \csc ^2(\theta )-4 e^{2 \beta_{0}} \beta_{0, \theta \theta}+2 e^{2 \beta_{0}} \gamma_{0, \theta \theta}+e^{2 \gamma_{0}} \gamma_{1,u})+\gamma_{0, \theta}^2 (-32 e^{4 \beta_{0}-4 \gamma_{0}} \beta_{0, \theta}^2+\\
&28 e^{4 \beta_{0}-4 \gamma_{0}} \text{ct}(\theta ) \beta_{0, \theta}-e^{2 \beta_{0}-4 \gamma_{0}} (-3 e^{2 \beta_{0}} \text{ct}^2(\theta )+4 e^{2 \beta_{0}}+9 e^{2 \beta_{0}} \csc ^2(\theta )+\\
&16 e^{2 \beta_{0}} \beta_{0, \theta \theta}+18 e^{2 \beta_{0}} \gamma_{0, \theta \theta}+4 e^{2 \gamma_{0}} \gamma_{1,u}))+(U_{0})^2 (\frac{1}{3} (-14) e^{2 \gamma_{0}} \gamma_{1}^4-\\
&\frac{7}{3} e^{2 \gamma_{0}-2 \beta_{0}} (U_{0, \theta}+2 \gamma_{0,u}) (\gamma_{1})^3+\frac{1}{2} (-19 \text{ct}^2(\theta )+60 \gamma_{0, \theta} \text{ct}(\theta )+10 \csc ^2(\theta )-\\
&8 \beta_{0, \theta}^2-32 \gamma_{0, \theta}^2+\beta_{0, \theta} (8 \gamma_{0, \theta}-17 \text{ct}(\theta ))-7 \beta_{0, \theta \theta}+20 \gamma_{0, \theta \theta}+8) (\gamma_{1})^2+\\
&\frac{1}{2} e^{-2 \beta_{0}} (2 e^{2 \gamma_{0}} W_{3}-e^{2 \beta_{0}} (8 e^{2 \gamma_{0}} \gamma_{3}+(51 \text{ct}(\theta )+30 \beta_{0, \theta}-56 \gamma_{0, \theta}) \gamma_{1, \theta}+9 \gamma_{1, \theta \theta})) \gamma_{1}+\\
&\frac{1}{2} e^{-2 \beta_{0}} (-13 e^{2 \beta_{0}} (\gamma_{1, \theta})^2+8 e^{2 (\beta_{0}+\gamma_{0})}\gamma_{4})+8 e^{2 \gamma_{0}} \gamma_{3} U_{0, \theta}+7 e^{2 \gamma_{0}} U_{3, \theta}+\\
&3 e^{2 \gamma_{0}} U_{3} (3 \text{ct}(\theta )-2 \beta_{0, \theta}+2 \gamma_{0, \theta})-8 e^{2 \gamma_{0}} \gamma_{3} \gamma_{0,u}+8 e^{2 \gamma_{0}} \gamma_{3, u}))+\\
&\gamma_{1, \theta} (-2 e^{2 \beta_{0}-2 \gamma_{0}} \beta_{0, \theta} (3 U_{0, \theta}-4 \gamma_{0,u})-\frac{1}{2} e^{2 \beta_{0}-2 \gamma_{0}} (13 \text{ct}(\theta ) U_{0, \theta}+2 U_{0, \theta \theta}+\\
&8 \beta_{0, u \theta}-4 \gamma_{0, u \theta}))+e^{2 \beta_{0}-4 \gamma_{0}} \beta_{0, \theta} (-e^{2 \beta_{0}} \text{ct}(\theta ) \csc ^2(\theta )-2 e^{2 \beta_{0}} \text{ct}(\theta )+3 e^{4 \gamma_{0}} U_{3}-\\
&16 e^{2 \beta_{0}} \text{ct}(\theta ) \beta_{0, \theta \theta}-10 e^{2 \beta_{0}} \text{ct}(\theta ) \gamma_{0, \theta \theta}-8 e^{2 \beta_{0}} \beta_{0, \theta \theta \theta}-2 e^{2 \beta_{0}} \gamma_{0, \theta \theta \theta}- \tag{A.2}\\
&10 e^{2 \gamma_{0}} \text{ct}(\theta ) \gamma_{1,u}-4 e^{2 \gamma_{0}} \gamma_{1, u \theta})+\gamma_{0, \theta} (32 e^{4 \beta_{0}-4 \gamma_{0}} (\beta_{0, \theta})^3+8 e^{4 \beta_{0}-4 \gamma_{0}} \text{ct}(\theta ) \beta_{0, \theta}^2+\\
&2 e^{2 \beta_{0}-4 \gamma_{0}} (-2 e^{2 \beta_{0}} \text{ct}^2(\theta )+6 e^{2 \beta_{0}}+3 e^{2 \beta_{0}} \csc ^2(\theta )+24 e^{2 \beta_{0}} \beta_{0, \theta \theta}+6 e^{2 \beta_{0}} \gamma_{0, \theta \theta}+\\
&4 e^{2 \gamma_{0}} \gamma_{1,u}) \beta_{0, \theta}+8 e^{2 \beta_{0}-2 \gamma_{0}} U_{0, \theta} \gamma_{1, \theta}+\frac{1}{2} e^{2 \beta_{0}-4 \gamma_{0}} (-3 e^{2 \beta_{0}} \text{ct}(\theta ) \csc ^2(\theta )+2 e^{2 \beta_{0}} \text{ct}(\theta )+\\
&8 e^{2 \beta_{0}} \text{ct}(\theta ) \beta_{0, \theta \theta}+30 e^{2 \beta_{0}} \text{ct}(\theta ) \gamma_{0, \theta \theta}+16 e^{2 \beta_{0}} \beta_{0, \theta \theta \theta}+10 e^{2 \beta_{0}} \gamma_{0, \theta \theta \theta}+\\
&12 e^{2 \gamma_{0}} \text{ct}(\theta ) \gamma_{1,u}+8 e^{2 \gamma_{0}} \gamma_{1, u \theta}))+U_{0} (10 e^{2 \beta_{0}} (\text{ct}(\theta )-\beta_{0, \theta}-\gamma_{0, \theta}) (\gamma_{1})^3+\\
&(U_{0, \theta} (-6 \text{ct}(\theta )-8 \beta_{0, \theta}+8 \gamma_{0, \theta})-e^{2 \beta_{0}} \gamma_{1, \theta}+U_{0, \theta \theta}+16 \text{ct}(\theta ) \gamma_{0,u}-4 \beta_{0, \theta} \gamma_{0,u}-\\
&16 \gamma_{0, \theta} \gamma_{0,u}-3 \beta_{0, u \theta}+8 \gamma_{0, u \theta}) (\gamma_{1})^2+e^{-2 \gamma_{0}} (-e^{2 \beta_{0}} \text{ct}^3(\theta )+3 e^{2 \beta_{0}} \gamma_{0, \theta} \text{ct}^2(\theta )+\\
&2 e^{2 \beta_{0}} \csc ^2(\theta ) \text{ct}(\theta )-14 e^{2 \beta_{0}} \gamma_{0, \theta}^2 \text{ct}(\theta )-9 e^{2 \beta_{0}} \beta_{0, \theta \theta} \text{ct}(\theta )+9 e^{2 \beta_{0}} \gamma_{0, \theta \theta} \text{ct}(\theta )-\\
&14 e^{2 \gamma_{0}} \gamma_{1,u} \text{ct}(\theta )+8 e^{2 \beta_{0}} (\gamma_{0, \theta})^3-4 e^{4 \gamma_{0}} U_{3}-4 e^{2 \beta_{0}} \gamma_{0, \theta}-10 e^{2 \beta_{0}} \csc ^2(\theta ) \gamma_{0, \theta}-  \\
&2 e^{2 \beta_{0}} \beta_{0, \theta}^2 (15 \text{ct}(\theta )-8 \gamma_{0, \theta})-2 e^{2 \gamma_{0}} U_{0, \theta} \gamma_{1, \theta}-16 e^{2 \beta_{0}} \gamma_{0, \theta} \gamma_{0, \theta \theta}+2 e^{2 \beta_{0}} \beta_{0, \theta \theta \theta}+\\
&4 e^{2 \beta_{0}} \gamma_{0, \theta \theta \theta}+8 e^{2 \gamma_{0}} \gamma_{1, \theta} \gamma_{0,u}+16 e^{2 \gamma_{0}} \gamma_{0, \theta} \gamma_{1,u}+\beta_{0, \theta} (-13 e^{2 \beta_{0}} \text{ct}^2(\theta )+\\
&60 e^{2 \beta_{0}} \gamma_{0, \theta} \text{ct}(\theta )+8 e^{2 \beta_{0}}+12 e^{2 \beta_{0}} \csc ^2(\theta )-32 e^{2 \beta_{0}} \gamma_{0, \theta}^2+8 e^{2 \beta_{0}} \beta_{0, \theta \theta}+\\
&20 e^{2 \beta_{0}} \gamma_{0, \theta \theta}-6 e^{2 \gamma_{0}} \gamma_{1,u})-3 e^{2 \gamma_{0}} \gamma_{1, u \theta}) \gamma_{1}+\frac{1}{2} e^{-2 (\beta_{0}+\gamma_{0})} (-7 e^{4 \beta_{0}} \gamma_{1, \theta} \text{ct}^2(\theta )-\\
&48 e^{4 \beta_{0}} \beta_{0, \theta} \gamma_{1, \theta} \text{ct}(\theta )+16 e^{4 \beta_{0}} \gamma_{0, \theta} \gamma_{1, \theta} \text{ct}(\theta )-7 e^{4 \beta_{0}} \gamma_{1, \theta \theta} \text{ct}(\theta )+8 e^{4 \gamma_{0}} U_{3} U_{0, \theta}-\\
&4 e^{2 (\beta_{0}+\gamma_{0})} W_{3, \theta}-e^{2 (\beta_{0}+\gamma_{0})} W_{3} (3 \text{ct}(\theta )+4 \beta_{0, \theta})+\\
&24 e^{4 \beta_{0}+2 \gamma_{0}} \gamma_{3} (\text{ct}(\theta )+\beta_{0, \theta}-\gamma_{0, \theta})+8 e^{4 \beta_{0}} \csc ^2(\theta ) \gamma_{1, \theta}-40 e^{4 \beta_{0}} \beta_{0, \theta}^2 \gamma_{1, \theta}-\\
&8 e^{4 \beta_{0}} \gamma_{0, \theta}^2 \gamma_{1, \theta}+64 e^{4 \beta_{0}} \beta_{0, \theta} \gamma_{0, \theta} \gamma_{1, \theta}+12 e^{4 \beta_{0}+2 \gamma_{0}} \gamma_{3, \theta}-12 e^{4 \beta_{0}} \gamma_{1, \theta} \beta_{0, \theta \theta}+\\
&8 e^{4 \beta_{0}} \gamma_{1, \theta} \gamma_{0, \theta \theta}-16 e^{4 \beta_{0}} \beta_{0, \theta} \gamma_{1, \theta \theta}+8 e^{4 \beta_{0}} \gamma_{0, \theta} \gamma_{1, \theta \theta}-2 e^{4 \beta_{0}} \gamma_{1, \theta \theta \theta}+6 e^{4 \gamma_{0}} U_{3, u}-\\
&12 e^{4 \gamma_{0}} U_{3} \beta_{0,u}+4 e^{4 \gamma_{0}} U_{3} \gamma_{0,u}-6 e^{2 (\beta_{0}+\gamma_{0})} \gamma_{1, \theta} \gamma_{1,u}))+\frac{1}{2} e^{-4 \gamma_{0}} (-2 e^{4 \beta_{0}} \beta_{0, \theta \theta} \text{ct}^2(\theta )-\\
&3 e^{4 \beta_{0}} \gamma_{0, \theta \theta} \text{ct}^2(\theta )-4 e^{2 (\beta_{0}+\gamma_{0})} \gamma_{1,u} \text{ct}^2(\theta )+3 e^{2 \beta_{0}+4 \gamma_{0}} U_{3} \text{ct}(\theta )-4 e^{4 \beta_{0}} \beta_{0, \theta \theta \theta} \text{ct}(\theta )-\\
&4 e^{4 \beta_{0}} \gamma_{0, \theta \theta \theta} \text{ct}(\theta )-6 e^{2 (\beta_{0}+\gamma_{0})} \gamma_{1, u \theta} \text{ct}(\theta )-8 e^{4 \beta_{0}} (\beta_{0, \theta \theta})^2+6 e^{4 \beta_{0}} (\gamma_{0, \theta \theta})^2+\\
&4 e^{4 \gamma_{0}} (\gamma_{1,u})^2+3 e^{2 \beta_{0}+4 \gamma_{0}} U_{3, \theta}-4 e^{4 \beta_{0}} \beta_{0, \theta \theta}+4 e^{4 \beta_{0}} \csc ^2(\theta ) \beta_{0, \theta \theta}+2 e^{4 \beta_{0}} \gamma_{0, \theta \theta}+\\
&6 e^{4 \beta_{0}} \csc ^2(\theta ) \gamma_{0, \theta \theta}+8 e^{4 \beta_{0}} \beta_{0, \theta \theta} \gamma_{0, \theta \theta}-5 e^{2 (\beta_{0}+\gamma_{0})} U_{0, \theta} \gamma_{1, \theta \theta}-2 e^{4 \beta_{0}} \beta_{0, \theta \theta \theta \theta}-\\
&e^{4 \beta_{0}} \gamma_{0, \theta \theta \theta \theta}-e^{4 \gamma_{0}} W_{3} (3 U_{0, \theta}-4 \beta_{0,u})+4 e^{2 (\beta_{0}+\gamma_{0})} \csc ^2(\theta ) \gamma_{1,u}+\\
&4 e^{2 (\beta_{0}+\gamma_{0})} \beta_{0, \theta \theta} \gamma_{1,u}+4 e^{2 (\beta_{0}+\gamma_{0})} \gamma_{0, \theta \theta} \gamma_{1,u}-2 e^{2 (\beta_{0}+\gamma_{0})} \gamma_{1, u \theta \theta}) 
\end{align*}
where we have used the abbreviations `ct$(\theta)$' to refer to the cotangent function and `csc$(\theta)$' for cosecant. These two equations are essential to check that $g_{(3)ab}$ satisfies the conservation property (\ref{eq: g_3_conditions}).

The supplementary conditions are enormously complicated by the presence of $\Lambda \neq 0$ with non-trivial coefficients $\gamma_0, \beta_0, U_0$. These formulae simplify significantly in the asymptotically (A)dS and asymptotically flat cases. 

Asymptotically (A)dS spacetimes in Bondi coordinates have $\gamma_0=\beta_0=U_0=0$ which gives $\gamma_1=0$ by equation (\ref{eq: gamma_1}). Setting these values in the supplementary equations above gives us
\begin{subequations}
\begin{align}
U_{3,u}&=\frac{1}{3}(4\Lambda \cot(\theta) \gamma_3 + W_{3, \theta}+ 2\Lambda \gamma_{3,\theta}) \tag{A.3a} \\
W_{3,u}&=-\frac{1}{2}\Lambda (\cot(\theta) U_3 + U_{3, \theta}) \tag{A.3b}
\end{align}
\end{subequations}
where we have reinstated the factors of $\Lambda$ using dimensional analysis. 

For the asymptotically flat supplementary conditions, we again have $\gamma_0=\beta_0=U_0=0$ as well as $\Lambda=0$ but now $\gamma_1 \neq 0$. As given in \cite{Bondi:1962px}, the asymptotically flat supplementary conditions are 
\begin{subequations}
\begin{align}
U_{3,u}&=\frac{1}{3} \left(7 \gamma_{1,\theta} \gamma_{1,u}+\gamma_1 \left(3 \gamma_{1,u\theta}+16 \cot (\theta ) \gamma_{1,u}\right)+W_{3, \theta}\right) \tag{A.4a} \\
W_{3,u}&=2 \left(\gamma_{1,u}\right)^2+2 \gamma_{1,u}-\gamma_{1,u\theta\theta}-3 \cot (\theta ) \gamma_{1,u\theta}. \tag{A.4b}
\end{align} 
\end{subequations}

\section{Intermediate pieces of the Fefferman-Graham transformation}  \label{sec: FG_appendix}

In this appendix we provide formulae for transforming the Bondi gauge metric into the Fefferman-Graham form. Expressions for the intermediate metric tensors are omitted for brevity.

\subsection{Vanishing of $g_{(1)}$}

In this section we demonstrate explicitly that $g_{(1)}$ vanishes. 
Note first that the Bondi metric (\ref{eq: Bondi_metric_zero_functions}) used to compute $g_{(0)}$  is insufficient for computing $g_{(1)}$: it only includes the solution to Einstein's equations at leading order but for $g_{(1)}$ we require $1/r \sim \rho$ contributions to the metric. To compute $g_{(1)}$ we therefore need to retain the following contributions to the metric functions
\begin{subequations}
\begin{align}
\gamma(u,r,\theta)&=\gamma_{0}(u,\theta)+ \frac{\gamma_{1} (u,\theta)}{r}  \\
\beta(u,r,\theta)&=\beta_{0}(u,\theta)  \\
U(u,r,\theta)&=U_{0}(u,\theta)+\frac{2}{r}e^{2(\beta_{0}(u,\theta)-\gamma_{0}(u,\theta))}\beta_{0,\theta}(u,\theta) \\
W(u,r,\theta)&=e^{2\beta_{0}(u,\theta)}+\frac{\cot(\theta)U_{0}(u,\theta)+U_{0,\theta}}{r}.
\end{align}
\end{subequations}
which are the solutions to the field equations (\ref{eq: AdS_me1}-\ref{eq: AdS_me4}) up to $\mathcal{O}(1/r)$. Note also that we use the normalisation $l=1$. 

As before, we begin with the Bondi metric in the form  (\ref{eq: Bondi_Sachs_Metric}) and transform into the coordinates $(t, r_*, \theta, \phi)$ using transformations  (\ref{eq: time_transformation}) and  (\ref{eq: leading_order_tortoise}) 
\begin{subequations} \label{eq: ttheta_transfromations_exact}
\begin{align}
u&=t-r_* \\
r&=\tan\left(r_*+ \frac{\pi}{2}\right)
\end{align}
\end{subequations}
where we have written the transformation  (\ref{eq: leading_order_tortoise}) in exact form. 

Next we extend the transformations (\ref{eq: tortoise_nbhd}, \ref{eq: theta_t_transformations}) to one order higher in $\rho$
\begin{subequations}
\begin{align}
&r_* \rightarrow \rho  + b_1(t,\theta) \rho^2 \\
&t \rightarrow t + \alpha_1(t,\theta)\rho + b_2(t,\theta) \rho^2 \\
&\theta \rightarrow \theta + \alpha_2(t,\theta)\rho + b_3(t,\theta) \rho^2 
\end{align}
\end{subequations}
where $\alpha_{1,2}$ are given in (\ref{eq: ttheta_trans}) and $b_{1,2,3}$ are to be determined. When considering how the differentials transform it will again be sufficient to consider the pieces which contribute to the metric at $\mathcal{O}(1/\rho)$
\begin{subequations}
\begin{align}
&dr_* \rightarrow d\rho +2b_1\rho d\rho \\
&dt \rightarrow dt + \alpha_1 d\rho + (\partial_t \alpha_1) \rho dt +  (\partial_{\theta} \alpha_1) \rho d\theta + 2\rho b_2 d\rho \\
&d\theta \rightarrow d\theta  + \alpha_2 d\rho + (\partial_t \alpha_2) \rho dt +  (\partial_{\theta} \alpha_2) \rho d\theta + 2\rho b_3 d\rho.
\end{align}
\end{subequations}
i.e. we do not need to include terms of $\mathcal{O}(\rho^2)$ or higher.

The final subtlety when applying this procedure is to take into account that the metric functions $(\gamma, \beta, U, W)$ are all functions of $(t-r_* , \theta)$ prior to applying these transformations. Terms up to $\mathcal{O}(\rho)$ need to be included in these arguments, i.e.
\begin{equation} 
t-r_* \rightarrow  \, t+\rho\alpha_1 - \rho+\mathcal{O}(\rho^2)=t+\rho(\alpha_1-1)+\mathcal{O}(\rho^2)
\end{equation}
and 
\begin{equation}
\theta \rightarrow  \theta +\alpha_2 \rho + \mathcal{O}(\rho^2),
\end{equation}
to calculate all terms contributing  at $\mathcal{O}(1/\rho)$.

 At order $1/\rho$ we are initially left with a seemingly non-zero term with dependence upon our three undetermined transformation coefficients $b_{1,2,3}$. To fix $b_{1,2,3}$ we enforce the following
\begin{equation} \label{eq: sim_eqns_bs}
g_{(1)\rho \rho}(b_1,b_2,b_3)=g_{(1)\rho t}(b_1,b_2,b_3)=g_{(1)\rho \theta}(b_1,b_2,b_3)=0 
\end{equation}
which gives us three equations for the three unknowns $b_{1,2,3}$ ($g_{(1) \rho \phi}$ vanishes automatically by the axi and reflection symmetry). It turns out that the $g_{\rho \rho}$ term is given by
\begin{equation}
g_{(1)\rho \rho}=g_{(1)\rho \rho}(b_1)=2b_1+e^{-2\hat{\beta}_0}\cot(\theta)\hat{U}_0+e^{-2\hat{\beta}_0}\hat{U}_{0,\theta}
\end{equation}
so we can solve $g_{(1)\rho \rho}=0$ for $b_1$ and then we will be left with two equations for the other two unknowns $b_{2,3}$. Solving $g_{(1)\rho \rho}=0$ gives us 
\begin{subequations}
\begin{equation} \label{eq: b_1}
b_1=-\frac{1}{2}e^{-2\hat{\beta}_{0}}(\hat{U}_{0,\theta}+\cot(\theta)\hat{U}_{0})
\end{equation}
using $b_1$ it is straightforward to now solve the remaining equations of (\ref{eq: sim_eqns_bs}), with solutions
\begin{align}
b_2&=-\frac{1}{2}e^{-4\hat{\beta}_{0}}(e^{2\hat{\beta}_{0}}(\hat{U}_{0,\theta}+\cot(\theta)\hat{U}_{0})+2(\hat{\beta}_{0,t}+\hat{\beta}_{0,\theta}\hat{U}_{0})) \\
b_3&=e^{-2\hat{\gamma}_{0}}\hat{\beta}_{0,\theta}+\frac{1}{2}e^{-4\hat{\beta}_{0}}(\hat{U}_{0,t}+\hat{U}_{0}(\hat{U}_{0,\theta}-2(\hat{\beta}_{0,t}+\hat{\beta}_{0,\theta}\hat{U}_{0}))). \label{eq: b_3}
\end{align}
\end{subequations}
where all function arguments are $(t,\theta)$. 

Enforcing equations (\ref{eq: b_1}-\ref{eq: b_3}) in the transformation should make all other coefficients at $\mathcal{O}(1/\rho)$ vanish. To check this we input the values of $b_{1,2,3}$ in (\ref{eq: b_1}-\ref{eq: b_3}). At $\mathcal{O}(1/\rho)$, the line element reduces to
\begin{align}
\begin{split} \label{eq: g_1_term}
ds_{(1)}^2=-\frac{1}{2}e^{-2(\hat{\beta}_{0}+\hat{\gamma}_{0})}&(2dt^2e^{4\hat{\gamma}_{0}}\hat{U}_{0}^2-4dtd\theta e^{4\hat{\gamma}_{0}}\hat{U}_{0}+2d\theta^2e^{4\hat{\gamma}_{0}}-2\sin^2(\theta)d\phi^2) \times \\
&(\hat{U}_{0,\theta}+2\hat{\gamma}_{0,t}-\cot(\theta)\hat{U}_{0}+2\hat{\gamma}_{0,\theta}\hat{U}_{0}+2e^{2\hat{\beta}_{0}}\hat{\gamma}_{1}).
\end{split}
\end{align}
From equation (\ref{eq: gamma_1})
\begin{equation}
\gamma_1=\frac{1}{2}e^{-2\beta_0}(\cot(\theta) U_0-U_{0,\theta}-2U_0 \gamma_{0,\theta}-2\gamma_{0,u})
\end{equation}
and thus the second line of (\ref{eq: g_1_term}) is precisely this Einstein equation (at the boundary), forcing equation (\ref{eq: g_1_term}) to vanish, as required. 

\subsection{Checking $g_{(2)}$}

The $g_{(2)}$ term in the Fefferman-Graham expansion is a useful consistency check as it must take the form
 \cite{Henningson:1998ey, deHaro:2000vlm}
\begin{equation} \label{eq: AdS_g_2_check}
g_{(2)ab}=-R_{(0)ab}+\frac{1}{4}R_{(0)}g_{(0)ab}
\end{equation}
where $R_{(0)ab}$ and $R_{(0)}$ are respectively the Ricci tensor and scalar of the boundary metric tensor $g_{(0)ab}$.

We now proceed to compute $g_{(2)}$ from the Fefferman-Graham expansion and check it via use of the formula above. The procedure for this step is the same as before with a term of one order higher added in each step. We impose the solutions to the Einstein equations as terms up to $\mathcal{O}(1/r^2)$. This procedure gives us the functions 
\begin{subequations}
\begin{align}
\gamma(u,r,\theta)&=\gamma_{0}+ \frac{\gamma_1}{r}\\
\beta(u,r,\theta)&=\beta_{0}-\frac{\gamma_1^2}{4r^2} \\
\begin{split}
U(u,r,\theta)&=U_{0}+\frac{2}{r}\beta_{0, \theta} e^{2( \beta_0- \gamma_0)} - \\
&\phantom{aaa} \frac{1}{r^2}e^{2 \beta_0-2 \gamma_0} (2 \beta_{0,\theta} \gamma_1-2 \gamma_{0,\theta}\gamma_1+\gamma_{1,\theta}+2 \cot(\theta ) \gamma_1)
\end{split}\\
\begin{split}
W(u,r,\theta)&=e^{2\beta_{0}}+\frac{1}{r}[\cot(\theta)U_{0}+U_{0,\theta}]+ \\
& \phantom{aaa} \frac{1}{2r^2}e^{2(\beta_{0}-\gamma_{0})}[2-3e^{2\gamma_0}\gamma_1^2+ 4\cot(\theta)\beta_{0,\theta}+8(\beta_{0,\theta})^2+6\cot(\theta)\gamma_{0,\theta}-\\
& \phantom{aaa  \frac{1}{2r^2}e^{2(\beta_{0}-\gamma_{0})}a} 8\beta_{0,\theta}\gamma_{0,\theta}-4(\gamma_{0,\theta})^2+4\beta_{0,\theta \theta}+2\gamma_{0,\theta \theta}]
\end{split}
\end{align}
\end{subequations}
where, as usual, all of the coefficient functions are taken to be functions of $(u,\theta)$. We will also make use of \eqref{eq: gamma_1} throughout. 

The full transformation is again performed by first using the transformations of (\ref{eq: ttheta_transfromations_exact}) to move into real time $t$ and tortoise coordinate $r_*$  before expanding our coordinates $(r_*, t, \theta)$ in a series in powers of $\rho$. In order to correctly compute $g_{(2)}$ these power series will include terms up to $\mathcal{O}(\rho^3)$. We use the choices of $\alpha_i$ and $\beta_i$ as before and introduce new unknown coefficients $c_i(t,\theta)$ at the next order
\begin{align}
\begin{split}
&r_* \rightarrow \rho  + b_1(t,\theta) \rho^2+c_1(t,\theta) \rho^3 \\
&t \rightarrow t + \alpha_1(t,\theta)\rho + b_2(t,\theta) \rho^2+c_2(t,\theta) \rho^3 \\
&\theta \rightarrow \theta + \alpha_2(t,\theta)\rho + b_3(t,\theta) \rho^2+c_3(t,\theta) \rho^3. 
\end{split}
\end{align}
The procedure for obtaining the $c_i$ is very similar to that for  $b_i$: we fix them by setting $g_{(2) \rho \rho}=g_{(2) \rho t}=g_{(2) \rho \theta}=0$. This gives
\begin{subequations}
\begin{align}
\begin{split}
8c_1(t,\theta)=\,&\frac{1}{3} e^{-4\hat{\beta}_{0}-2 \hat{\gamma}_{0}} (-8 e^{4\hat{\beta}_{0}+2 \hat{\gamma}_{0}}-12 e^{4\hat{\beta}_{0}} (\hat{\gamma}_{0, \theta})^2-24 e^{4\hat{\beta}_{0}}\hat{\beta}_{0, \theta} \hat{\gamma}_{0, \theta}+\\
&6 e^{4\hat{\beta}_{0}} \hat{\gamma}_{0, \theta \theta}+18 \cot (\theta ) e^{4\hat{\beta}_{0}} \hat{\gamma}_{0, \theta}+6 e^{4\hat{\beta}_{0}}+24 e^{4\hat{\beta}_{0}} (\hat{\beta}_{0, \theta})^2+12 e^{4\hat{\beta}_{0}}\hat{\beta}_{0, \theta \theta}+\\
&12 \cot (\theta ) e^{4\hat{\beta}_{0}}\hat{\beta}_{0, \theta}-6 e^{2 \hat{\gamma}_{0}} (\hat{\gamma}_{0, t})^2-12\hat{\beta}_{0, \theta} e^{2 \hat{\gamma}_{0}} \hat{U}_{0,\theta} \hat{U}_{0}-12\hat{\beta}_{0, t} e^{2 \hat{\gamma}_{0}} \hat{U}_{0,\theta}-\\
&12 \cot (\theta )\hat{\beta}_{0, \theta} e^{2 \hat{\gamma}_{0}} (\hat{U}_{0})^2-12 \cot (\theta )\hat{\beta}_{0, t} e^{2 \hat{\gamma}_{0}} \hat{U}_{0}-6 e^{2 \hat{\gamma}_{0}} (\hat{U}_{0})^2-\\
&6 e^{2 \hat{\gamma}_{0}} (\hat{\gamma}_{0, \theta})^2 (\hat{U}_{0})^2-6 e^{2 \hat{\gamma}_{0}} \hat{\gamma}_{0, \theta} \hat{U}_{0,\theta} \hat{U}_{0}+6 e^{2 \hat{\gamma}_{0}} \hat{U}_{0,\theta \theta} \hat{U}_{0}-\\
&12 e^{2 \hat{\gamma}_{0}} \hat{\gamma}_{0, \theta} \hat{\gamma}_{0, t} \hat{U}_{0}+3 e^{2 \hat{\gamma}_{0}} (\hat{U}_{0,\theta})^2-6 e^{2 \hat{\gamma}_{0}} \hat{\gamma}_{0, t} \hat{U}_{0,\theta}+6 e^{2 \hat{\gamma}_{0}} \hat{U}_{0,t \theta}-\\
&3 \cot ^2(\theta ) e^{2 \hat{\gamma}_{0}} (\hat{U}_{0})^2+6 \cot (\theta ) e^{2 \hat{\gamma}_{0}} \hat{\gamma}_{0, \theta} (\hat{U}_{0})^2+18 \cot (\theta ) e^{2 \hat{\gamma}_{0}} \hat{U}_{0,\theta} \hat{U}_{0}+\\
&6 \cot (\theta ) e^{2 \hat{\gamma}_{0}} \hat{\gamma}_{0, t} \hat{U}_{0}+6 \cot (\theta ) e^{2 \hat{\gamma}_{0}} \hat{U}_{0,t})
\end{split}
\end{align}
\begin{align}
\begin{split}
8c_2(t, \theta) =\, & -\frac{1}{3} e^{-6\hat{\beta}_{0}-2 \hat{\gamma}_{0}} (2 e^{2 \hat{\gamma}_{0}} (\hat{U}_{0})^2+6 e^{2\hat{\beta}_{0}+2 \hat{\gamma}_{0}} (\hat{U}_{0})^2+4 e^{2 \hat{\gamma}_{0}} \cot ^2(\theta ) (\hat{U}_{0})^2+\\
&3 e^{2\hat{\beta}_{0}+2 \hat{\gamma}_{0}} \cot ^2(\theta ) (\hat{U}_{0})^2-4 e^{2 \hat{\gamma}_{0}} \csc ^2(\theta ) (\hat{U}_{0})^2+32 e^{2 \hat{\gamma}_{0}} (\hat{\beta}_{0, \theta})^2 (\hat{U}_{0})^2+\\
&6 e^{2 \hat{\gamma}_{0}} (\hat{\gamma}_{0, \theta})^2 (\hat{U}_{0})^2+6 e^{2\hat{\beta}_{0}+2 \hat{\gamma}_{0}} (\hat{\gamma}_{0, \theta})^2 (\hat{U}_{0})^2-4 e^{2 \hat{\gamma}_{0}} \cot (\theta )\hat{\beta}_{0, \theta} (\hat{U}_{0})^2+\\
&12 e^{2\hat{\beta}_{0}+2 \hat{\gamma}_{0}} \cot (\theta )\hat{\beta}_{0, \theta} (\hat{U}_{0})^2-6 e^{2 \hat{\gamma}_{0}} \cot (\theta ) \hat{\gamma}_{0, \theta} (\hat{U}_{0})^2-\\
&6 e^{2\hat{\beta}_{0}+2 \hat{\gamma}_{0}} \cot (\theta ) \hat{\gamma}_{0, \theta} (\hat{U}_{0})^2-8 e^{2 \hat{\gamma}_{0}}\hat{\beta}_{0, \theta \theta} (\hat{U}_{0})^2-\\
&18 e^{2\hat{\beta}_{0}+2 \hat{\gamma}_{0}} \cot (\theta ) \hat{U}_{0, \theta} \hat{U}_{0}-12 e^{2 \hat{\gamma}_{0}} \hat{U}_{0, \theta}\hat{\beta}_{0, \theta} \hat{U}_{0}+12 e^{2\hat{\beta}_{0}+2 \hat{\gamma}_{0}} \hat{U}_{0, \theta}\hat{\beta}_{0, \theta} \hat{U}_{0}+\\
&6 e^{2 \hat{\gamma}_{0}} \hat{U}_{0, \theta} \hat{\gamma}_{0, \theta} \hat{U}_{0}+6 e^{2\hat{\beta}_{0}+2 \hat{\gamma}_{0}} \hat{U}_{0, \theta} \hat{\gamma}_{0, \theta} \hat{U}_{0}+2 e^{2 \hat{\gamma}_{0}} \hat{U}_{0, \theta \theta} \hat{U}_{0}-\\
&6 e^{2\hat{\beta}_{0}+2 \hat{\gamma}_{0}} \hat{U}_{0, \theta \theta} \hat{U}_{0}-4 e^{2 \hat{\gamma}_{0}} \cot (\theta )\hat{\beta}_{0, t} \hat{U}_{0}+12 e^{2\hat{\beta}_{0}+2 \hat{\gamma}_{0}} \cot (\theta )\hat{\beta}_{0, t} \hat{U}_{0}+\\
&64 e^{2 \hat{\gamma}_{0}}\hat{\beta}_{0, \theta}\hat{\beta}_{0, t} \hat{U}_{0}-6 e^{2 \hat{\gamma}_{0}} \cot (\theta ) \hat{\gamma}_{0, t} \hat{U}_{0}-6 e^{2\hat{\beta}_{0}+2 \hat{\gamma}_{0}} \cot (\theta ) \hat{\gamma}_{0, t} \hat{U}_{0}+\\
&12 e^{2 \hat{\gamma}_{0}} \hat{\gamma}_{0, \theta} \hat{\gamma}_{0, t} \hat{U}_{0}+12 e^{2\hat{\beta}_{0}+2 \hat{\gamma}_{0}} \hat{\gamma}_{0, \theta} \hat{\gamma}_{0, t} \hat{U}_{0}-16 e^{2 \hat{\gamma}_{0}}\hat{\beta}_{0, t \theta} \hat{U}_{0}-2 e^{4\hat{\beta}_{0}}-\\
&6 e^{6\hat{\beta}_{0}}+8 e^{6\hat{\beta}_{0}+2 \hat{\gamma}_{0}}+2 e^{2 \hat{\gamma}_{0}} (\hat{U}_{0, \theta})^2-3 e^{2\hat{\beta}_{0}+2 \hat{\gamma}_{0}} (\hat{U}_{0, \theta})^2-24 e^{4\hat{\beta}_{0}} (\hat{\beta}_{0, \theta})^2-\\
&24 e^{6\hat{\beta}_{0}} (\hat{\beta}_{0, \theta})^2+4 e^{4\hat{\beta}_{0}} (\hat{\gamma}_{0, \theta})^2+12 e^{6\hat{\beta}_{0}} (\hat{\gamma}_{0, \theta})^2+32 e^{2 \hat{\gamma}_{0}} (\hat{\beta}_{0, t})^2+\\
&6 e^{2 \hat{\gamma}_{0}} (\hat{\gamma}_{0, t})^2+6 e^{2\hat{\beta}_{0}+2 \hat{\gamma}_{0}} (\hat{\gamma}_{0, t})^2-4 e^{4\hat{\beta}_{0}} \cot (\theta )\hat{\beta}_{0, \theta}-12 e^{6\hat{\beta}_{0}} \cot (\theta )\hat{\beta}_{0, \theta}-\\
&6 e^{4\hat{\beta}_{0}} \cot (\theta ) \hat{\gamma}_{0, \theta}-18 e^{6\hat{\beta}_{0}} \cot (\theta ) \hat{\gamma}_{0, \theta}+8 e^{4\hat{\beta}_{0}}\hat{\beta}_{0, \theta} \hat{\gamma}_{0, \theta}+24 e^{6\hat{\beta}_{0}}\hat{\beta}_{0, \theta} \hat{\gamma}_{0, \theta}-\\
&4 e^{4\hat{\beta}_{0}}\hat{\beta}_{0, \theta \theta}-12 e^{6\hat{\beta}_{0}}\hat{\beta}_{0, \theta \theta}-2 e^{4\hat{\beta}_{0}} \hat{\gamma}_{0, \theta \theta}-6 e^{6\hat{\beta}_{0}} \hat{\gamma}_{0, \theta \theta}+2 e^{2 \hat{\gamma}_{0}} \cot (\theta ) \hat{U}_{0, t}-\\
&6 e^{2\hat{\beta}_{0}+2 \hat{\gamma}_{0}} \cot (\theta ) \hat{U}_{0, t}-8 e^{2 \hat{\gamma}_{0}}\hat{\beta}_{0, \theta} \hat{U}_{0, t}-4 e^{2 \hat{\gamma}_{0}} \hat{U}_{0, \theta}\hat{\beta}_{0, t}+\\
&12 e^{2\hat{\beta}_{0}+2 \hat{\gamma}_{0}} \hat{U}_{0, \theta}\hat{\beta}_{0, t}+6 e^{2 \hat{\gamma}_{0}} \hat{U}_{0, \theta} \hat{\gamma}_{0, t}+6 e^{2\hat{\beta}_{0}+2 \hat{\gamma}_{0}} \hat{U}_{0, \theta} \hat{\gamma}_{0, t}+2 e^{2 \hat{\gamma}_{0}} \hat{U}_{0, t \theta}-\\
&6 e^{2\hat{\beta}_{0}+2 \hat{\gamma}_{0}} \hat{U}_{0, t \theta}-8 e^{2 \hat{\gamma}_{0}}\hat{\beta}_{0, t t})
\end{split}
\end{align}
\begin{align}
\begin{split}
8c_3(t,\theta)=\,&-\frac{2}{3} e^{-6\hat{\beta}_{0}-2 \hat{\gamma}_{0}} (e^{2 \hat{\gamma}_{0}} (\hat{U}_{0})^3+2 e^{2 \hat{\gamma}_{0}} \cot ^2(\theta ) (\hat{U}_{0})^3-2 e^{2 \hat{\gamma}_{0}} \csc ^2(\theta ) (\hat{U}_{0})^3+\\
&16 e^{2 \hat{\gamma}_{0}} (\hat{\beta}_{0, \theta})^2 (\hat{U}_{0})^3+3 e^{2 \hat{\gamma}_{0}} (\hat{\gamma}_{0, \theta})^2 (\hat{U}_{0})^3-2 e^{2 \hat{\gamma}_{0}} \cot (\theta )\hat{\beta}_{0, \theta} (\hat{U}_{0})^3-\\
&3 e^{2 \hat{\gamma}_{0}} \cot (\theta ) \hat{\gamma}_{0, \theta} (\hat{U}_{0})^3-4 e^{2 \hat{\gamma}_{0}}\hat{\beta}_{0, \theta \theta} (\hat{U}_{0})^3-18 e^{2 \hat{\gamma}_{0}} \hat{U}_{0, \theta}\hat{\beta}_{0, \theta} (\hat{U}_{0})^2+\\
&3 e^{2 \hat{\gamma}_{0}} \hat{U}_{0, \theta} \hat{\gamma}_{0, \theta} (\hat{U}_{0})^2+3 e^{2 \hat{\gamma}_{0}} \hat{U}_{0, \theta \theta} (\hat{U}_{0})^2-2 e^{2 \hat{\gamma}_{0}} \cot (\theta )\hat{\beta}_{0, t} (\hat{U}_{0})^2+\\
&32 e^{2 \hat{\gamma}_{0}}\hat{\beta}_{0, \theta}\hat{\beta}_{0, t} (\hat{U}_{0})^2-3 e^{2 \hat{\gamma}_{0}} \cot (\theta ) \hat{\gamma}_{0, t} (\hat{U}_{0})^2+6 e^{2 \hat{\gamma}_{0}} \hat{\gamma}_{0, \theta} \hat{\gamma}_{0, t} (\hat{U}_{0})^2-\\
&8 e^{2 \hat{\gamma}_{0}}\hat{\beta}_{0, t \theta} (\hat{U}_{0})^2-4 e^{4\hat{\beta}_{0}} \cot ^2(\theta ) \hat{U}_{0}+4 e^{4\hat{\beta}_{0}} \csc ^2(\theta ) \hat{U}_{0}+3 e^{2 \hat{\gamma}_{0}} (\hat{U}_{0, \theta})^2 \hat{U}_{0}-\\
&12 e^{4\hat{\beta}_{0}} (\hat{\beta}_{0, \theta})^2 \hat{U}_{0}-6 e^{4\hat{\beta}_{0}} (\hat{\gamma}_{0, \theta})^2 \hat{U}_{0}+16 e^{2 \hat{\gamma}_{0}} (\hat{\beta}_{0, t})^2 \hat{U}_{0}+3 e^{2 \hat{\gamma}_{0}} (\hat{\gamma}_{0, t})^2 \hat{U}_{0}-\\
&e^{4\hat{\beta}_{0}} \hat{U}_{0}+6 e^{4\hat{\beta}_{0}} \cot (\theta )\hat{\beta}_{0, \theta} \hat{U}_{0}+9 e^{4\hat{\beta}_{0}} \cot (\theta ) \hat{\gamma}_{0, \theta} \hat{U}_{0}-12 e^{4\hat{\beta}_{0}}\hat{\beta}_{0, \theta} \hat{\gamma}_{0, \theta} \hat{U}_{0}+\\
&6 e^{4\hat{\beta}_{0}}\hat{\beta}_{0, \theta \theta} \hat{U}_{0}+3 e^{4\hat{\beta}_{0}} \hat{\gamma}_{0, \theta \theta} \hat{U}_{0}+e^{2 \hat{\gamma}_{0}} \cot (\theta ) \hat{U}_{0, t} \hat{U}_{0}-16 e^{2 \hat{\gamma}_{0}}\hat{\beta}_{0, \theta} \hat{U}_{0, t} \hat{U}_{0}-\\
&14 e^{2 \hat{\gamma}_{0}} \hat{U}_{0, \theta}\hat{\beta}_{0, t} \hat{U}_{0}+3 e^{2 \hat{\gamma}_{0}} \hat{U}_{0, \theta} \hat{\gamma}_{0, t} \hat{U}_{0}+5 e^{2 \hat{\gamma}_{0}} \hat{U}_{0, t \theta} \hat{U}_{0}-4 e^{2 \hat{\gamma}_{0}}\hat{\beta}_{0, t t} \hat{U}_{0}+\\
&12 e^{4\hat{\beta}_{0}} \hat{U}_{0, \theta}\hat{\beta}_{0, \theta}+2 e^{2 \hat{\gamma}_{0}} \hat{U}_{0, \theta} \hat{U}_{0, t}-12 e^{2 \hat{\gamma}_{0}} \hat{U}_{0, t}\hat{\beta}_{0, t}+8 e^{4\hat{\beta}_{0}} \cot (\theta ) \hat{\gamma}_{0, t}-\\
&16 e^{4\hat{\beta}_{0}}\hat{\beta}_{0, \theta} \hat{\gamma}_{0, t}-8 e^{4\hat{\beta}_{0}} \hat{\gamma}_{0, \theta} \hat{\gamma}_{0, t}+8 e^{4\hat{\beta}_{0}}\hat{\beta}_{0, t \theta}+4 e^{4\hat{\beta}_{0}} \hat{\gamma}_{0, t \theta}+2 e^{2 \hat{\gamma}_{0}} \hat{U}_{0, t t})
\end{split}
\end{align}
\end{subequations}
Using the $c_i$ coefficients above, we obtain the following components for $g_{(2)}$:
\begin{subequations}
\begin{align}
\begin{split} \label{eq: g_2tt}
g_{(2) tt}=\,&\frac{1}{2} e^{2 \hat{\gamma}_{0}-4 \hat{\beta}_{0}} ((\hat{\gamma}_{0, \theta})^2-3 \cot (\theta ) \hat{\gamma}_{0, \theta}-2 \hat{\beta}_{0, \theta} (\cot (\theta )-2 \hat{\gamma}_{0, \theta})-2 \hat{\gamma}_{0, \theta \theta}-\\
&1) (\hat{U}_{0})^4+\frac{1}{2} e^{2 \hat{\gamma}_{0}-4 \hat{\beta}_{0}} (\hat{U}_{0, \theta} (2 \hat{\beta}_{0, \theta}-3 \hat{\gamma}_{0, \theta})-\hat{U}_{0, \theta \theta}-2 \cot (\theta ) \hat{\beta}_{0, t}+\\
&4 \hat{\gamma}_{0, \theta} \hat{\beta}_{0, t}-3 \cot (\theta ) \hat{\gamma}_{0, t}+4 \hat{\beta}_{0, \theta} \hat{\gamma}_{0, t}+2 \hat{\gamma}_{0, \theta} \hat{\gamma}_{0, t}-4 \hat{\gamma}_{0, t \theta}) (\hat{U}_{0})^3+\\
&\frac{1}{2} e^{-4 \hat{\beta}_{0}} (-e^{2 \hat{\gamma}_{0}} (\hat{U}_{0, \theta})^2+e^{2 \hat{\gamma}_{0}} (2 \hat{\beta}_{0, t}-\hat{\gamma}_{0, t}) \hat{U}_{0, \theta}+2 e^{4 \hat{\beta}_{0}}+4 e^{4 \hat{\beta}_{0}} (\hat{\beta}_{0, \theta})^2-\\
&3 e^{4 \hat{\beta}_{0}} (\hat{\gamma}_{0, \theta})^2+e^{2 \hat{\gamma}_{0}} (\hat{\gamma}_{0, t})^2+6 e^{4 \hat{\beta}_{0}} \cot (\theta ) \hat{\gamma}_{0, \theta}+4 e^{4 \hat{\beta}_{0}} \hat{\beta}_{0, \theta} (\cot (\theta )-2 \hat{\gamma}_{0, \theta})+\\
&2 e^{4 \hat{\beta}_{0}} \hat{\beta}_{0, \theta \theta}+3 e^{4 \hat{\beta}_{0}} \hat{\gamma}_{0, \theta \theta}+e^{2 \hat{\gamma}_{0}} \cot (\theta ) \hat{U}_{0, t}-2 e^{2 \hat{\gamma}_{0}} \hat{\gamma}_{0, \theta} \hat{U}_{0, t}+4 e^{2 \hat{\gamma}_{0}} \hat{\beta}_{0, t} \hat{\gamma}_{0, t}-\\
&e^{2 \hat{\gamma}_{0}} \hat{U}_{0, t \theta}-2 e^{2 \hat{\gamma}_{0}} \hat{\gamma}_{0, t t}) (\hat{U}_{0})^2+\frac{1}{2} (\hat{U}_{0, \theta} (3 \hat{\gamma}_{0, \theta}-2 \hat{\beta}_{0, \theta})+\hat{U}_{0, \theta \theta}-\\
&2 \cot (\theta ) \hat{\beta}_{0, t}+5 \cot (\theta ) \hat{\gamma}_{0, t}-8 \hat{\beta}_{0, \theta} \hat{\gamma}_{0, t}-2 \hat{\gamma}_{0, \theta} \hat{\gamma}_{0, t}+4 \hat{\gamma}_{0, t \theta}) \hat{U}_{0}+\\
&\frac{1}{2} e^{-2 \hat{\gamma}_{0}} (e^{2 \hat{\gamma}_{0}} (\hat{U}_{0, \theta})^2-e^{2 \hat{\gamma}_{0}} (2 \hat{\beta}_{0, t}-3 \hat{\gamma}_{0, t}) \hat{U}_{0, \theta}-e^{4 \hat{\beta}_{0}}-4 e^{4 \hat{\beta}_{0}} (\hat{\beta}_{0, \theta})^2+\\
&2 e^{4 \hat{\beta}_{0}} (\hat{\gamma}_{0, \theta})^2+3 e^{2 \hat{\gamma}_{0}} (\hat{\gamma}_{0, t})^2-3 e^{4 \hat{\beta}_{0}} \cot (\theta ) \hat{\gamma}_{0, \theta}-2 e^{4 \hat{\beta}_{0}} \hat{\beta}_{0, \theta} (\cot (\theta )-2 \hat{\gamma}_{0, \theta})-\\
&2 e^{4 \hat{\beta}_{0}} \hat{\beta}_{0, \theta \theta}-e^{4 \hat{\beta}_{0}} \hat{\gamma}_{0, \theta \theta}+e^{2 \hat{\gamma}_{0}} \cot (\theta ) \hat{U}_{0, t}+e^{2 \hat{\gamma}_{0}} \hat{U}_{0, t \theta})
\end{split}
\end{align}
\begin{align}
\begin{split}
g_{(2) t \theta}=\,&2 \hat{\gamma}_{0, t} (\hat{\beta}_{0, \theta}+\hat{\gamma}_{0, \theta}-\cot (\theta ))-\hat{\gamma}_{0, t \theta}+\frac{1}{2} (\hat{U}_{0})^3 e^{2 \hat{\gamma}_{0}-4 \hat{\beta}_{0}} (2 \hat{\beta}_{0, \theta} (\cot (\theta )-\\
&2 \hat{\gamma}_{0, \theta})-(\hat{\gamma}_{0, \theta})^2+2 \hat{\gamma}_{0, \theta \theta}+3 \cot (\theta ) \hat{\gamma}_{0, \theta}+1)+\frac{1}{2} (\hat{U}_{0})^2 e^{2 \hat{\gamma}_{0}-4 \hat{\beta}_{0}} (-4 \hat{\beta}_{0, t} \hat{\gamma}_{0, \theta}-\\
&4 \hat{\beta}_{0, \theta} \hat{\gamma}_{0, t}+2 \cot (\theta ) \hat{\beta}_{0, t}-2 \hat{\gamma}_{0, \theta} \hat{\gamma}_{0, t}+4 \hat{\gamma}_{0, t \theta}+3 \cot (\theta ) \hat{\gamma}_{0, t}+\hat{U}_{0, \theta} (3 \hat{\gamma}_{0, \theta}-\\
&2 \hat{\beta}_{0, \theta})+\hat{U}_{0, \theta \theta})+\frac{1}{2} e^{-4 \hat{\beta}_{0}} \hat{U}_{0} (2 e^{4 \hat{\beta}_{0}} (\hat{\gamma}_{0, \theta})^2-e^{4 \hat{\beta}_{0}} \hat{\gamma}_{0, \theta \theta}-4 \hat{\beta}_{0, t} e^{2 \hat{\gamma}_{0}} \hat{\gamma}_{0, t}-\\
&3 \cot (\theta ) e^{4 \hat{\beta}_{0}} \hat{\gamma}_{0, \theta}-2 e^{4 \hat{\beta}_{0}} \hat{\beta}_{0, \theta} (\cot (\theta )-2 \hat{\gamma}_{0, \theta})-e^{4 \hat{\beta}_{0}}-4 e^{4 \hat{\beta}_{0}} (\hat{\beta}_{0, \theta})^2-\\
&2 e^{4 \hat{\beta}_{0}} \hat{\beta}_{0, \theta \theta}-e^{2 \hat{\gamma}_{0}} (\hat{\gamma}_{0, t})^2+2 e^{2 \hat{\gamma}_{0}} \hat{\gamma}_{0, t t}-e^{2 \hat{\gamma}_{0}} \hat{U}_{0, \theta} (2 \hat{\beta}_{0, t}-\hat{\gamma}_{0, t})+e^{2 \hat{\gamma}_{0}} (\hat{U}_{0, \theta})^2+\\
&2 e^{2 \hat{\gamma}_{0}} \hat{\gamma}_{0, \theta} \hat{U}_{0, t}+e^{2 \hat{\gamma}_{0}} \hat{U}_{0, t \theta}-\cot (\theta ) e^{2 \hat{\gamma}_{0}} \hat{U}_{0, t})
\end{split}
\end{align} 
\begin{align}
\begin{split}
g_{(2) \theta \theta}=\,&\frac{1}{2} (\hat{U}_{0})^2 e^{2 \hat{\gamma}_{0}-4 \hat{\beta}_{0}} (-2 \hat{\beta}_{0, \theta} (\cot (\theta )-2 \hat{\gamma}_{0, \theta})+(\hat{\gamma}_{0, \theta})^2-2 \hat{\gamma}_{0, \theta \theta}-\\
&3 \cot (\theta ) \hat{\gamma}_{0, \theta}-1)+\frac{1}{2} \hat{U}_{0} e^{2 \hat{\gamma}_{0}-4 \hat{\beta}_{0}} (4 \hat{\beta}_{0, t} \hat{\gamma}_{0, \theta}+4 \hat{\beta}_{0, \theta} \hat{\gamma}_{0, t}-2 \cot (\theta ) \hat{\beta}_{0, t}+\\
&2 \hat{\gamma}_{0, \theta} \hat{\gamma}_{0, t}-4 \hat{\gamma}_{0, t \theta}-3 \cot (\theta ) \hat{\gamma}_{0, t}+\hat{U}_{0, \theta} (2 \hat{\beta}_{0, \theta}-3 \hat{\gamma}_{0, \theta})-\hat{U}_{0, \theta \theta})+\\
&\frac{1}{2} e^{-4 \hat{\beta}_{0}} (2 e^{4 \hat{\beta}_{0}} (\hat{\gamma}_{0, \theta})^2-e^{4 \hat{\beta}_{0}} \hat{\gamma}_{0, \theta \theta}+4 \hat{\beta}_{0, t} e^{2 \hat{\gamma}_{0}} \hat{\gamma}_{0, t}-3 \cot (\theta ) e^{4 \hat{\beta}_{0}} \hat{\gamma}_{0, \theta}-e^{4 \hat{\beta}_{0}}+\\
&4 e^{4 \hat{\beta}_{0}} (\hat{\beta}_{0, \theta})^2+2 e^{4 \hat{\beta}_{0}} \hat{\beta}_{0, \theta \theta}-2 \cot (\theta ) e^{4 \hat{\beta}_{0}} \hat{\beta}_{0, \theta}+e^{2 \hat{\gamma}_{0}} (\hat{\gamma}_{0, t})^2-2 e^{2 \hat{\gamma}_{0}} \hat{\gamma}_{0, t t}+\\
&e^{2 \hat{\gamma}_{0}} \hat{U}_{0, \theta} (2 \hat{\beta}_{0, t}-\hat{\gamma}_{0, t})-e^{2 \hat{\gamma}_{0}} (\hat{U}_{0, \theta})^2-\\
&2 e^{2 \hat{\gamma}_{0}} \hat{\gamma}_{0, \theta} \hat{U}_{0, t}-e^{2 \hat{\gamma}_{0}} \hat{U}_{0, t \theta}+\cot (\theta ) e^{2 \hat{\gamma}_{0}} \hat{U}_{0, t})
\end{split}
\end{align}
\begin{align}
\begin{split} \label{eq: g_2phiphi}
g_{(2) \phi \phi}=\,&\frac{1}{2} \sin (\theta ) (\hat{U}_{0})^2  e^{-2 (2 \hat{\beta}_{0}+\hat{\gamma}_{0})} (2 \hat{\beta}_{0, \theta} (\cos (\theta )-2 \sin (\theta ) \hat{\gamma}_{0, \theta})+\sin (\theta ) (\hat{\gamma}_{0, \theta})^2+\\
&2 \sin (\theta ) \hat{\gamma}_{0, \theta \theta}+\cos (\theta ) \hat{\gamma}_{0, \theta}+\sin (\theta ))-\\
&\frac{1}{2} \sin (\theta ) \hat{U}_{0}  e^{-2 (2 \hat{\beta}_{0}+\hat{\gamma}_{0})} (4 \sin (\theta ) \hat{\beta}_{0, t} \hat{\gamma}_{0, \theta}+4 \sin (\theta ) \hat{\beta}_{0, \theta} \hat{\gamma}_{0, t}-2 \cos (\theta ) \hat{\beta}_{0, t}-\\
&2 \sin (\theta ) \hat{\gamma}_{0, \theta} \hat{\gamma}_{0, t}-4 \sin (\theta ) \hat{\gamma}_{0, t \theta}-\cos (\theta ) \hat{\gamma}_{0, t}+\hat{U}_{0, \theta} (2 \sin (\theta ) \hat{\beta}_{0, \theta}-\\
&5 \sin (\theta ) \hat{\gamma}_{0, \theta}+2 \cos (\theta ))-\sin (\theta ) \hat{U}_{0, \theta \theta})+\\
&\frac{1}{2} \sin (\theta ) e^{-4 (\hat{\beta}_{0}+\hat{\gamma}_{0})} (2 \sin (\theta ) e^{4 \hat{\beta}_{0}} (\hat{\gamma}_{0, \theta})^2-\sin (\theta ) e^{4 \hat{\beta}_{0}} \hat{\gamma}_{0, \theta \theta}-\\
&4 \sin (\theta ) \hat{\beta}_{0, t} e^{2 \hat{\gamma}_{0}} \hat{\gamma}_{0, t}-3 \cos (\theta ) e^{4 \hat{\beta}_{0}} \hat{\gamma}_{0, \theta}-4 \sin (\theta ) e^{4 \hat{\beta}_{0}} (\hat{\beta}_{0, \theta})^2-\sin (\theta ) e^{4 \hat{\beta}_{0}}-\\
&2 \sin (\theta ) e^{4 \hat{\beta}_{0}} \hat{\beta}_{0, \theta \theta}+2 \cos (\theta ) e^{4 \hat{\beta}_{0}} \hat{\beta}_{0, \theta}+\sin (\theta ) e^{2 \hat{\gamma}_{0}} (\hat{\gamma}_{0, t})^2+\\
&2 \sin (\theta ) e^{2 \hat{\gamma}_{0}} \hat{\gamma}_{0, t t}-\sin (\theta ) e^{2 \hat{\gamma}_{0}} \hat{U}_{0, \theta} (2 \hat{\beta}_{0, t}-3 \hat{\gamma}_{0, t})+\sin (\theta ) e^{2 \hat{\gamma}_{0}} (\hat{U}_{0, \theta})^2+\\
&2 \sin (\theta ) e^{2 \hat{\gamma}_{0}} \hat{\gamma}_{0, \theta} \hat{U}_{0, t}+\sin (\theta ) e^{2 \hat{\gamma}_{0}} \hat{U}_{0, t \theta}-\cos (\theta ) e^{2 \hat{\gamma}_{0}} \hat{U}_{0, t}).
\end{split}
\end{align}
\end{subequations}
Given these $g_{(2)}$ coefficients, we can use formula  (\ref{eq: AdS_g_2_check}) as a consistency check, using the non-zero coefficients of the Ricci tensor, $R_{(0) ab}$, and the Ricci scalar, $R_{(0)}$, of the boundary metric, given below. \\

\noindent \textbf{Ricci Tensor}

\begin{subequations}
\begin{align}
\begin{split}
R_{(0)tt}=\, &e^{-2 (2 \hat{\beta}_{0}+\hat{\gamma}_{0})} (e^{4 \hat{\gamma}_{0}} (\hat{U}_{0})^4 (\hat{\gamma}_{0, \theta} (\cot (\theta )-2 \hat{\beta}_{0, \theta})+\hat{\gamma}_{0, \theta \theta})+\\
&e^{4 \hat{\gamma}_{0}} (\hat{U}_{0})^3 (-2 \hat{\beta}_{0, t} \hat{\gamma}_{0, \theta}-2 \hat{\beta}_{0, \theta} \hat{\gamma}_{0, t}+2 \hat{\gamma}_{0, t \theta}+\cot (\theta ) \hat{\gamma}_{0, t}+\hat{U}_{0, \theta} (-2 \hat{\beta}_{0, \theta}+\\
&2 \hat{\gamma}_{0, \theta}+\cot (\theta ))+\hat{U}_{0, \theta \theta})+e^{2 \hat{\gamma}_{0}} (\hat{U}_{0})^2 (-e^{4 \hat{\beta}_{0}} \hat{\gamma}_{0, \theta \theta}-2 \hat{\beta}_{0, t} e^{2 \hat{\gamma}_{0}} \hat{\gamma}_{0, t}-\\
&\cot (\theta ) e^{4 \hat{\beta}_{0}} \hat{\gamma}_{0, \theta}-2 e^{4 \hat{\beta}_{0}} \hat{\beta}_{0, \theta} (\cot (\theta )-3 \hat{\gamma}_{0, \theta})-4 e^{4 \hat{\beta}_{0}} (\hat{\beta}_{0, \theta})^2-2 e^{4 \hat{\beta}_{0}} \hat{\beta}_{0, \theta \theta}+\\
&e^{2 \hat{\gamma}_{0}} \hat{\gamma}_{0, t t}-e^{2 \hat{\gamma}_{0}} \hat{U}_{0, \theta} (2 \hat{\beta}_{0, t}-\hat{\gamma}_{0, t})+e^{2 \hat{\gamma}_{0}} (\hat{U}_{0, \theta})^2+e^{2 \hat{\gamma}_{0}} \hat{\gamma}_{0, \theta} \hat{U}_{0, t}+e^{2 \hat{\gamma}_{0}} \hat{U}_{0, t \theta})-\\
&\hat{U}_{0} e^{4 \hat{\beta}_{0}+2 \hat{\gamma}_{0}} (2 (\hat{\gamma}_{0, t} (\cot (\theta )-2 \hat{\beta}_{0, \theta})-\cot (\theta ) \hat{\beta}_{0, t}+\hat{\gamma}_{0, t \theta})+\\
&\hat{U}_{0, \theta} (-2 \hat{\beta}_{0, \theta}+2 \hat{\gamma}_{0, \theta}+\cot (\theta ))+\hat{U}_{0, \theta \theta})-e^{4 \hat{\beta}_{0}} (-2 e^{4 \hat{\beta}_{0}} \hat{\beta}_{0, \theta} (\cot (\theta )-\\
&2 \hat{\gamma}_{0, \theta})-4 e^{4 \hat{\beta}_{0}} (\hat{\beta}_{0, \theta})^2-2 e^{4 \hat{\beta}_{0}} \hat{\beta}_{0, \theta \theta}+2 e^{2 \hat{\gamma}_{0}} (\hat{\gamma}_{0, t})^2-2 e^{2 \hat{\gamma}_{0}} \hat{U}_{0, \theta} (\hat{\beta}_{0, t}-\hat{\gamma}_{0, t})+\\
&e^{2 \hat{\gamma}_{0}} (\hat{U}_{0, \theta})^2+e^{2 \hat{\gamma}_{0}} \hat{U}_{0, t \theta}+\cot (\theta ) e^{2 \hat{\gamma}_{0}} \hat{U}_{0, t}))
\end{split}
\end{align}
\begin{align}
\begin{split}
R_{(0)t \theta}=R_{(0)\theta t}=\,&e^{-4 \hat{\beta}_{0}} (e^{4 \hat{\beta}_{0}} (2 \hat{\gamma}_{0, t} (-\hat{\beta}_{0, \theta}-\hat{\gamma}_{0, \theta}+\cot (\theta ))+\hat{\gamma}_{0, t \theta})-\\
&e^{2 \hat{\gamma}_{0}} (\hat{U}_{0})^3 (\hat{\gamma}_{0, \theta} (\cot (\theta )-2 \hat{\beta}_{0, \theta})+\hat{\gamma}_{0, \theta \theta})-\\
&e^{2 \hat{\gamma}_{0}} (\hat{U}_{0})^2 (-2 \hat{\beta}_{0, t} \hat{\gamma}_{0, \theta}-2 \hat{\beta}_{0, \theta} \hat{\gamma}_{0, t}+2 \hat{\gamma}_{0, t \theta}+\cot (\theta ) \hat{\gamma}_{0, t}+\\
&\hat{U}_{0, \theta} (-2 \hat{\beta}_{0, \theta}+2 \hat{\gamma}_{0, \theta}+\cot (\theta ))+\hat{U}_{0, \theta \theta})-\hat{U}_{0} (-2 \hat{\beta}_{0, t} e^{2 \hat{\gamma}_{0}} \hat{\gamma}_{0, t}-\\
&2 e^{4 \hat{\beta}_{0}} \hat{\beta}_{0, \theta} (\cot (\theta )-2 \hat{\gamma}_{0, \theta})-4 e^{4 \hat{\beta}_{0}} (\hat{\beta}_{0, \theta})^2-2 e^{4 \hat{\beta}_{0}} \hat{\beta}_{0, \theta \theta}+\\
&e^{2 \hat{\gamma}_{0}} \hat{\gamma}_{0, t t}-e^{2 \hat{\gamma}_{0}} \hat{U}_{0, \theta} (2 \hat{\beta}_{0, t}-\hat{\gamma}_{0, t})+e^{2 \hat{\gamma}_{0}} (\hat{U}_{0, \theta})^2+\\
&e^{2 \hat{\gamma}_{0}} \hat{\gamma}_{0, \theta} \hat{U}_{0, t}+e^{2 \hat{\gamma}_{0}} \hat{U}_{0, t \theta}))
\end{split}
\end{align}
\begin{align}
\begin{split}
R_{(0) \theta \theta}=\,&e^{-4 \hat{\beta}_{0}} (-2 e^{4 \hat{\beta}_{0}} (\hat{\gamma}_{0, \theta})^2+2 e^{4 \hat{\beta}_{0}} \hat{\beta}_{0, \theta} \hat{\gamma}_{0, \theta}+e^{4 \hat{\beta}_{0}} \hat{\gamma}_{0, \theta \theta}-2 \hat{\beta}_{0, t} e^{2 \hat{\gamma}_{0}} \hat{\gamma}_{0, t}+\\
&3 \cot (\theta ) e^{4 \hat{\beta}_{0}} \hat{\gamma}_{0, \theta}+e^{4 \hat{\beta}_{0}}-4 e^{4 \hat{\beta}_{0}} (\hat{\beta}_{0, \theta})^2-2 e^{4 \hat{\beta}_{0}} \hat{\beta}_{0, \theta \theta}+e^{2 \hat{\gamma}_{0}} \hat{\gamma}_{0, t t}-\\
&e^{2 \hat{\gamma}_{0}} \hat{U}_{0, \theta} (2 \hat{\beta}_{0, t}-\hat{\gamma}_{0, t})+e^{2 \hat{\gamma}_{0}} (\hat{U}_{0})^2 (\hat{\gamma}_{0, \theta} (\cot (\theta )-2 \hat{\beta}_{0, \theta})+\hat{\gamma}_{0, \theta \theta})+\\
&e^{2 \hat{\gamma}_{0}} \hat{U}_{0} (-2 \hat{\beta}_{0, t} \hat{\gamma}_{0, \theta}-2 \hat{\beta}_{0, \theta} \hat{\gamma}_{0, t}+2 \hat{\gamma}_{0, t \theta}+\cot (\theta ) \hat{\gamma}_{0, t}+\\
&\hat{U}_{0, \theta} (-2 \hat{\beta}_{0, \theta}+2 \hat{\gamma}_{0, \theta}+\cot (\theta ))+\hat{U}_{0, \theta \theta})+\\
&e^{2 \hat{\gamma}_{0}} (\hat{U}_{0, \theta})^2+e^{2 \hat{\gamma}_{0}} \hat{\gamma}_{0, \theta} \hat{U}_{0, t}+e^{2 \hat{\gamma}_{0}} \hat{U}_{0, t \theta})
\end{split}
\end{align}
\begin{align}
\begin{split}
R_{(0) \phi \phi}=\,&\sin (\theta ) e^{-4 (\hat{\beta}_{0}+\hat{\gamma}_{0})} (-2 \sin (\theta ) e^{4 \hat{\beta}_{0}} (\hat{\gamma}_{0, \theta})^2+\sin (\theta ) e^{4 \hat{\beta}_{0}} \hat{\gamma}_{0, \theta \theta}+\\
&2 \sin (\theta ) \hat{\beta}_{0, t} e^{2 \hat{\gamma}_{0}} \hat{\gamma}_{0, t}+3 \cos (\theta ) e^{4 \hat{\beta}_{0}} \hat{\gamma}_{0, \theta}+2 e^{4 \hat{\beta}_{0}} \hat{\beta}_{0, \theta} (\sin (\theta ) \hat{\gamma}_{0, \theta}-\cos (\theta ))+\\
&\sin (\theta ) e^{4 \hat{\beta}_{0}}-\sin (\theta ) e^{2 \hat{\gamma}_{0}} \hat{\gamma}_{0, t t}+e^{2 \hat{\gamma}_{0}} (\hat{U}_{0})^2 (2 \hat{\beta}_{0, \theta} (\sin (\theta ) \hat{\gamma}_{0, \theta}-\cos (\theta ))-\\
&\sin (\theta ) (\hat{\gamma}_{0, \theta \theta}+1)-\cos (\theta ) \hat{\gamma}_{0, \theta})-e^{2 \hat{\gamma}_{0}} \hat{U}_{0} (-2 \sin (\theta ) \hat{\beta}_{0, \theta} \hat{\gamma}_{0, t}+\\
&2 \hat{\beta}_{0, t} (\cos (\theta )-\sin (\theta ) \hat{\gamma}_{0, \theta})+2 \sin (\theta ) \hat{\gamma}_{0, t \theta}+\cos (\theta ) \hat{\gamma}_{0, t}+\\
&2 \hat{U}_{0, \theta} (\sin (\theta ) \hat{\gamma}_{0, \theta}-\cos (\theta )))-\sin (\theta ) e^{2 \hat{\gamma}_{0}} \hat{\gamma}_{0, \theta} \hat{U}_{0, t}-\\
&\sin (\theta ) e^{2 \hat{\gamma}_{0}} \hat{\gamma}_{0, t} \hat{U}_{0, \theta}+\cos (\theta ) e^{2 \hat{\gamma}_{0}} \hat{U}_{0, t})
\end{split}
\end{align}
\end{subequations}

\noindent \textbf{Ricci Scalar}

\begin{align}
\begin{split}
R_{(0)}=\,&2 e^{ -2 (2 \hat{\beta}_{0}+\hat{\gamma}_{0})} (-2 e^{4 \hat{\beta}_{0}} (\hat{\gamma}_{0, \theta})^2+4 e^{4 \hat{\beta}_{0}} \hat{\beta}_{0, \theta} \hat{\gamma}_{0, \theta}+e^{4 \hat{\beta}_{0}} \hat{\gamma}_{0, \theta \theta}+\\
&3 \cot (\theta ) e^{4 \hat{\beta}_{0}} \hat{\gamma}_{0, \theta}+e^{4 \hat{\beta}_{0}}-4 e^{4 \hat{\beta}_{0}} (\hat{\beta}_{0, \theta})^2-2 e^{4 \hat{\beta}_{0}} \hat{\beta}_{0, \theta \theta}-2 \cot (\theta ) e^{4 \hat{\beta}_{0}} \hat{\beta}_{0, \theta}+\\
&e^{2 \hat{\gamma}_{0}} (\hat{\gamma}_{0, t})^2-e^{2 \hat{\gamma}_{0}} \hat{U}_{0, \theta} (2 \hat{\beta}_{0, t}-\hat{\gamma}_{0, t})-e^{2 \hat{\gamma}_{0}} (\hat{U}_{0})^2 (2 \cot (\theta ) \hat{\beta}_{0, \theta}-(\hat{\gamma}_{0, \theta})^2+\\
&\cot (\theta ) \hat{\gamma}_{0, \theta}+1)+e^{2 \hat{\gamma}_{0}} \hat{U}_{0} (-2 \cot (\theta ) \hat{\beta}_{0, t}+2 \hat{\gamma}_{0, \theta} \hat{\gamma}_{0, t}-\cot (\theta ) \hat{\gamma}_{0, t}+\\
&\hat{U}_{0, \theta} (-2 \hat{\beta}_{0, \theta}+\hat{\gamma}_{0, \theta}+2 \cot (\theta ))+\hat{U}_{0, \theta \theta})+\\
&e^{2 \hat{\gamma}_{0}} (\hat{U}_{0, \theta})^2+e^{2 \hat{\gamma}_{0}} \hat{U}_{0, t \theta}+\cot (\theta ) e^{2 \hat{\gamma}_{0}} \hat{U}_{0, t}).
\end{split}
\end{align}

\subsection{Explicit expressions for $g_{(3)}$} \label{g3-expressions}

Finally, we want to obtain $g_{(3)}$. To do this we extend our transformation in the coordinates to $\mathcal{O}(\rho^4)$
\begin{align}
\begin{split}
&r_* \rightarrow \rho  + b_1(t,\theta) \rho^2+c_1(t,\theta) \rho^3+d_1(t,\theta) \rho^4 \\
&t \rightarrow t + \alpha_1(t,\theta)\rho + b_2(t,\theta) \rho^2+c_2(t,\theta) \rho^3+d_2(t,\theta) \rho^4  \\
&\theta \rightarrow \theta + \alpha_2(t,\theta)\rho + b_3(t,\theta) \rho^2+c_3(t,\theta) \rho^3+d_3(t,\theta) \rho^4,
\end{split}
\end{align}
where $\alpha_i, b_i, c_i$ are the functions already obtained from previous orders.  As in the previous orders, we obtain $d_{1,2,3}$ by forcing the vanishing of the $d\rho$ terms, now at $\mathcal{O}(1/\rho)$. The expressions for $d_i$ are too long to be reported here but they can be found in Mathematica file included in the arXiv submission of this paper.
Using this transformation we can finally extract $g_{(3)ab}$, which may be manipulated to the form (\ref{g3-rel}) with ${\cal U}_{3}, {\cal W}_{3}, {\cal G}_{3}$ given by
\begin{align}
\begin{split} \label{eq: U_3_EM_tensor}
{\cal U}_{3}=&\frac{1}{12} e^{-2 (\hat{\beta}_{0}+\hat{\gamma}_{0})} [-12 \cot ^3(\theta )+15 \csc ^2(\theta ) \cot (\theta )+72 \hat{\gamma}_{0, \theta}^2 \cot (\theta )-\\
&8 \hat{\beta}_{0, \theta \theta} \cot (\theta )-30 \hat{\gamma}_{0, \theta \theta} \cot (\theta )-24 \cot (\theta )-32 \hat{\gamma}_{0, \theta}^3+\\
&32 \hat{\beta}_{0, \theta}^2 (\cot (\theta )-2 \hat{\gamma}_{0, \theta})+\hat{\gamma}_{0, \theta} (-2 (13 \cos (2 \theta )+2) \csc ^2(\theta )+16 \hat{\beta}_{0, \theta \theta}+\\
&36 \hat{\gamma}_{0, \theta \theta})+\hat{\beta}_{0, \theta} (-7 \cot ^2(\theta )+92 \hat{\gamma}_{0, \theta} \cot (\theta )-60 \hat{\gamma}_{0, \theta}^2+48 \hat{\gamma}_{0, \theta \theta}+24)-\\
&8 \hat{\gamma}_{0, \theta \theta \theta} ] \hat{U}_{0}^2+\frac{1}{12} e^{-2 (\hat{\beta}_{0}+\hat{\gamma}_{0})} [-16 \hat{\beta}_{0, t}  \cot ^2(\theta )-16 \hat{\gamma}_{0, t}  \cot ^2(\theta )-9 \hat{U}_{0, \theta \theta} \cot (\theta )+\\
&32 \hat{\beta}_{0, \theta} \hat{\beta}_{0, t}  \cot (\theta )+48 \hat{\gamma}_{0, \theta} \hat{\beta}_{0, t}  \cot (\theta )+76 \hat{\beta}_{0, \theta} \hat{\gamma}_{0, t}  \cot (\theta )+104 \hat{\gamma}_{0, \theta} \hat{\gamma}_{0, t}  \cot (\theta )-\\
&8 \hat{\beta}_{0, t \theta}  \cot (\theta )-46 \hat{\gamma}_{0, t \theta}  \cot (\theta )+24 \hat{\beta}_{0, \theta} \hat{U}_{0, \theta \theta}+6 \hat{\gamma}_{0, \theta} \hat{U}_{0, \theta \theta}-\\
&\hat{U}_{0, \theta} (\cot ^2(\theta )+12 \hat{\gamma}_{0, \theta} \cot (\theta )+5 \csc ^2(\theta )+32 \hat{\beta}_{0, \theta}^2-12 \hat{\gamma}_{0, \theta}^2+\\
&2 \hat{\beta}_{0, \theta} (\cot (\theta )-18 \hat{\gamma}_{0, \theta})-8 \hat{\beta}_{0, \theta \theta}+18 \hat{\gamma}_{0, \theta \theta}+2)-4 \hat{U}_{0, \theta \theta \theta} +8 \csc ^2(\theta ) \hat{\beta}_{0, t} -\\
&32 \hat{\gamma}_{0, \theta}^2 \hat{\beta}_{0, t} -64 \hat{\beta}_{0, \theta} \hat{\gamma}_{0, \theta} \hat{\beta}_{0, t} +16 \hat{\gamma}_{0, \theta \theta} \hat{\beta}_{0, t} -10 \csc ^2(\theta ) \hat{\gamma}_{0, t} -\\
&64 \hat{\beta}_{0, \theta}^2 \hat{\gamma}_{0, t} -64 \hat{\gamma}_{0, \theta}^2 \hat{\gamma}_{0, t} -88 \hat{\beta}_{0, \theta} \hat{\gamma}_{0, \theta} \hat{\gamma}_{0, t} +16 \hat{\beta}_{0, \theta \theta} \hat{\gamma}_{0, t} +20 \hat{\gamma}_{0, \theta \theta} \hat{\gamma}_{0, t} +24 \hat{\gamma}_{0, t} +\\
&16 \hat{\gamma}_{0, \theta} \hat{\beta}_{0, t \theta} +80 \hat{\beta}_{0, \theta} \hat{\gamma}_{0, t \theta} +52 \hat{\gamma}_{0, \theta} \hat{\gamma}_{0, t \theta} -16 \hat{\gamma}_{0, t \theta \theta}] \hat{U}_{0}+\\
&\frac{1}{12} e^{-2 (\hat{\beta}_{0}+2 \hat{\gamma}_{0})} [3 e^{2 \hat{\gamma}_{0}} (\cot (\theta )+3 \hat{\beta}_{0, \theta}-2 \hat{\gamma}_{0, \theta}) \hat{U}_{0, \theta}^2-\\
&e^{2 \hat{\gamma}_{0}} (3 \hat{U}_{0, \theta \theta}-2 (4 (\cot (\theta )-4 \hat{\beta}_{0, \theta}) \hat{\beta}_{0, t} +(5 \cot (\theta )+2 \hat{\beta}_{0, \theta}-2 \hat{\gamma}_{0, \theta}) \hat{\gamma}_{0, t} +\\
&4 \hat{\beta}_{0, t \theta} -5 \hat{\gamma}_{0, t \theta} )) \hat{U}_{0, \theta}+2 (8 e^{4 \hat{\beta}_{0}} \hat{\gamma}_{0, \theta}^3+\\
&(8 e^{2 \hat{\gamma}_{0}} \hat{U}_{0, t} -12 e^{4 \hat{\beta}_{0}} \cot (\theta )) \hat{\gamma}_{0, \theta}^2-2 (3 e^{4 \hat{\beta}_{0}} \csc ^2(\theta )+2 e^{4 \hat{\beta}_{0}}+8 e^{2 \hat{\gamma}_{0}} \hat{\gamma}_{0, t} ^2+\\
&8 e^{4 \hat{\beta}_{0}} \hat{\beta}_{0, \theta \theta}+6 e^{4 \hat{\beta}_{0}} \hat{\gamma}_{0, \theta \theta}+6 e^{2 \hat{\gamma}_{0}} \cot (\theta ) \hat{U}_{0, t} +8 e^{2 \hat{\gamma}_{0}} \hat{\beta}_{0, t}  \hat{\gamma}_{0, t} -4 e^{2 \hat{\gamma}_{0}} \hat{\gamma}_{0, tt}) \hat{\gamma}_{0, \theta}+\\
&16 e^{2 \hat{\gamma}_{0}} \cot (\theta ) \hat{\gamma}_{0, t} ^2+16 e^{4 \hat{\beta}_{0}} \hat{\beta}_{0, \theta}^2 (\cot (\theta )-2 \hat{\gamma}_{0, \theta})+4 e^{4 \hat{\beta}_{0}} \cot (\theta ) \hat{\beta}_{0, \theta \theta}+\\
&6 e^{4 \hat{\beta}_{0}} \cot (\theta ) \hat{\gamma}_{0, \theta \theta}+4 e^{4 \hat{\beta}_{0}} \hat{\beta}_{0, \theta \theta \theta} +2 e^{4 \hat{\beta}_{0}} \hat{\gamma}_{0, \theta \theta \theta} +4 e^{2 \hat{\gamma}_{0}} \cot ^2(\theta ) \hat{U}_{0, t} -\\
&2 e^{2 \hat{\gamma}_{0}} \csc ^2(\theta ) \hat{U}_{0, t} -4 e^{2 \hat{\gamma}_{0}} \hat{\gamma}_{0, \theta \theta} \hat{U}_{0, t} +4 e^{2 \hat{\gamma}_{0}} \hat{U}_{0, \theta \theta} \hat{\beta}_{0, t} +3 e^{2 \hat{\gamma}_{0}} \hat{U}_{0, \theta \theta} \hat{\gamma}_{0, t} +\\
&16 e^{2 \hat{\gamma}_{0}} \cot (\theta ) \hat{\beta}_{0, t}  \hat{\gamma}_{0, t} -2 e^{2 \hat{\gamma}_{0}} \cot (\theta ) \hat{U}_{0, t \theta} +8 e^{2 \hat{\gamma}_{0}} \hat{\gamma}_{0, t}  \hat{\beta}_{0, t \theta} +8 e^{2 \hat{\gamma}_{0}} \hat{\beta}_{0, t}  \hat{\gamma}_{0, t \theta} +\\
&10 e^{2 \hat{\gamma}_{0}} \hat{\gamma}_{0, t}  \hat{\gamma}_{0, t \theta} -2 e^{2 \hat{\gamma}_{0}} \hat{U}_{0, t \theta \theta}-8 e^{2 \hat{\gamma}_{0}} \cot (\theta ) \hat{\gamma}_{0, tt}-2 \hat{\beta}_{0, \theta} (2 e^{4 \hat{\beta}_{0}} \csc ^2(\theta )-4 e^{4 \hat{\beta}_{0}}+\\
&7 e^{2 \hat{\gamma}_{0}} \hat{\gamma}_{0, t} ^2-8 e^{4 \hat{\beta}_{0}} \hat{\beta}_{0, \theta \theta}+4 e^{2 \hat{\gamma}_{0}} \cot (\theta ) \hat{U}_{0, t} -8 \hat{\gamma}_{0, \theta} (e^{4 \hat{\beta}_{0}} \cot (\theta )+e^{2 \hat{\gamma}_{0}} \hat{U}_{0, t} )+\\
&16 e^{2 \hat{\gamma}_{0}} \hat{\beta}_{0, t}  \hat{\gamma}_{0, t} -4 e^{2 \hat{\gamma}_{0}} \hat{U}_{0, t \theta} -8 e^{2 \hat{\gamma}_{0}} \hat{\gamma}_{0, tt})-4 e^{2 \hat{\gamma}_{0}} \hat{\gamma}_{0, tt \theta})]
\end{split}
\end{align}
\begin{align}
\begin{split}
{\cal W}_{3}=&-\frac{1}{8} e^{-4 \hat{\beta}_{0}} (\cot (\theta )-2 \hat{\gamma}_{0, \theta}) [-\cot ^2(\theta )+4 \hat{\gamma}_{0, \theta} \cot (\theta )+3 \csc ^2(\theta )+\\
&8 \hat{\beta}_{0, \theta} (\cot (\theta )-2 \hat{\gamma}_{0, \theta})+8 \hat{\gamma}_{0, \theta \theta}+1] \hat{U}_{0}^3 +
\\
& \frac{1}{4} e^{-4 \hat{\beta}_{0}} [
\hat{U}_{0, \theta} (2 \cot ^2(\theta )-8 \hat{\gamma}_{0, \theta} \cot (\theta )+\csc ^2(\theta )+12 \hat{\gamma}_{0, \theta}^2+\\
&8 \hat{\beta}_{0, \theta} (\cot (\theta )-2 \hat{\gamma}_{0, \theta})+4 \hat{\gamma}_{0, \theta \theta}+1)+2 \{2 (-2 \hat{\gamma}_{0, t \theta} (\cot (\theta )-2 \hat{\gamma}_{0, \theta})-\\
&(\cot (\theta )-2 \hat{\gamma}_{0, \theta})^2 \hat{\beta}_{0, t }+(2 \cot (\theta ) \hat{\gamma}_{0, \theta}+4 \hat{\beta}_{0, \theta} (\cot (\theta )-2 \hat{\gamma}_{0, \theta})+\\
&2 \hat{\gamma}_{0, \theta \theta}+1) \hat{\gamma}_{0, t })-(\cot (\theta )-2 \hat{\gamma}_{0, \theta}) \hat{U}_{0, \theta \theta}\}] \hat{U}_{0}^2+\\
&\frac{1}{4} e^{-2 (2 \hat{\beta}_{0}+\hat{\gamma}_{0})} [-e^{2 \hat{\gamma}_{0}} (3 \cot (\theta )+4 \hat{\beta}_{0, \theta}-8 \hat{\gamma}_{0, \theta}) \hat{U}_{0, \theta}^2+\\
&2 e^{2 \hat{\gamma}_{0}} (\hat{U}_{0, \theta \theta}+4 (\cot (\theta )-2 \hat{\gamma}_{0, \theta}) \hat{\beta}_{0, t }-2 \cot (\theta ) \hat{\gamma}_{0, t }-8 \hat{\beta}_{0, \theta} \hat{\gamma}_{0, t }+8 \hat{\gamma}_{0, \theta} \hat{\gamma}_{0, t }+\\
&4 \hat{\gamma}_{0, t \theta}) \hat{U}_{0, \theta}+2 \{8 e^{4 \hat{\beta}_{0}} (\cot (\theta )-2 \hat{\gamma}_{0, \theta}) \hat{\beta}_{0, \theta}^2-\\
&2 (e^{4 \hat{\beta}_{0}} \cot ^2(\theta )-12 e^{4 \hat{\beta}_{0}} \hat{\gamma}_{0, \theta} \cot (\theta )-3 e^{4 \hat{\beta}_{0}}+e^{4 \hat{\beta}_{0}} \csc ^2(\theta )+8 e^{4 \hat{\beta}_{0}} \hat{\gamma}_{0, \theta}^2+\\
&4 e^{2 \hat{\gamma}_{0}} \hat{\gamma}_{0, t }^2-4 e^{4 \hat{\beta}_{0}} \hat{\gamma}_{0, \theta \theta}) \hat{\beta}_{0, \theta}+e^{2 \hat{\gamma}_{0}} (\hat{U}_{0, t } (\cot (\theta )-2 \hat{\gamma}_{0, \theta})^2+2 \cot (\theta ) \hat{\gamma}_{0, t }^2+\\
&2 \hat{U}_{0, \theta \theta} \hat{\gamma}_{0, t }+8 \cot (\theta ) \hat{\beta}_{0, t } \hat{\gamma}_{0, t }-16 \hat{\gamma}_{0, \theta} \hat{\beta}_{0, t } \hat{\gamma}_{0, t }-\cot (\theta ) \hat{U}_{0, t \theta}+2 \hat{\gamma}_{0, \theta} \hat{U}_{0, t \theta}+\\
&8 \hat{\gamma}_{0, t } \hat{\gamma}_{0, t \theta}-2 \cot (\theta ) \hat{\gamma}_{0, t t}+4 \hat{\gamma}_{0, \theta} \hat{\gamma}_{0, t t})\}] \hat{U}_{0}+\frac{1}{4} e^{-2 (2 \hat{\beta}_{0}+\hat{\gamma}_{0})} [e^{2 \hat{\gamma}_{0}} \hat{U}_{0, \theta}^3-\\
&4 e^{2 \hat{\gamma}_{0}} (\hat{\beta}_{0, t }-\hat{\gamma}_{0, t }) \hat{U}_{0, \theta}^2+2 (-8 e^{4 \hat{\beta}_{0}} \hat{\beta}_{0, \theta}^2+4 e^{4 \hat{\beta}_{0}} \cot (\theta ) \hat{\beta}_{0, \theta}+e^{2 \hat{\gamma}_{0}} (2 \hat{\gamma}_{0, t }^2-\\
&8 \hat{\beta}_{0, t } \hat{\gamma}_{0, t }-(\cot (\theta )-2 \hat{\gamma}_{0, \theta}) \hat{U}_{0, t }+\hat{U}_{0, t \theta}+2 \hat{\gamma}_{0, t t})) \hat{U}_{0, \theta} -\\
&4 \{8 e^{4 \hat{\beta}_{0}} \hat{\gamma}_{0, t } \hat{\beta}_{0, \theta}^2-2 e^{4 \hat{\beta}_{0}} (\hat{U}_{0, \theta \theta}+2 (2 (\cot (\theta )-\hat{\gamma}_{0, \theta}) \hat{\gamma}_{0, t }+\hat{\gamma}_{0, t \theta})) \hat{\beta}_{0, \theta}+\\
&e^{2 \hat{\gamma}_{0}} \hat{\gamma}_{0, t } ((\cot (\theta )-2 \hat{\gamma}_{0, \theta}) \hat{U}_{0, t }+4 \hat{\beta}_{0, t } \hat{\gamma}_{0, t }-\hat{U}_{0, t \theta}-2 \hat{\gamma}_{0, t t})\}]
\end{split}
\end{align}
\begin{align}
\begin{split} \label{eq: gamma_3_EM_tensor}
{\cal G}_{3}=&\frac{1}{48} e^{-6 \hat{\beta}_{0}} [-19 \cot ^3(\theta )+18 \csc ^2(\theta ) \cot (\theta )+36 \hat{\gamma}_{0, \theta}^2 \cot (\theta )+16 \hat{\beta}_{0, \theta \theta} \cot (\theta )+\\
&24 \hat{\gamma}_{0, \theta \theta} \cot (\theta )-18 \cot (\theta )-24 \hat{\gamma}_{0, \theta}^3-64 \hat{\beta}_{0, \theta}^2 (\cot (\theta )-2 \hat{\gamma}_{0, \theta})-\\
&2 \hat{\gamma}_{0, \theta} (7 \cot ^2(\theta )+2 \csc ^2(\theta )+16 \hat{\beta}_{0, \theta \theta}+2)-8 \hat{\beta}_{0, \theta} (\cot ^2(\theta )+8 \hat{\gamma}_{0, \theta} \cot (\theta )+\\
&\csc ^2(\theta )+12 \hat{\gamma}_{0, \theta \theta}+5)+16 \hat{\gamma}_{0, \theta \theta \theta} ] \hat{U}_{0}^3+\\
&\frac{1}{48} e^{-6 \hat{\beta}_{0}} [\hat{U}_{0, \theta} (-7 \cot ^2(\theta )+84 \hat{\gamma}_{0, \theta} \cot (\theta )+10 \csc ^2(\theta )+\\
&64 \hat{\beta}_{0, \theta}^2-36 \hat{\gamma}_{0, \theta}^2+64 \hat{\beta}_{0, \theta} (\cot (\theta )-3 \hat{\gamma}_{0, \theta})-16 \hat{\beta}_{0, \theta \theta}+72 \hat{\gamma}_{0, \theta \theta}+22)+\\
&2 (-4 \hat{\beta}_{0, t}  \cot ^2(\theta )-7 \hat{\gamma}_{0, t}  \cot ^2(\theta )+16 \hat{\beta}_{0, t \theta}  \cot (\theta )+24 \hat{\gamma}_{0, t \theta}  \cot (\theta )+4 \hat{U}_{0, \theta \theta \theta} -\\
&4 \csc ^2(\theta ) \hat{\beta}_{0, t} -48 \hat{\gamma}_{0, \theta \theta} \hat{\beta}_{0, t} -20 \hat{\beta}_{0, t} -2 \csc ^2(\theta ) \hat{\gamma}_{0, t} +64 \hat{\beta}_{0, \theta}^2 \hat{\gamma}_{0, t} -\\
&36 \hat{\gamma}_{0, \theta}^2 \hat{\gamma}_{0, t} -16 \hat{\beta}_{0, \theta \theta} \hat{\gamma}_{0, t} -2 \hat{\gamma}_{0, t} +4 \hat{\gamma}_{0, \theta} (3 \hat{U}_{0, \theta \theta}-8 \cot (\theta ) \hat{\beta}_{0, t} +9 \cot (\theta ) \hat{\gamma}_{0, t} -\\
&8 \hat{\beta}_{0, t \theta} )-8 \hat{\beta}_{0, \theta} (3 \hat{U}_{0, \theta \theta}+4 (2 (\cot (\theta )-2 \hat{\gamma}_{0, \theta}) \hat{\beta}_{0, t} +\cot (\theta ) \hat{\gamma}_{0, t} +3 \hat{\gamma}_{0, t \theta} ))+\\
&24 \hat{\gamma}_{0, t \theta \theta})] \hat{U}_{0}^2+\\
&\frac{1}{48} e^{-2 (3 \hat{\beta}_{0}+\hat{\gamma}_{0})} [-3 e^{2 \hat{\gamma}_{0}} (\cot (\theta )+16 \hat{\beta}_{0, \theta}-10 \hat{\gamma}_{0, \theta}) \hat{U}_{0, \theta}^2+4 e^{2 \hat{\gamma}_{0}} (6 \hat{U}_{0, \theta \theta}+\\
&4 (3 \cot (\theta )+8 \hat{\beta}_{0, \theta}-10 \hat{\gamma}_{0, \theta}) \hat{\beta}_{0, t} +15 \cot (\theta ) \hat{\gamma}_{0, t} -24 \hat{\beta}_{0, \theta} \hat{\gamma}_{0, t} -18 \hat{\gamma}_{0, \theta} \hat{\gamma}_{0, t} -\\
&8 \hat{\beta}_{0, t \theta} +24 \hat{\gamma}_{0, t \theta} ) \hat{U}_{0, \theta}-4 (8 e^{4 \hat{\beta}_{0}} \hat{\gamma}_{0, \theta}^3-12 e^{4 \hat{\beta}_{0}} \cot (\theta ) \hat{\gamma}_{0, \theta}^2+\\
&2 (2 e^{4 \hat{\beta}_{0}} \cot ^2(\theta )-4 e^{2 \hat{\gamma}_{0}} \hat{U}_{0, t}  \cot (\theta )-5 e^{4 \hat{\beta}_{0}} \csc ^2(\theta )-16 e^{2 \hat{\gamma}_{0}} \hat{\beta}_{0, t} ^2+\\
&9 e^{2 \hat{\gamma}_{0}} \hat{\gamma}_{0, t} ^2-8 e^{4 \hat{\beta}_{0}} \hat{\beta}_{0, \theta \theta}-6 e^{4 \hat{\beta}_{0}} \hat{\gamma}_{0, \theta \theta}-5 e^{2 \hat{\gamma}_{0}} \hat{U}_{0, t \theta} +4 e^{2 \hat{\gamma}_{0}} \hat{\beta}_{0, tt}) \hat{\gamma}_{0, \theta}+\\
&16 e^{2 \hat{\gamma}_{0}} \cot (\theta ) \hat{\beta}_{0, t} ^2-9 e^{2 \hat{\gamma}_{0}} \cot (\theta ) \hat{\gamma}_{0, t} ^2+16 e^{4 \hat{\beta}_{0}} \hat{\beta}_{0, \theta}^2 (\cot (\theta )-2 \hat{\gamma}_{0, \theta})+\\
&4 e^{4 \hat{\beta}_{0}} \cot (\theta ) \hat{\beta}_{0, \theta \theta}+6 e^{4 \hat{\beta}_{0}} \cot (\theta ) \hat{\gamma}_{0, \theta \theta}+4 e^{4 \hat{\beta}_{0}} \hat{\beta}_{0, \theta \theta \theta} +2 e^{4 \hat{\beta}_{0}} \hat{\gamma}_{0, \theta \theta \theta} -5 e^{2 \hat{\gamma}_{0}} \hat{U}_{0, t} -\\
&e^{2 \hat{\gamma}_{0}} \cot ^2(\theta ) \hat{U}_{0, t} -e^{2 \hat{\gamma}_{0}} \csc ^2(\theta ) \hat{U}_{0, t} -12 e^{2 \hat{\gamma}_{0}} \hat{\gamma}_{0, \theta \theta} \hat{U}_{0, t} +12 e^{2 \hat{\gamma}_{0}} \hat{U}_{0, \theta \theta} \hat{\beta}_{0, t} -\\
&2 e^{2 \hat{\gamma}_{0}} \hat{U}_{0, \theta \theta} \hat{\gamma}_{0, t} +16 e^{2 \hat{\gamma}_{0}} \cot (\theta ) \hat{\beta}_{0, t}  \hat{\gamma}_{0, t} +2 e^{2 \hat{\gamma}_{0}} \cot (\theta ) \hat{U}_{0, t \theta} +16 e^{2 \hat{\gamma}_{0}} \hat{\gamma}_{0, t}  \hat{\beta}_{0, t \theta} +\\
&48 e^{2 \hat{\gamma}_{0}} \hat{\beta}_{0, t}  \hat{\gamma}_{0, t \theta} -4 e^{2 \hat{\gamma}_{0}} \hat{U}_{0, t \theta \theta}-4 e^{2 \hat{\gamma}_{0}} \cot (\theta ) \hat{\beta}_{0, tt}-6 e^{2 \hat{\gamma}_{0}} \cot (\theta ) \hat{\gamma}_{0, tt}-\\
&2 \hat{\beta}_{0, \theta} (7 e^{4 \hat{\beta}_{0}} \cot ^2(\theta )+8 e^{2 \hat{\gamma}_{0}} \hat{U}_{0, t}  \cot (\theta )+3 e^{4 \hat{\beta}_{0}}-5 e^{4 \hat{\beta}_{0}} \csc ^2(\theta )-8 e^{4 \hat{\beta}_{0}} \hat{\beta}_{0, \theta \theta}-\\
&8 \hat{\gamma}_{0, \theta} (e^{4 \hat{\beta}_{0}} \cot (\theta )+2 e^{2 \hat{\gamma}_{0}} \hat{U}_{0, t} )+32 e^{2 \hat{\gamma}_{0}} \hat{\beta}_{0, t}  \hat{\gamma}_{0, t} -6 e^{2 \hat{\gamma}_{0}} \hat{U}_{0, t \theta} -12 e^{2 \hat{\gamma}_{0}} \hat{\gamma}_{0, tt})-\\
&12 e^{2 \hat{\gamma}_{0}} \hat{\gamma}_{0, tt \theta})] \hat{U}_{0}+\\
&\frac{1}{48} e^{-2 (3 \hat{\beta}_{0}+\hat{\gamma}_{0})} [e^{2 \hat{\gamma}_{0}} (\hat{U}_{0, \theta}^3-2 (16 \hat{\beta}_{0, t} +5 \hat{\gamma}_{0, t} ) \hat{U}_{0, \theta}^2-4 (-16 \hat{\beta}_{0, t} ^2+\\
&16 \hat{\gamma}_{0, t}  \hat{\beta}_{0, t} +9 \hat{\gamma}_{0, t} ^2+2 (2 \cot (\theta )+2 \hat{\beta}_{0, \theta}-5 \hat{\gamma}_{0, \theta}) \hat{U}_{0, t} -4 \hat{U}_{0, t \theta} +4 \hat{\beta}_{0, tt}-\\
&6 \hat{\gamma}_{0, tt}) \hat{U}_{0, \theta}+8 (-3 \hat{\gamma}_{0, t} ^3+16 \hat{\beta}_{0, t} ^2 \hat{\gamma}_{0, t} +\hat{U}_{0, t \theta}  \hat{\gamma}_{0, t} -4 \hat{\beta}_{0, tt} \hat{\gamma}_{0, t} +\\
&\hat{U}_{0, \theta \theta} \hat{U}_{0, t} -6 \hat{\beta}_{0, t}  \hat{U}_{0, t \theta} +\hat{U}_{0, t}  (6 (\cot (\theta )-2 \hat{\gamma}_{0, \theta}) \hat{\beta}_{0, t} +\\
&(\cot (\theta )-4 \hat{\beta}_{0, \theta}) \hat{\gamma}_{0, t} +6 \hat{\gamma}_{0, t \theta} )-\cot (\theta ) \hat{U}_{0, tt}+2 \hat{\gamma}_{0, \theta} \hat{U}_{0, tt}-12 \hat{\beta}_{0, t}  \hat{\gamma}_{0, tt}+\hat{U}_{0, tt \theta}+\\
&2 \hat{\gamma}_{0, ttt}))-4 e^{4 \hat{\beta}_{0}} \{-2 (4 (\csc ^2(\theta )+4 \hat{\beta}_{0, \theta}^2-\hat{\gamma}_{0, \theta}^2-2 \cot (\theta ) \hat{\beta}_{0, \theta}+\\
&\cot (\theta ) \hat{\gamma}_{0, \theta}+2 \hat{\beta}_{0, \theta \theta}+\hat{\gamma}_{0, \theta \theta}) \hat{\gamma}_{0, t} +2 (\cot (\theta )-4 \hat{\beta}_{0, \theta}) \hat{\beta}_{0, t \theta} -\\
&\cot (\theta ) \hat{\gamma}_{0, t \theta} +2 \hat{\gamma}_{0, \theta} \hat{\gamma}_{0, t \theta} -2 \hat{\beta}_{0, t \theta \theta}-\hat{\gamma}_{0, t \theta \theta})\}]
\end{split}
\end{align}
where all of the metric coefficients are functions of $(t, \theta)$ (as indicated by the hats over the functions). 

\section{Logarithmic terms in the presence of matter} \label{logs_appendix}

In this appendix we explore how matter can affect the asymptotic expansions, inducing logarithmic terms that are related to conformal anomalies. The latter is well-understood within the context of holography (see \cite{deHaro:2000vlm}). When logarithmic terms in the asymptotic solutions appear then the on-shell gravitational action also has closely logarithmic divergences\footnote{The logarithmic terms both in the on-shell action and the asymptotic solution are local functions of the fields specifying the boundary conditions for gravity coupled to matter. The logarithmic term in the asymptotic solution of a given field is given by the functional derivative of the on-shell logarithmic term w.r.t. the corresponding boundary condition \cite{deHaro:2000vlm}.}. The presence of such divergences implies that the theory depends not only on the conformal class fixed at the conformal boundary but also on the specific representative picked: there is a conformal anomaly. Via AdS/CFT this anomaly should match a corresponding quantum anomaly in the dual QFT (and it does \cite{Henningson:1998gx, deHaro:2000vlm}).
In the context of Bondi gauge analysis for $\Lambda \neq 0$, it was noted in \cite{Chrusciel:2016oux} that the metric functions acquire logarithmic contributions given specific fall-off conditions on the bulk stress energy tensor, and we now explain how such terms emerge.

Using the Fefferman-Graham gauge \eqref{eq: FG gauge}, the fall-off conditions on the bulk stress energy tensor that lead to logarithmic terms in the metric expansions in \cite{Chrusciel:2016oux} are
\begin{equation}
{\cal T}_{\rho \rho} \sim \rho;
\qquad
{\cal T}_{ab} \sim \rho.
\end{equation}
This can be understood easily from the Einstein equations in this gauge. The $(\rho \rho)$ equation is
\begin{equation}
- \frac{\rho}{4} {\rm Tr} (g^{-1} g_{,\rho} )^2 + \frac{\rho}{2} {\rm Tr} (g^{-1} g_{,\rho \rho}) - \frac{1}{2} {\rm Tr} (g^{-1} g_{,\rho}) = \rho  \bar{\cal T}_{\rho \rho} \label{Fgrr}
\end{equation}
where $\bar{\cal T}_{\mu \nu}$ is the trace adjusted bulk stress tensor and the subscript denotes a derivative; the trace is over the 
indices $(a,b)$. The $(a b)$ equations are 
\begin{eqnarray}
 && - \frac{1}{2} {\rm Tr} (g^{-1} g_{,\rho}) g_{ab} - (g_{ab})_{,\rho}  \label{Fgab} \\
 && + \rho \left ( \frac{1}{2} (g_{ab})_{,\rho \rho} -  \frac{1}{2} (g_{,\rho} g^{-1} g_{,\rho})_{ab} 
 - \hat{R}_{ab}  + - \frac{1}{4} {\rm Tr} (g^{-1} g_{,\rho} ) (g_{ab})_{,\rho} \right )   =  \rho \bar{\cal T}_{ab},  \nonumber
\end{eqnarray}
where $\hat{R}_{ab}$ is the Ricci curvature of $g_{ab}$. We do not give the $(\rho a)$ equations as we will not need them below. 

In the absence of a bulk stress tensor these equations admit asymptotic solutions with 
\begin{equation}
g_{ab} = g_{(0) ab} + g_{(2) ab} \rho^2 + g_{(3) ab} \rho^3 + \cdots 
\end{equation}
where $g_{(2)}$ is determined by the curvature of $g_{(0)}$ and $g_{(3)}$ is traceless and divergenceless. (The tracelessness of $g_{(3)}$ follows from differentiating \eqref{Fgrr} and \eqref{Fgab} with respect to $\rho$ and then setting $\rho \rightarrow 0$.)  If we now impose the falloff  conditions above:
\begin{equation}
\bar{\cal T}_{\rho \rho} = \bar{\cal T}_{(1) \rho \rho} \rho + \cdots  \qquad
\bar{\cal T}_{a b } = \bar{\cal T}_{(1) ab} \rho + \cdots \label{falloff-c}
\end{equation}
then the asymptotic expansion is modified to 
\begin{equation}
g_{ab} = g_{(0) ab} + g_{(2) ab} \rho^2 + ( g_{(3) ab} + h_{(3) ab} \log \rho) \rho^3 + \cdots 
\end{equation}
with 
\begin{equation}
{\rm Tr} (g_{(0)}^{-1} h_{(3)} )= 0; \qquad {\rm Tr} (g_{(0)}^{-1} g_{(3)} ) = \frac{2}{3} \bar{\cal T}_{(1) \rho \rho}
\end{equation}
and
\begin{equation}
h_{(3) ab} = \frac{2}{3} \bar{\cal T}_{(1) \rho \rho} g_{(0) ab} + \frac{2}{3} \bar{\cal T}_{(1) ab}.
\end{equation}
Note that self consistency requires that 
\begin{equation}
\bar{\cal T}_{(1) \rho \rho} + \frac{1}{3} g_{(0)}^{ab} \bar{\cal T}_{(1) ab} = 0. 
\end{equation}
The $(\rho a)$ equations determine the divergence of $g_{(3)}$ and $h_{(3)}$; apart from the trace and divergence constraints, $g_{(3)}$ remains undetermined by the field equations and describes the energy momentum tensor of the dual theory. 

Thus the falloff conditions \eqref{falloff-c} imposed on the bulk stress tensor induce logarithmic terms in the asymptotic expansion, along with non-zero trace and divergence of $g_{(3)}$. Such effects are associated with conformal anomalies. 

An explicit example of bulk matter that induces such a conformal anomaly is the following. Consider a bulk scalar field $\phi$ of mass $m^2 = -2$, corresponding to a scalar operator of dimension two in the conformal field theory, and let the field have a cubic interaction i.e. the field equation is
\begin{equation}
( \Box + 2 ) \phi = \lambda \phi^2
\end{equation}
where $\lambda$ is the cubic coupling. The asymptotic expansion of the field $\phi$ is of the form
\begin{equation}
\phi = \phi_{(1)} \rho + \cdots 
\end{equation}
where $\phi_{(1)}(x)$ is the source for the dual operator in the field theory. The cubic interaction induces terms of the form \eqref{falloff-c} in the bulk stress tensor, and hence logarithmic terms $h_{3)}$ and non-zero trace and divergence of $g_{(3)}$. These are associated with a conformal anomaly in the dual stress energy tensor of the form
\begin{equation}
g_{(0)}^{ab} \langle T_{ab} \rangle \sim \lambda \phi_{(1)}^3.
\end{equation}
It follows that there is a conformal anomaly associated with the 3-point function of the operator of dimension 2, in agreement with the QFT analysis in \cite{Bzowski:2015pba}. 

\section{Equivalence of Bondi and Abbott-Deser masses in asymptotically AdS spacetimes} \label{AD_mass_appendix}

In this appendix we will show that our candidate for the Bondi mass (\ref{eq: mass_formula}) agrees with the well-known Abbott-Deser mass \cite{Abbott:1981ff} in asymptotically AdS spacetime. We recall that the Abbott-Deser mass is defined relative to a reference background spacetimes which
for asymptotically AdS spacetimes is taken to be pure AdS. Specifically, we write the spacetime metric $g_{\mu \nu}$ as 
\begin{equation}
g_{\mu \nu}=\bar{g}_{\mu \nu}+h_{\mu \nu} 
\end{equation} 
where $\bar{g}_{\mu \nu}$ is the metric of pure AdS$_{4}$ and $h_{\mu \nu}$ is a perturbation chosen such that $g_{\mu \nu}$ solves (\ref{eq: EE_lambda}) and $h_{\mu \nu}$ vanishes at $\mathscr{I}$. Note that the vanishing condition at $\mathscr{I}$ ensures that $g_{\mu \nu}$ is asymptotically AdS as it has the same conformal structure induced at $\mathscr{I}$ as $\bar{g}_{\mu \nu}$, the metric for pure AdS$_{4}$. In this appendix we will restrict our attention to  $\Lambda < 0$ and $T_{\mu \nu}=0$. We will use the normalisation of $l=1$ ($\Lambda =- 3$) which can of course be reintroduced via dimensional analysis. 

We first recall the definition of the Abbott-Deser energy-momentum for asymptotically AdS spacetimes as given in \cite{Abbott:1981ff} (using units where $G=1$):
\begin{equation} \label{eq: AD_mass}
E[\bar{\xi}]= \frac{1}{8\pi } \lim_{S_a \rightarrow \mathscr{I}_a} \oint dS_a \sqrt{-\bar{g}} [ \bar{D}_{\beta} K^{t a \nu \beta} -  K^{t b \nu a}\bar{D}_b]\bar{\xi}_{\nu} 
\end{equation}
where the integral is taken over a spacelike 2-surface at the conformal boundary $\mathscr{I}$. $\bar{\xi}$ is a Killing vector associated with the background metric $\bar{g}_{\mu \nu}$ (which is also used to raise and lower indices) and $\bar{D}_{\mu}$ its associated covariant derivative operator. In the equation above we continue to use the convention that Greek indices $\beta, \nu$ run over all spacetime values and Roman indices $a, b$ over spatial values (the index $t$ is of course the time coordinate). The rank four tensor $K$ is known as the \textit{superpotential} and is given by 
\begin{equation} \label{eq: superpotential}
K^{\mu \alpha \nu \beta} = \frac{1}{2}[\bar{g}^{\nu \beta} H^{\nu \alpha}+\bar{g}^{\nu \alpha} H^{\mu \beta}-\bar{g}^{\mu \nu} H^{\alpha \beta}-\bar{g}^{\alpha \beta} H^{\mu \nu}]
\end{equation}
where
\begin{equation} \label{eq: superpotential_H}
H^{\mu \nu}=h^{\mu \nu}-\frac{1}{2} g^{\mu \nu} h^{\alpha}_{\hspace{0.5em} \alpha}.
\end{equation}

In order to compute the Abbott-Deser mass we follow the prescription of \cite{Abbott:1981ff} and evaluate (\ref{eq: AD_mass}) when $\bar{\xi}$ is a timelike Killing vector, namely 
\begin{equation}
\bar{\xi}^{\mu}=-\left(\frac{\partial}{\partial t}\right)^{\mu}=(-1,0).
\end{equation}
To evaluate this integrand (and to make connection with our earlier discussion of the Bondi mass) we will work in the Fefferman-Graham gauge. We note that we have 
\begin{equation} \label{eq: AdS_in_FG}
\bar{g}_{\mu \nu} dx^{\mu} dx^{\nu} = \frac{d\rho^2}{\rho^2}+\frac{1}{\rho^2}\left(g_{(0)ab}+\rho^2 g_{(2)ab}+\rho^4 g_{(4)ab} \right)dx^a dx^b
\end{equation}
where the terms in the expansion on the RHS have the line-elements
\begin{align} \label{eq: AdS_FG_terms}
\begin{split}
ds_{(0)}^2 &=-dt^2+d\Omega^2 \\
ds_{(2)}^2 &=\frac{1}{2}(-dt^2-d\Omega^2) \\
ds_{(4)}^2 &=\frac{1}{16}(-dt^2+d\Omega^2)
\end{split}
\end{align} 
($g_{(0)}$ and $g_{(2)}$ were already given in equations (\ref{Ein_Un}) and  (\ref{eq: pure_AdS_g_2}) respectively). Enforcing the requirements that $g_{\mu \nu}$ solves the field equations and $h_{\mu \nu}$ vanishes at $\mathscr{I}$, the most general form for $h_{\mu \nu}$ is 
\begin{equation} \label{eq: FG_Perturbation}
h_{ab}dx^{a}dx^{b} = \rho g_{(3)ab}dx^a dx^b + \mathcal{O}(\rho^2)  
\end{equation}
where $g_{(3)}$ is given by (\ref{eq: g_3_asym_AdS}) and we note that the Fefferman-Graham gauge forces $h_{\rho \mu}=0$. The higher order terms do not contribute to the Abbott-Deser mass (they vanish in the limit to $\mathscr{I}$), so we focus on the $g_{(3)}$ term.

With the coordinates, timelike Killing vector and perturbation specified, we are ready to compute the Abbott-Deser mass. In Fefferman-Graham coordinates the limit in (\ref{eq: AD_mass}) simply becomes $\rho \rightarrow 0$ (recall $\mathscr{I}=\{\rho=0\}$) and we can apply formulae (\ref{eq: AdS_in_FG})-(\ref{eq: AdS_FG_terms}) for the background metric and (\ref{eq: FG_Perturbation}) for the perturbation in order to write the superpotential (\ref{eq: superpotential}) and thus the Abbott-Deser mass. Explicitly the Abbott-Deser mass, $\mathcal{M}_{AD}$, is given by
\begin{equation} \label{eq: AD_mass_2}
\mathcal{M}_{AD} = \frac{1}{8\pi} \lim_{S_a \rightarrow \mathscr{I}_a} \oint dS_a m^a 
\end{equation}
with 
\begin{equation}
m^a =  \sqrt{-\bar{g}} [ \bar{D}_{\beta} K^{t a \nu \beta} -  K^{t b \nu a}\bar{D}_b]\bar{\xi}_{\nu}.
\end{equation}
Given that we are working in the Fefferman-Graham gauge, the only component which we will need in order to compute $\mathcal{M}_{AD}$ is $m^{\rho}$:
\begin{equation}
m^{\rho} = \frac{\left(\rho ^4+48\right) \sin \theta  \hat{W}_3(t,\theta )}{12 \left(\rho ^2-4\right)}
\left(1+ \mathcal{O}(\rho)\right) 
\end{equation}
Taking the limit to $\mathscr{I}$ gives 
\begin{align}
\begin{split}
\mathcal{M}_{AD}&=\frac{1}{8\pi} \lim_{\rho \rightarrow 0} \oint dS_{\rho} m^{\rho} \\
&=- \frac{1}{8\pi} \int_0^{2\pi} d\phi \int_{0}^{\pi} d\theta \, \hat{W}_3(t, \theta) \sin \theta \\
&=\frac{1}{4\pi} \int_{S^2} m_B (t, \theta)= \mathcal{M}_B 
\end{split}
\end{align} 
where in going from the second to the third line we have used the relationship (\ref{eq: g_3_tt_asym_AdS}) to rewrite the integral in terms of the Bondi mass aspect. Thus we have shown that the Bondi and Abbott-Deser masses are the same for asymptotically AdS spacetimes.
\bibliography{Master_Bibliography}{}
\bibliographystyle{JHEP}
\end{document}